\documentclass[letterpaper]{aa} % for a referee version use [referee]
\usepackage{graphicx}
\usepackage{txfonts}
\usepackage{aalongtable}
\def\muas {\hbox{$\mu$as}}
\def\Ftwo {\hbox{$F2$}}
\def\Pchi {\hbox{$P(\chi^2)$}}
 
\def\PF {\hbox{$P(F)$}}
\begin{document}
\title{Double-blind test program for astrometric 
planet detection with Gaia
%\thanks{Based on observations made at the Observatoire de 
%Haute-Provence (CNRS, France)
%        and at the W. M Keck Observatory, 
%which is operated as a scientific partnership
%        among the California Institute of Technology, 
%the University of California and
%        the National Aeronautics and Space Administration. 
%The Observatory was made
%        possible by the generous financial support from 
%the W. M. Keck Foundation.}
}

%   \subtitle{I. Overviewing the $\kappa$-mechanism}

\author{S. Casertano\inst{1} \and M. G. Lattanzi\inst{2} \and 
A. Sozzetti\inst{2,3} \and A. Spagna\inst{2} \and S. Jancart\inst{4} \and 
R. Morbidelli\inst{2} \and R. Pannunzio\inst{2} \and D. Pourbaix\inst{4}  
\and D. Queloz\inst{5}}

\offprints{A. Sozzetti, \\\email{sozzetti@oato.inaf.it}}

\institute{Space Telescope Science Institute, 3700 San Martin Drive, 
     Baltimore, MD 21218, USA 
\and INAF - Osservatorio Astronomico di Torino, via Osservatorio 20, 
     10025 Pino Torinese, Italy
\and Harvard-Smithsonian Center for Astrophysics, 60 Garden Street, 
     Cambridge, MA 02138, USA
\and Institut d'Astronomie et d'Astrophysique, 
     Universit\'e Libre de Bruxelles, CP. 226, Boulevard du Triomphe, 
     1050 Bruxelles, Belgium
\and Observatoire de Gen\`eve, 51 Ch. de Maillettes, 1290 Sauveny, 
     Switzerland}

\date{Received ......; accepted ......}

  \abstract
    {}
   {The scope of this paper is twofold. First, it describes the simulation scenarios 
   and the results of a large-scale, double-blind test campaign carried out 
   to estimate the potential of Gaia for detecting and measuring planetary systems. 
   The identified capabilities are then put in context by highlighting the 
   unique contribution that the Gaia exoplanet discoveries will be able to bring to 
   the science of extrasolar planets in the next decade.}
   {We use detailed simulations of the Gaia observations of synthetic planetary systems 
   and develop and utilize independent software codes in double-blind mode to analyze 
   the data, including statistical tools for planet detection and different algorithms for 
   single and multiple Keplerian orbit fitting that use no a priori knowledge of the 
   true orbital parameters of the systems.}
   {1) Planets with astrometric signatures $\alpha\simeq 3$ times the assumed 
   single-measurement error $\sigma_\psi$ and period $P\leq 5$ yr 
   can be detected reliably and consistently, with a very small number of false positives. 
   2) At twice the detection limit, uncertainties in orbital parameters and masses are 
   typically $15\%-20\%$. 3) Over 70\% of two-planet systems with well-separated periods 
   in the range $0.2\leq P\leq 9$ yr, astrometric signal-to-noise ratio 
   $2\leq\alpha/\sigma_\psi\leq 50$, and eccentricity $e\leq 0.6$ are correctly identified. 
   4) Favorable orbital configurations (both planets with $P\leq 4$ yr and $\alpha/\sigma_\psi\geq 10$, 
   redundancy over a factor of 2 in the number of observations) 
   have orbital elements measured to better than 10\% accuracy $> 90\%$ 
   of the time, and the value of the mutual inclination 
   angle $i_\mathrm{rel}$ determined with uncertainties $\leq 10^{\degr}$. 
   5) Finally, nominal uncertainties obtained from the 
   fitting procedures are a good estimate of the actual errors in the orbit reconstruction. 
   Extrapolating from the present-day statistical properties of the exoplanet sample, the results imply 
   that a Gaia with $\sigma_\psi$ = 8 $\mu$as, in its unbiased and complete magnitude-limited 
   census of planetary systems, will discover and 
   measure several thousands of giant planets out to 3-4 AUs from stars within 200 pc, and 
   will characterize hundreds of multiple-planet systems, including meaningful coplanarity tests. 
   Finally, we put Gaia's planet discovery potential into context, identifying several areas of 
   planetary-system science (statistical properties and correlations, 
   comparisons with predictions from theoretical models of formation 
   and evolution, interpretation of direct detections) in which Gaia can be expected, 
   on the basis of our results, to have a relevant impact, 
   when combined with data coming from other ongoing and future planet search programs.}
   {}


\titlerunning{DBT Campaign for Planet Detection with Gaia}
\authorrunning{S. Casertano et al.}
\maketitle

%
%________________________________________________________________

\section{Introduction}

The continuously increasing catalog of extrasolar planets is 
today surpassing the threshold of 270 planets announced
\footnote{See, for example, Jean Schneider's Extrasolar Planet 
Encyclopedia at http://exoplanet.eu/}. 
Most of the nearby ($d \lesssim 200-300$ pc) exoplanet candidates have 
been detected around F-G-K-M dwarfs by long-term, 
high-precision ($1-5$ m s$^{-1}$) Doppler search programs (e.g., Butler et al. 2006, 
and references therein; Udry et al. 2007, and references therein). Over a dozen 
of these are `hot Jupiters' with orbital periods $P\simeq 1-20$ days discovered to 
cross the disk of their relatively bright ($V < 13$) parent stars thanks to 
high-cadence, milli-mag photometric measurements
\footnote{For a review, see Charbonneau et al. 2007, and references therein. 
For an updated list, see for example obswww.unige.ch/~pont/TRANSITS.htm, and references 
therein}. An additional dozen or so extrasolar planets have been found 
residing at $d > 300$ pc, thanks to both transit photometry 
(e.g., Konacki et al. 2003, 2005; Bouchy et al. 2004; Pont et al. 2004, 2007; 
Udalski et al. 2007; Collier Cameron et al. 2006; Mandushev et al. 2007; 
Kov\'acs et al. 2007; Bakos et al. 2007) 
as well as microlensing  surveys in the Galactic bulge 
(e.g., Bond et al. 2004; Udalski et al. 2005; Gould et al. 2006; Bealieu et al. 2006). 

The sample of nearby exoplanets and their host stars is amenable to 
follow-up studies with a variety of indirect and direct techniques, 
such as high-resolution (visible-light and infrared) imaging and 
stellar spectroscopy, and photometric transit timing (for a review, 
see for example Charbonneau et al. 2007, and references therein). 
Milli-arcsecond (mas) astrometry for bright planet hosts within 200-300 pc 
provides precise distance estimates thanks to Hipparcos 
parallaxes (Perryman et al. 1997). However, despite a few important 
successes (Benedict et al. 2002, 2006; McArthur et al. 2004; Bean et al. 2007), 
astrometric measurements with mas precision have so far
proved of limited utility when employed as either a follow-up tool 
or to independently search for planetary mass companions orbiting 
nearby stars (for a review, see for example Sozzetti 2005, and 
references therein). 

In several past exploratory works (Casertano et al. 1996; Casertano \& Sozzetti 1999; 
Lattanzi et al. 1997, 2000a, 2000b, 2002; 
Sozzetti et al 2001, 2002, 2003a, 2003b), we have shown in some detail what 
space-borne astrometric observatories with micro-arcsecond ($\mu$as)-level precision, 
such as Gaia (Perryman et al. 2001) and SIM PlanetQuest (Unwin et al. 2007), 
can achieve in terms of
search, detection and measurement of extrasolar planets of mass 
ranging from Jupiter-like to Earth-like. 
In those studies we adopted a qualitatively 
correct description of the
measurements that each mission will carry out, and we estimated 
detection probabilities and orbital parameters using realistic, non-linear least-square 
fits to those measurements. For Gaia, we used
the then-current scanning law and error model; for SIM, we 
included reference stars, as well as the
target, and adopted realistic observational overheads and 
signal-to-noise estimates as provided by 
the SIM Project. Other, more recent studies (Ford \& Tremaine 2003; 
Ford 2004, 2006; Catanzarite et al. 2006) have focused on 
evaluating the potential of astrometric planet searches with SIM, 
revisiting and essentially confirming the findings of our previous works.

Although valid and useful, the studies currently available need
updating and improving. In the specific case of planet detection 
and measurement with Gaia, we have thus far largely neglected the
difficult problem of selecting adequate starting values for the
non-linear fits, using perturbed starting values instead.  The
study of multiple-planet systems, and in particular the determination
of whether the planets are coplanar---within suitable tolerances---is
incomplete.  The characteristics of Gaia have changed, in some ways
substantially, since our last work on the subject (Sozzetti et al 2003a).  
Last but not least, in order to render the analysis truly independent 
from the simulations, these studies should be carried out in 
double-blind mode. 

We present here a substantial program of double-blind tests for 
planet detection with Gaia (preliminary findings were recently 
presented by Lattanzi et al. (2005)). The results expected from this study include: 
a) an improved, more realistic assessment of the detectability
and measurability of single and multiple planets under a variety of
conditions, parametrized by the sensitivity of Gaia; 
b) an assessment of the impact of Gaia in critical areas of
planet research, in dependence on its expected capabilities; 
and c) the establishment of several Centers with a 
high level of readiness for the analysis of Gaia observations 
relevant to the study of exoplanets.

This paper is arranged as follows. In Section 2 we describe 
our simulation setup and clearly state the working assumptions 
adopted (the relaxation of some of which might have a non-negligible 
impact on Gaia's planet-hunting capabilities). In Section 3 
we present the bulk of the results obtained during our extensive 
campaign of double-blind tests. Section 4 attempts to put Gaia's 
potential for planet detection and measurement in context, by 
identifying several areas of planetary science  in which 
Gaia can be expected, on the basis of 
our results, to have a dominant impact, and by delineating a 
small number of recommended research 
programs that can be conducted successfully by the mission as planned. 
In section 5 we summarize our findings and provide concluding remarks.

\section{Protocol definition and simulation setup}

\subsection{Double-blind tests protocol}

For the purpose of this study, we have devised a specific protocol for 
the double-blind tests campaign. Initially, three groups of participants 
are identified: 
1) the {\it Simulators} define and generate the simulated observations, 
assuming specific characteristics of the Gaia satellite;
simulators also define the type of results that are expected for each set
of simulations; 2) the {\it Solvers} receive the simulated data and 
produce ``solutions''---as defined by the simulators; solvers define 
the criteria they adopt in answering the questions posed by the simulators; 
3) the {\it Evaluators} receive both the ``truth''---i.e.,
the input parameters---from the simulators and the solutions from the
solvers, compare the two, and draw a set of conclusions on the
process. 

A sequence of tasks, each with well-defined goals and
time scales, has been established. Each task requires a separate set of 
simulations, and is carried out in several steps:

\begin {enumerate}

\item {Simulation:} The Simulators make the required set of simulations
available to the Solvers, together with a clear definition of the
required solutions.

\item {Clarification:} A short period after the
simulations are made available in which the Solvers request any
necessary clarification on the simulations themselves and on the
required solutions; after the clarification period, there is no
contact between Simulators and Solvers until the Discussion step.

\item {Delivery:} On a specified deadline, the Simulators deliver the
input parameters for the simulations to the Evaluator, and the Solvers
deliver their solutions together with a clear explanation of the
criteria they used---e.g., the statistical meaning of ``detection'',
or how uncertainties on estimated parameters were defined.

\item {Evaluation:} The Evaluators compare input parameters and solutions
and carry out any statistical tests they find useful, both to establish
the quality of the solutions and to interpret their results in terms
of the capabilities of Gaia, if applicable.

\item {Discussion:} The Evaluators publicize their initial results.
All participants are given access to input parameters and all solutions, 
and the Evaluators' results are discussed and modified as needed.

\end {enumerate}

\subsection{Observing scenario}

The simulations were provided by the group at the Torino 
Observatory. The simulations were made available via WWW 
as plain text files. A detailed description of the code for 
the simulation of Gaia astrometric observations 
can be found in our previous exploratory works (Lattanzi et al. 2000a; 
Sozzetti et al. 2001). We summarize here its main features. 

Each simulation consists of a time series of observations 
(with a nominal mission lifetime set to 5 years) of 
a sample of stars with given (catalog) low-accuracy astrometric 
parameters (positions, proper motions, and parallax). 
Each observation consists of a one-dimensional 
coordinate in the along-scan direction of the instantaneous 
great circle followed by Gaia at that instant. The initially 
unperturbed photocenter position of a star is computed on 
the basis of its five astrometric parameters, 
which are drawn from simple distributions, not resembling any 
specific galaxy model. The distribution of two-dimensional positions is random,
uniform. The distribution of proper motions is Gaussian, with dispersion equal to a value
of transverse velocity $V_T = 15$ km sec$^{-1}$, typical of the solar neighborhood.
The photocenter position can then be corrected for the 
gravitational perturbation of one or more planetary mass 
companions. The Keplerian motion of each orbiting planet is 
described via the full set of seven orbital elements. For 
simplicity, all experiments described here were produced assuming 
stellar mass $M_\star=1$ $M_\odot$. The resulting astrometric 
signature (in arcsec) is then expressed as 
$\alpha=(M_p/M_\star)\times(a_p/d)$, where $M_p$ is the planet 
mass, $a_p$ is the planet orbital semi-major axis and $d$ is 
the distance to the system (in units of $M_\odot$, AU, and pc, 
respectively). Simulated observations are generated by adding the appropriate 
astrometric noise, as described in the next section. 

Finally, the Gaia scanning law 
has been updated to the most recent expectations 
(precession angle around the Sun direction $\xi=50^{\degr}$, precession 
speed of the satellite's spin axis $v_p=5.22$ rev yr$^{-1}$, 
spin axis rotation speed $v_r=60$ arcsec s$^{-1}$), 
which result in fewer observations 
and possibly less ability to disentangle near-degenerate 
solutions than with the original scanning law (e.g., Lindegren \& de Bruijne 2005). 
Each star is observed on average $N_\mathrm{obs} = 43$ times
\footnote{We define as elementary observation the successive 
crossing of the two fields-of-view of the satellite, separated 
by the basic angle $\gamma = 106.5^{\degr}$.}; note that the simplest 
star+planet solution has 12 parameters, and therefore the 
redundancy of the information is not very high. 

\subsection{Assumptions and caveats}

The simulated data described above indicate that a number of 
working assumptions have been made. In particular, a variety 
of physical effects that can affect stellar positions have 
not been included, and a number of instrumental 
as well as astrophysical noise sources have not been considered 
(for a detailed review, see for example Sozzetti 2005). 
Our main simplifying assumptions are summarized below.
%\footnote{}
 
\begin{itemize}

\item[1)] the position of a star at a 
given time is described in Euclidean space. A general 
relativistic formulation of Gaia-like global astrometric 
observations, which has been the subject of several 
studies in the recent past (Klioner \& Kopeikin 1992; 
de Felice et al. 1998, 2001, 2004; Vecchiato et al. 2003; 
Klioner 2003, 2004), has not been taken into account;

\item[2)] we assume that the reconstruction and calibration of 
individual great circles have been carried out, 
with nominal zero errors (i.e., knowledge of the spacecraft 
attitude is assumed perfect). We refer the reader to e.g. Sozzetti (2005), 
and references therein, for a summary of issues related to the complex problem of 
accurately calibrating out attitude errors (due to, e.g., particle radiation, 
thermal drifts, and spacecraft jitter) in space-borne astrometric measurements;

\item[3)] the abscissa is only 
affected by random errors, and no systematic effects are 
considered (e.g., zero-point errors, chromaticity, radiation 
damage, etc...). 
A simple Gaussian measurement error model is implemented, 
with standard deviation $\sigma_\psi = 8$ $\mu$as, which 
applies to bright targets ($V < 13$). In this context, 
the projected end-of-mission accuracy on astrometric parameters 
is 4 $\mu$as. Recently, Gaia has successfully passed the 
Preliminary Design Review and entered phase C/D of the mission 
development. ESA has selected EADS-Astrium as Prime Contractor 
for the realization of the satellite. Scanning law and astrometric 
section of the selected payload, the only of relevance here, 
remain largely consistent with the assumptions adopted in our 
simulations. However, very recent estimates of the error budget indicate 
a possible degradation of $35\%-40\%$ (i.e., $\sim 11$ $\mu$as) 
in the single-measurement precision, corresponding to a typical final 
accuracy of $5-5.5$ $\mu$as for objects in the above 
magnitude range, with some dependence on spectral type (red 
objects performing closer to specifications). We will address 
in the discussion section the possible impact of such performance 
degradation on Gaia's planet-hunting capabilities;

\item[4)] light aberration, light deflection, and other 
apparent effects (e.g., perspective acceleration) are 
considered as perfectly removed from the observed 
along-scan measurements; 

\item[5)] when multi-component systems are generated around a 
target, the resulting astrometric signal is the superposition of 
two strictly non-interacting Keplerian orbits. It is 
recognized that gravitational interactions among planets 
can result in significant deviations from purely Keplerian 
motion (such as the case of the GJ 876 planetary system, 
e.g. Laughlin et al. 2005). 
However, most of the multiple-planet systems discovered 
to-date by radial-velocity techniques can be well modeled 
by planets on independent Keplerian orbits, at least 
for time-scales comparable to Gaia's expected mission 
duration; 

\item[6)] a number of potentially important 
sources of `astrophysical' noise, due to the environment or 
intrinsic to the target, have not been included in 
the simulations. In particular, we have not considered 
a) the dynamical effect induced by long-period 
stellar companions to the targets, b) the possible shifts in 
the stellar photocenter due to the presence of circumstellar 
disks with embedded planets (Rice et al. 2003a; Takeuchi et al. 2005), 
and c) variations in the apparent stellar position produced by 
surface temperature inhomogeneities, such as spots and flares 
(e.g, Sozzetti 2005, and references therein; Eriksson \& Lindegren 2007). 

\end{itemize}

The geometric model of the measurement process is described in detail in the Appendix.

\section{Results}

The double-blind test campaign encompassed a set of 
experiments that were necessary to
establish a reliable estimate of the planet search and measurement
capabilities of Gaia under realistic analysis procedures, albeit 
in the presence of an idealized measurement process. 
In particular, a number of different 
tasks were designed, such that the participating groups would be able 
a) to analyze data produced by a nominal satellite, without taking into account the
imperfections due to measurement biases, non-Gaussian error
distributions, imperfections in the sphere solution, and so on; 
b) to convert any Gaussian error model for
Gaia measurements into expected detection probability---including
completeness and false positives---and accuracy in orbital parameters
that can be achieved within the mission; c) to assess
to what extent, and with what reliability, coplanarity of multiple
planets can be determined, and how the presence of a planet can
degrade the orbital solution for another. 

We broke down the work plan into three tests: T1, T2,
and T3, whose main results are presented below. 
To facilitate reading, we have chosen to provide the summaries of 
the results concerning each of the three tests at the beginning 
of the corresponding sub-sections.

\subsection{Test T1: Planet detection}

 \begin{figure*}
\centering
$\begin{array}{cc}
\includegraphics[width=0.4\textwidth]{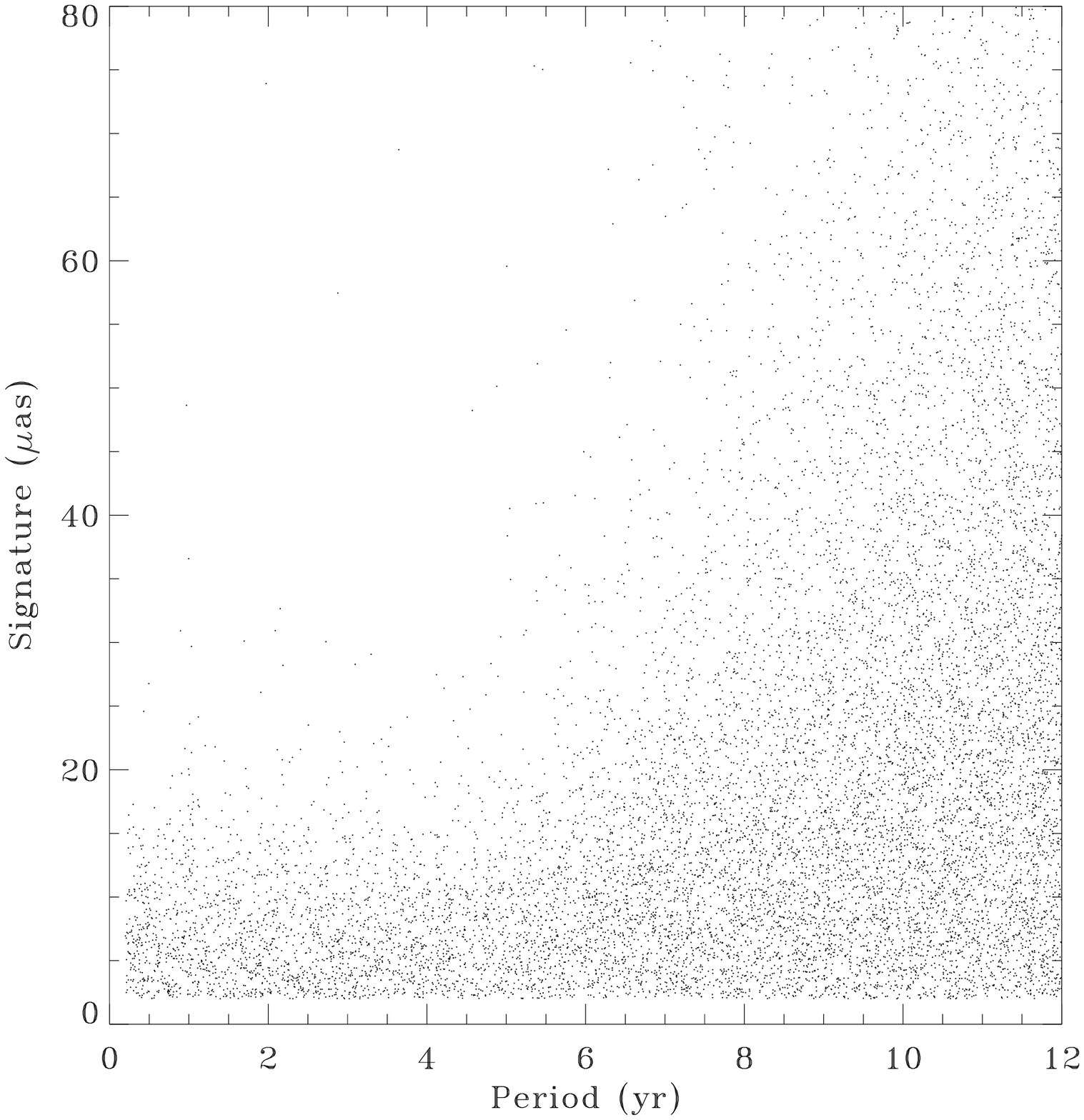} & 
\includegraphics[width=0.4\textwidth]{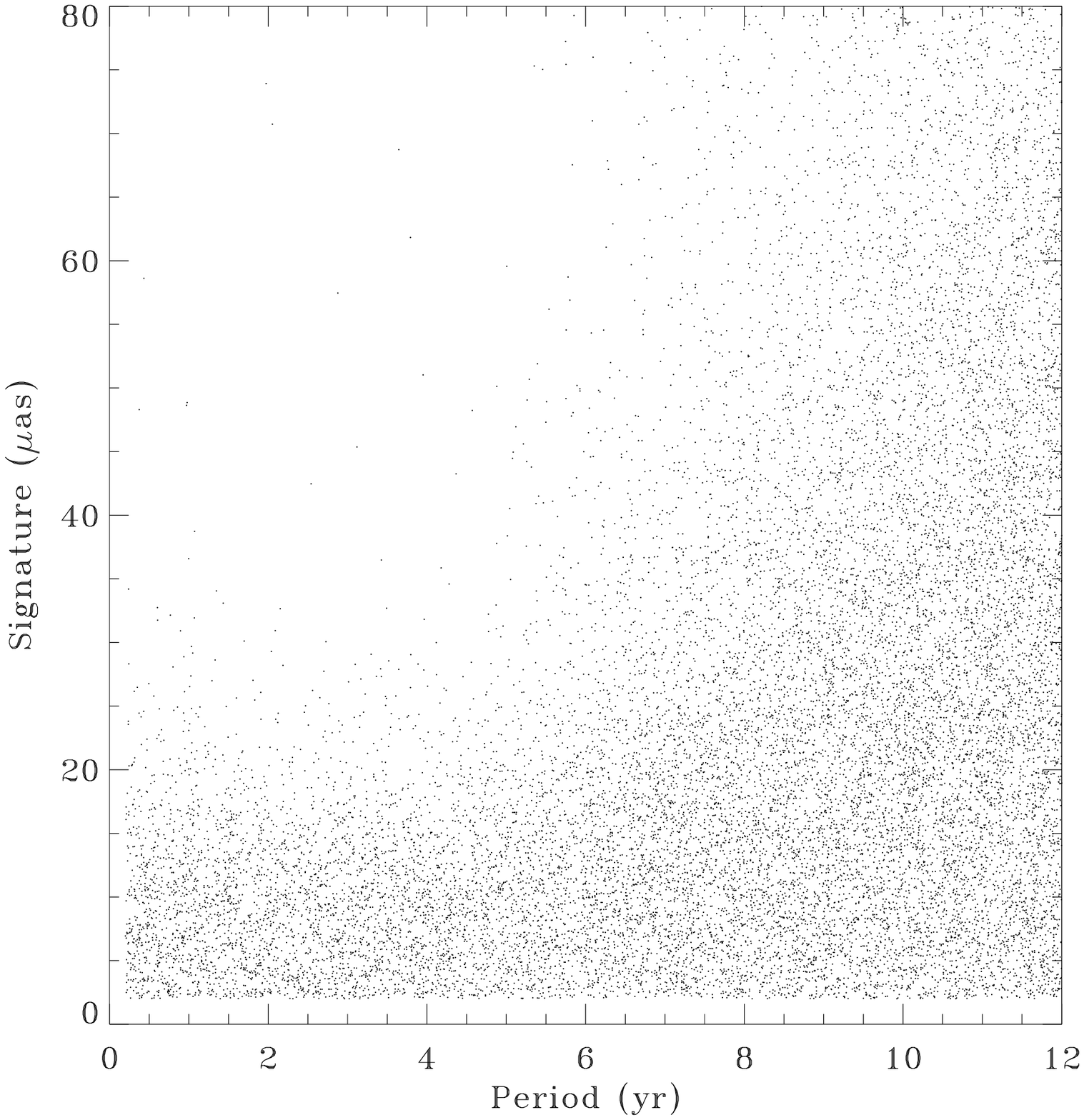} \\
\end{array} $
 \caption{Left: Distribution of period and signature for the planets missed by
Solver A's broad criterion (A1).  If more than one planet is present,
the one with the largest signature is plotted. Right: distribution of period and 
signature for the planets missed by criterion B1.}
\label{fig1}
\end{figure*}

 \begin{figure}
\centering
\includegraphics[width=0.4\textwidth]{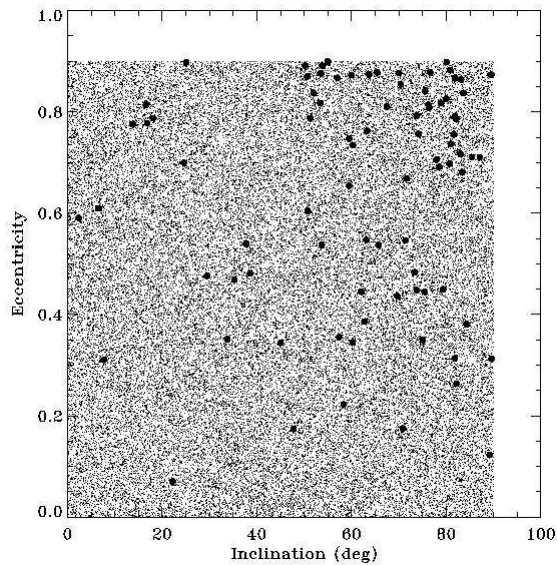}
\caption{Inclination and eccentricity of the planets simulated for the T1 experiment. 
Black dots are planets with $\alpha < 15$ $\mu$as and $P < 6$ yr not detected by 
the A1 criterion.}
\label{fig2}
 \end{figure}

Test T1 is  designed primarily to establish the reliability and
completeness of planet detections for single-planet systems based on
simulated data, with full {\it a priori} understanding of their noise
characteristics.  Simulated data were prepared for 100,000 stars. 
Of these, 45,202 have no planets, 49,870 one, 3878 two, and
1050 have three planets.  The astrometric signature of each planet
ranges from 0.8 to 80 $\mu$as ($0.1\leq\alpha/\sigma_\psi\leq 10$), 
the period $P$ from 0.2 to 12 years, while eccentricities are drawn 
from a random distribution within the range $0.0\leq e\leq 0.9$. 
All other orbital elements (inclination $i$, longitude of pericenter 
$\omega$, pericenter epoch $\tau$, and position angle of the nodes $\Omega$) 
were distributed randomly as follows: $1^{\degr}\leq i\leq 90^{\degr}$, 
$0^{\degr}\omega\leq 360^{\degr}$, $0\leq\tau\leq P$, and $0{\degr}\leq\Omega\leq 180^{\degr}$. 
For systems with multiple planets, there was no specific relationship between
periods, phases, or amplitudes of the planetary signatures. 
The distribution of planetary signatures was unknown to
the solvers. 

\begin{figure*}
\centering
$\begin{array}{cc}
\includegraphics[width=0.4\textwidth]{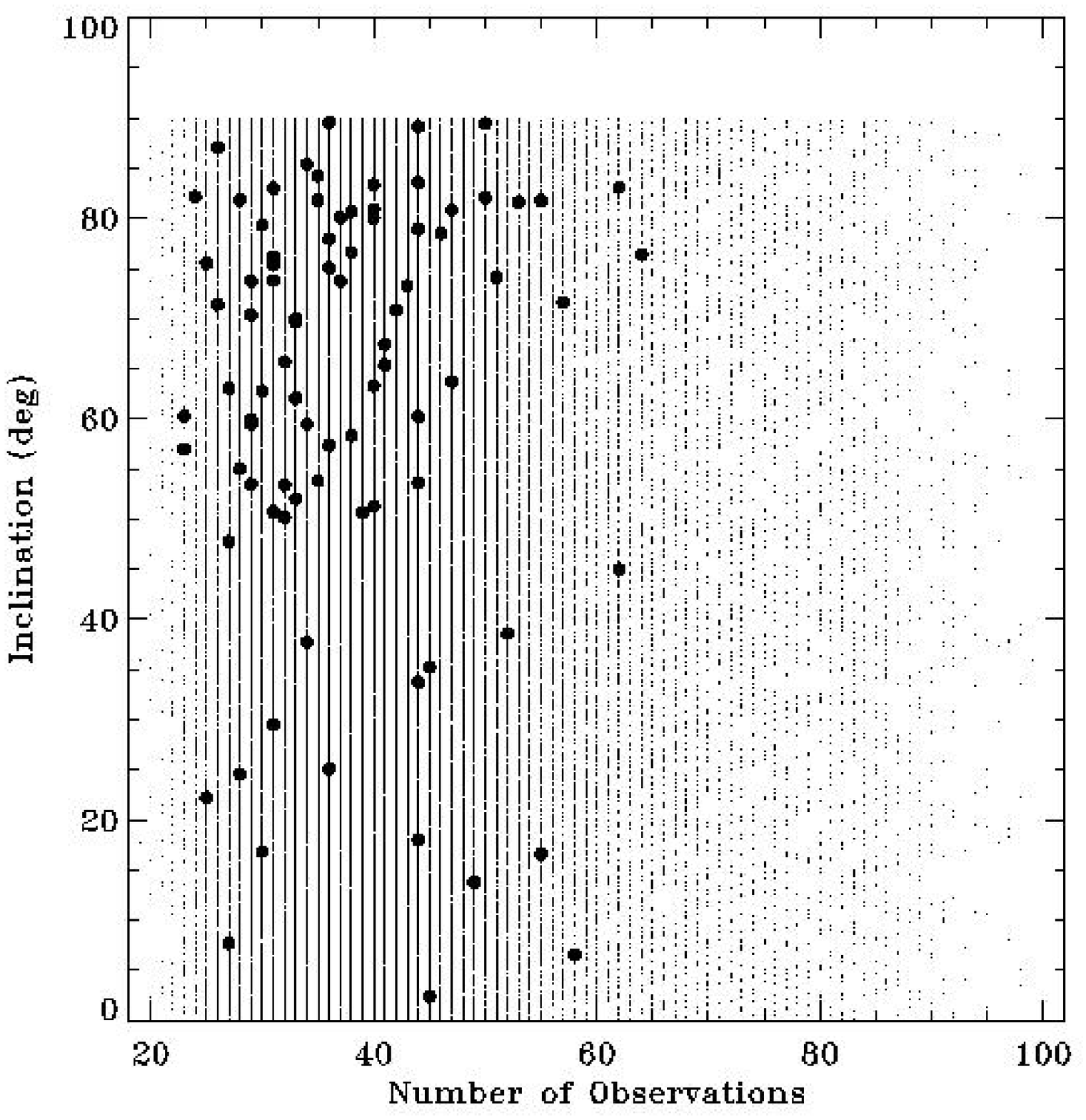} & 
\includegraphics[width=0.4\textwidth]{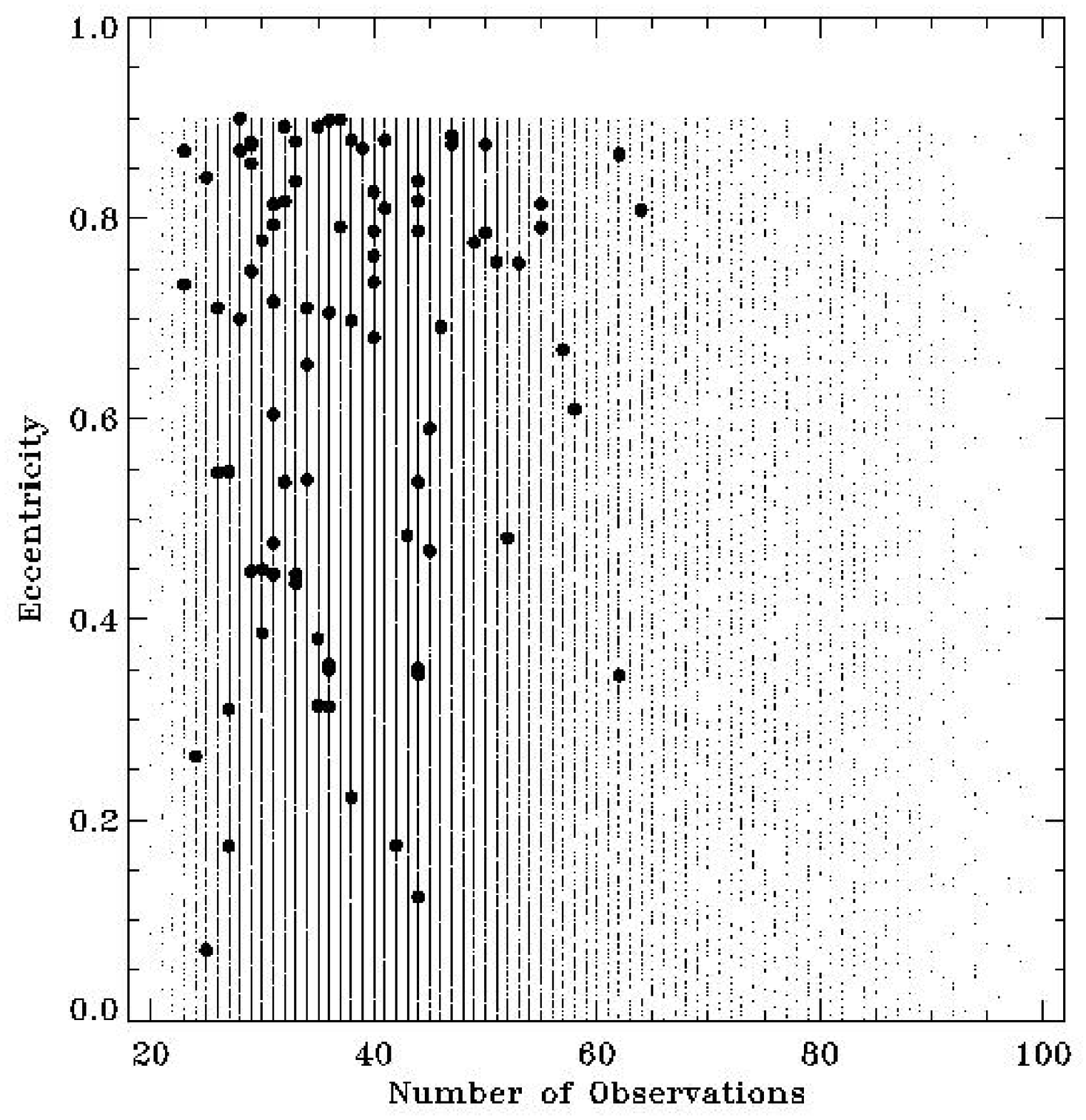} \\
\end{array} $
 \caption{Left: same as Figure~\ref{fig2}, but here 
 the results are expressed in terms of inclination angle and 
 number of observations. Right: same as Figure~\ref{fig2}, 
 in the $e$-$N_\mathrm{obs}$ plane.}
\label{fig3}
 \end{figure*}

On this dataset, solvers were asked to carry out two tests.  Test T1
consisted of identifying the likely planet detections, based on a
single-star analysis and criteria of the Solvers' own choosing.  Test
T1b gave the opportunity to the solvers to improve on their planet
detection on the basis of an orbital fit, i.e., using the knowledge that
the deviations from a single-star model were in fact expected to have the
signature of a star-planet system. 
Two Solvers participated in this step and provided completely
independent solutions. Solver A attempted to improve
the quality of planet detection using orbital fits, in the spirit of the
T1b test; Solver B was satisfied with the quality of the
detection achieved from the statistical properties of the residuals to
the single-star fit.  Although the solvers used different detection
criteria and post-processing analysis, both ultimately achieved good
(and comparable) detection quality, indicating that the procedures they
used are robust and consistent. 
In particular, the T1 experiment has shown that, at least for the cases
under consideration, detection tests (e.g., $\chi^2$ or {\Ftwo}) based on 
deviations from the single-star astrometric solution  perform as well as can be expected. 
Planets down to astrometric signature $\alpha\simeq 2\sigma_\psi$ can be detected
reliably and consistently, with a very small number of false positives. 
Even better, the choice of the detection threshold is an effective way
to distinguish between highly reliable and marginal candidates.  Under
the assumptions of this test, which is based on an idealized, perfectly
known noise model, potential planet-bearing stars can be identified and
screened reliably. Refinements of the detection criteria based on additional
considerations, e.g., the quality of the orbital fit, can potentially
make an improvement in the fitting procedure.  However, the performance
of a straight $ \chi^2 $ or {\Ftwo} test is already extremely good; such
tests, if properly applied, can yield candidates with the expected range
of sensitivity and with a powerful discrimination against false
positives.

\subsubsection{First-pass detection}

Both solvers approach test T1 on the basis of the quality of the
single-star, five-parameter solution for the astrometric measurements.

Solver A adopts two criteria to identify candidate planets, one broad, aimed
at detecting as many candidates as reasonable, and one strict, designed
to reduce the number of false positives.  Specifically, Solver A uses
{\Pchi}, the probability that the observed $ \chi^2 $ of the single-star
solution is as bad or worse than the value observed in the presence of
pure measurement errors, and {\PF}, the F-test probability on the same
fit.  A large value of $ \chi^2 $ or of the $ F $ statistic can readily
arise if the deviations due to the presence of a planet are much larger
than the expected measurement errors, and thus a low value of {\Pchi}
and {\PF} signifies likely planet (and unlikely false positive). 

The broad criterion, A1, requires only that $ \Pchi < 0.05 $, and
favors completeness over reliability: many more marginal candidates are
included, but false positives will be more numerous.  The strict
criterion, A2, requires both $ \Pchi < 0.0001 $ and $ \PF < 0.0001 $,
and favors reliability over completeness: candidates satisfying this
criterion have a small probability of being false positives, but many
marginal cases will be missed. 

Criterion A1 identifies 44,914 candidates, of which 42,810 are indeed
planets and 2,104 are false positives, close to the 5\% expected from the
criterion.  On the other hand, 11,988 planets are missed by this
criterion. Typically, the planets missed have signature smaller than 15
{\muas} or period longer than 5 years (Figure~\ref{fig1}, left panel), 
high eccentricity and/or close to edge-on orbits (Figure~\ref{fig2}), 
and relatively small numbers of observations ($N_\mathrm{obs} < 40$, 
Figure~\ref{fig3}).  The performance of this
and other criteria discussed here is summarized in Table~\ref{detrecap}.

Criterion A2 yields only 28,655 detections, with {\it no} false
positives, but misses 26,143 planets---only half of the true planets are
found.  Because of the more demanding criteria, planets with signature
up to 30 {\muas} can be missed by this criterion, regardless of period. 
Nonetheless, the dramatic drop in false positives is very important, and
would probably favor the stricter criterion. 

\begin{table*}
 \caption {Summary of detection probability}
\centering
 \begin{tabular}{ccrrrr}
\hline\hline
 Name & Criterion & \multicolumn{3}{c}{Detections} & Missed \\
           &           & Total & True & False \\
\hline
 A1 & $\Pchi<0.05 $                              & 44\,914 & 42\,810 & 2\,104 & 11\,988 \\
 A2 & $\Pchi<0.0001,\,\PF<0.0001 $               & 28\,655 & 28\,655 &     0 & 26\,143 \\
 A3 & $\Pchi<0.0001 $ or                         & 40\,196 & 39\,630 &   566 & 15\,168 \\
     & $\Pchi<0.05,\,\Pchi_{\rm orbital}>0.2 $ \\
 A4 & $\Pchi<0.0001 $ or                         & 39\,957 & 39\,768 &   189 & 15\,030 \\
     & $\Pchi<0.05,\,P(\Delta \chi^2)<0.001 $ \\
 B1 & $|\Ftwo|>3 $                               & 37\,749 & 37\,643 &   106 & 17\,155 \\
 C1 & $\Pchi<0.001 $                             & 37\,782 & 37\,714 &    68 & 17\,084 \\
\hline
 \end{tabular}
\label{detrecap}
\end{table*}

A further refinement of Solver A's search criterion is discussed below. 
However, it is worth noting that a criterion based purely on $ \Pchi <
0.0001 $, without the {\PF} requirement, would detect 34,918 planets,
only 4 of which are false positives, and miss 19,880---a substantially
better performance at the cost of a modest number of false positives. 

Solver B adopts a similar method, using specifically the {\Ftwo} indicator
(see the Hipparcos Catalogue, vol.~1, p.~112), which is expected to
follow a normal distribution with mean 0 and dispersion 1.  His
criterion, B1, requires $ |{\Ftwo}| > 3 $, which in essence is a
3-sigma criterion.  With this criterion, Solver B identifies 37,643 correctly
as having a planet (or more), while 17,155 are missed and 106 (0.2\% of
the no-star sample) are false positives.  Similarly to A1, the missed planets
mostly have signature smaller than 20 {\muas} or period longer than 5 years 
(Figure~\ref{fig1}, right panel), The overall distribution is similar
to that of planets missed by A1, although more marginal cases are excluded---and
fewer false positives are included.

Criterion B1 appears to be preferable to A1, which
finds 5,000 more planets at the cost of nearly 2,000 false positives. 
If a 0.2\% incidence of false positives is considered acceptable, the
performance is also better than that of A2, with nearly 9,000 more
planets found at a modest cost in false positives.  However, the simple
$ \Pchi < 0.0001 $ criterion finds nearly as many planets, with a much
smaller fraction of false positives. In practice, the choice between
these criteria would depend on the specific application and sample
properties. For example, for the simulated data studied here, a fine-tuned $
\Pchi $ test, e.g., with threshold set at 0.001 (C1), would find 37,714
valid candidates (about as many as B1 and 2,000 fewer than A4, discussed 
below) and only 68 false positives. 

\subsubsection{Refining the detection criteria}

Realizing that his strict criterion (which requires both $ \Pchi <
0.0001 $ and $ \PF < 0.0001 $) may be too stringent, while the simple $
\Pchi < 0.05 $ criterion is expected to allow too many false positives,
Solver A attempts a detection refinement based on the quality of the orbital
fit, in the spirit of the T1b test. 

For the purpose of this test, Solver A considers the ``marginal'' candidates
with $ 0.0001 < \Pchi < 0.05 $; of these, 
2,100 are in fact false positives, while 7,892 have a real planet.  In
this case, there are 34,918 non-marginal detections---those with $ \Pchi
< 0.0001 $---of which only 4 are false detections. 

The first refinement (A3) is based on the quality of the orbital fit: a
marginal candidate passes if the $ \chi^2 $ statistics of the residuals
after the orbital fit improves to $ \Pchi > 0.2 $ (a minimum factor 4
improvement).  A total of 5,274 marginal candidates pass this test; of
these, 11\% are false detections.  Of the marginal
candidates that do not pass the refinement, 33\% are false
positives.  Thus, this orbital refinement does improve the probability
that the candidate is real, and can in fact increase the sample of
possible candidates (see Table~\ref{detrecap}).

The second refinement for the marginal candidates (A4) is based on the
likelihood ratio test applied to the two fits, with or without the
planet.  For a candidate to pass, the fit with the planet is required to
improve the $ \chi^2 $ with a probability better than 0.001, i.e., $
P(\Delta \chi^2) < 0.001 $.  Of the 5,035 stars that pass, 96\% do 
in fact have a planet; only 185 are false positives.  The
likelihood ratio improvement appears to perform significantly better
than the simpler test based on the new $ \chi^2 $ probability (see Table~\ref{detrecap}).

The refined criteria, especially A4, do improve substantially on A1,
bringing its performance in line with that of B1. A4 finds about
2,000 more candidates than B1, but 83 more false positives. B1 is
simpler to apply, and the expected distribution of the {\Ftwo} statistic
is well-defined in the case of stars without planets; this makes it
possible to clearly label those candidates that are most likely to be
false positives, and therefore to derive samples with different levels
of confidence for different purposes.  On the other hand, A4 offers the
potential to detect more stars, including potentially some stars with
relatively small signatures but a good orbital fit, without an excessive
increase in the number of false positives.  Neither approach offers the
freedom from false detections of A2, which however comes at the cost of
fewer candidates. 

It may be worthwhile considering orbital fit criteria as a means to
improve the detection statistics for a more tightly selected initial
sample.  For example, one could consider a likelihood ratio threshold
that depends on the original $ \Pchi $, so that more marginal candidates
(with a greater probability of being false positives) are held to a stricter
likelihood ratio requirement.  Conceivably, such 
requirements could achieve a better combination of sensitivity and
reliability than straight $ \chi^2 $ or {\Ftwo} tests.  However, their
investigation is beyond the scope of this analysis; a new set of tests would
be needed to assess such techniques in true double-blind fashion.

\subsection{Test T2: Single-planet orbit determination}

The T2 experiment is designed primarily to establish the accuracy of the orbital
determination for single planets with solidly detected signatures, under
the assumption that the noise characteristics of the data are fully
understood. Solvers knew that each star had one planet, but
did not know the distribution of signatures and periods.
The T2 test determines how well the orbital parameters of
a single planet can be measured for a variety of signature
significance, period, inclination, and other parameters.
Simulated data were prepared for 50,000 stars, each with
exactly one planet with signatures ranging between 16 $\mu$as 
(astrometric signal-to-noise $\alpha/\sigma_\psi=$~2) and 1.6 mas 
($\alpha/\sigma_\psi=$~200) and periods between 0.2 and 12 years; 
all other orbital parameters were randomly distributed with the same 
prescriptions of Test T1.  

Each solver was asked to carry out a full orbital reconstruction
analysis for each star, beginning from the period search and including
error estimates for each of the orbital parameters.  As for the T1 test,
two solvers, A and B, participated in this test, each with their independently developed
numerical code. The first, obvious conclusion is that both solvers 
achieve very good results, recovering very solidly
the orbital parameters of the vast majority of `good' cases - those with
high astrometric signature and period shorter than the mission duration.
In addition, their results are extremely consistent, 
indicating the robustness of the procedures they developed and of the overall approach.

Both Solvers run their respective pipelines, consisting of detection,
initial parameter determination, and orbital reconstruction, on each of
the 50,000 simulated time series provided by the Simulators.  They have
no a priori knowledge of the orbital properties of each planet, although
they do know that each star is expected to have one and only one planet. 

In both cases, solvers use the equivalent of a least-squares algorithm
to fit the astrometric data for each planet; they need to solve for the
star's basic astrometric information (position, parallax, proper
motion), for which only low-accuracy catalog parameters are provided, as
well as for the parameters of the reflex motion. Solver B provides orbital 
solutions expressed in terms of $ P $, $ e $, $ \tau $, and the 
four Thiele-Innes parameters $ A, B, F, G$ (e.g., Green 1985). 
He provides also estimated uncertainties for each parameter
and the full covariance matrix. Solver A also provides $P$, $e$, and $\tau$, 
but instead of the Thiele-Innes parameters, he returns $a$, $ i $, 
$ \Omega $, and $ \omega $.  He computes formal errors for each 
parameter, but not the covariance matrix. 

Solver B reports no solution for 521 stars, about 1\% of the total. Solver A
reports a solution for all stars, but 69 are invalid as the estimated
error in the orbital parameters is undefined; we exclude these objects
from further consideration.  In addition, a few tens of objects have
very large errors, and may not be meaningful.  It is important to note
that for both solvers the number of such cases is very small, and---as
they are identified during the solution process---they present no risk
of contaminating the search for planets; they simply reflect the limitations
of the observations. 

\subsubsection{Retrieving orbital parameters}

\begin{figure*}
\centering
$\begin{array}{cc}
\includegraphics[width=0.4\textwidth]{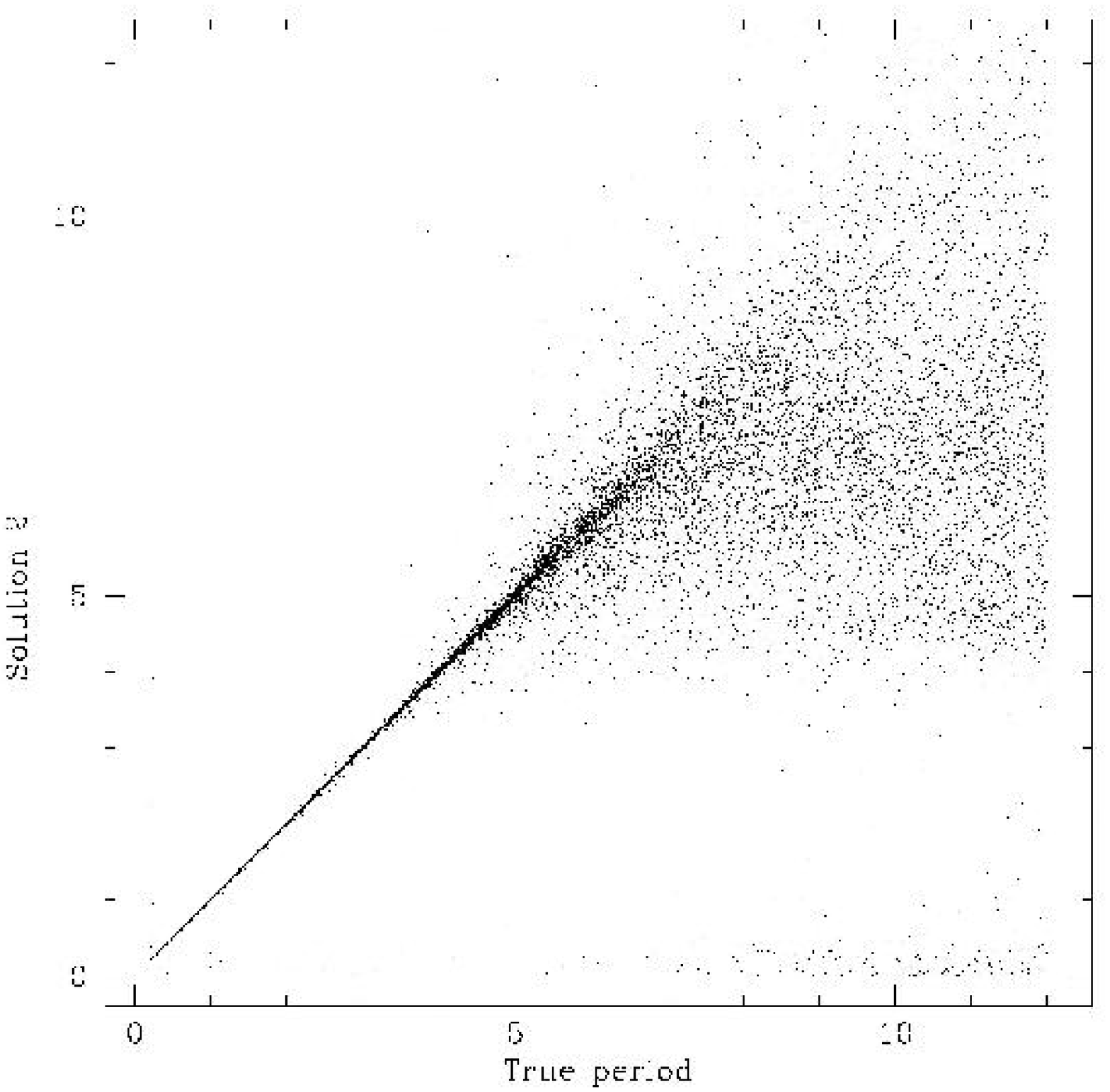} & 
\includegraphics[width=0.4\textwidth]{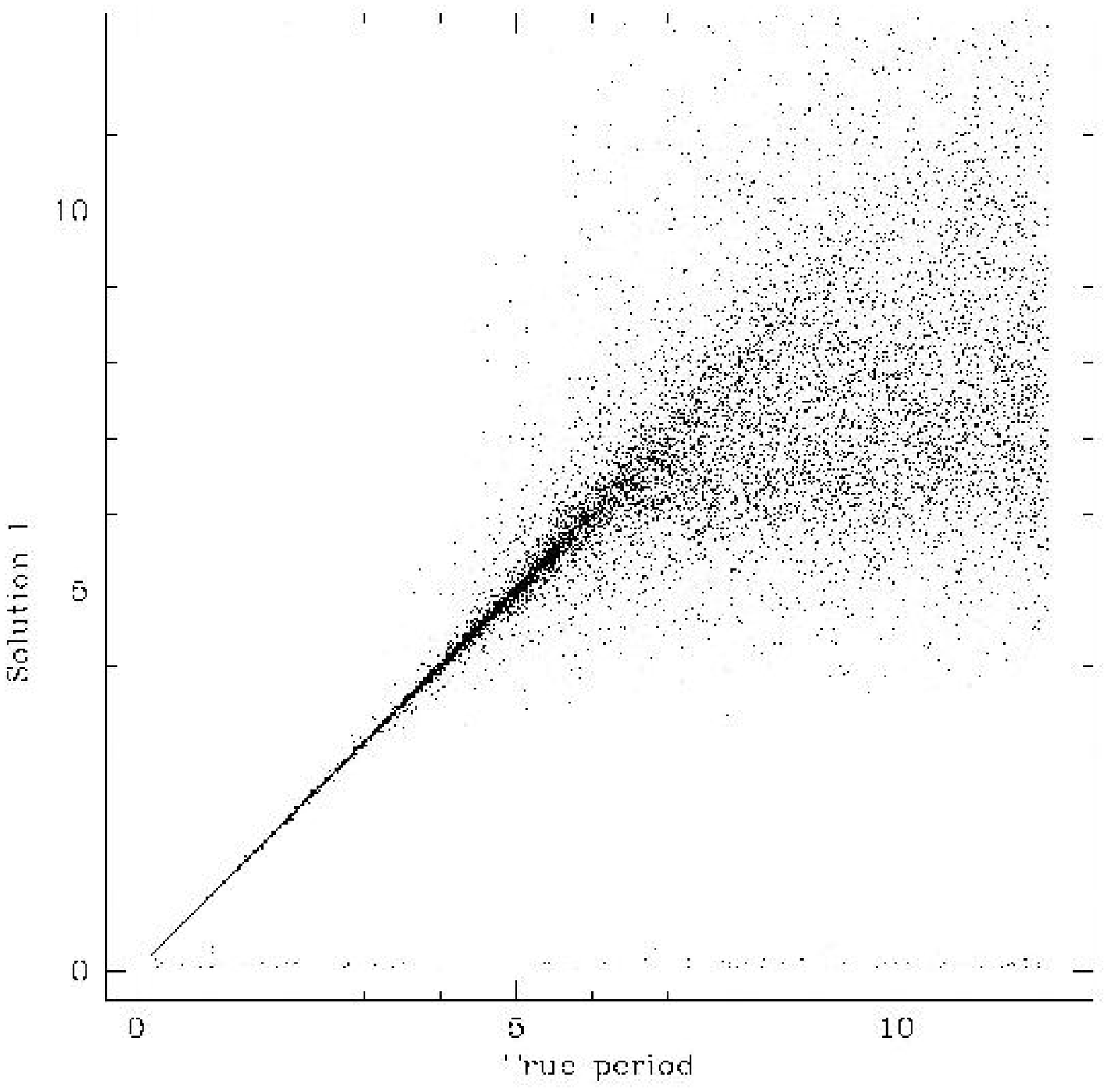} \\
\end{array} $
 \caption{Distribution of fitted period as a function of true period for
Solver A (left) and B (right).}
\label{fig4}
 \end{figure*}

 \begin{figure*}
\centering
$\begin{array}{cc}
\includegraphics[width=0.4\textwidth]{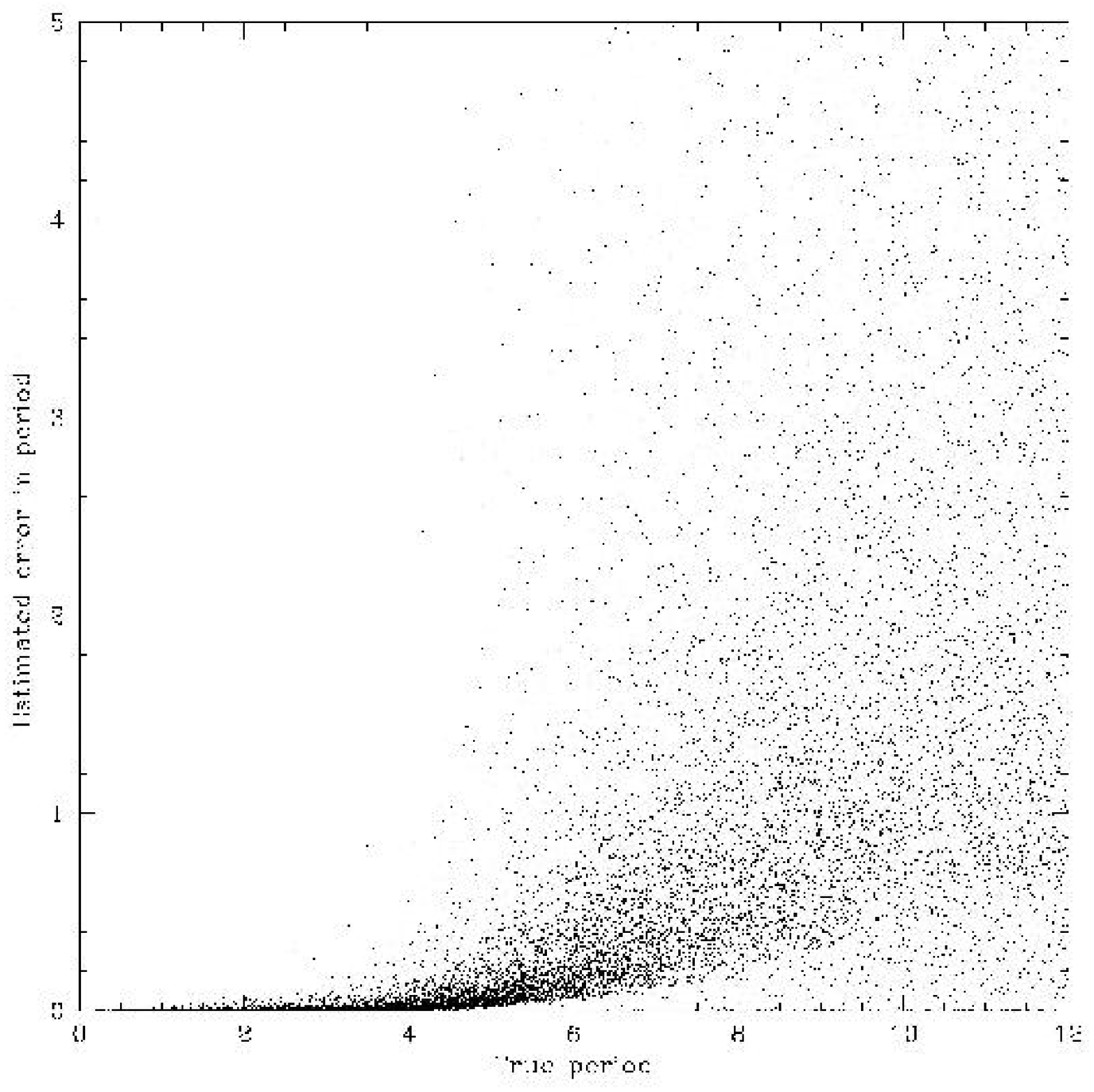} & 
\includegraphics[width=0.9\columnwidth]{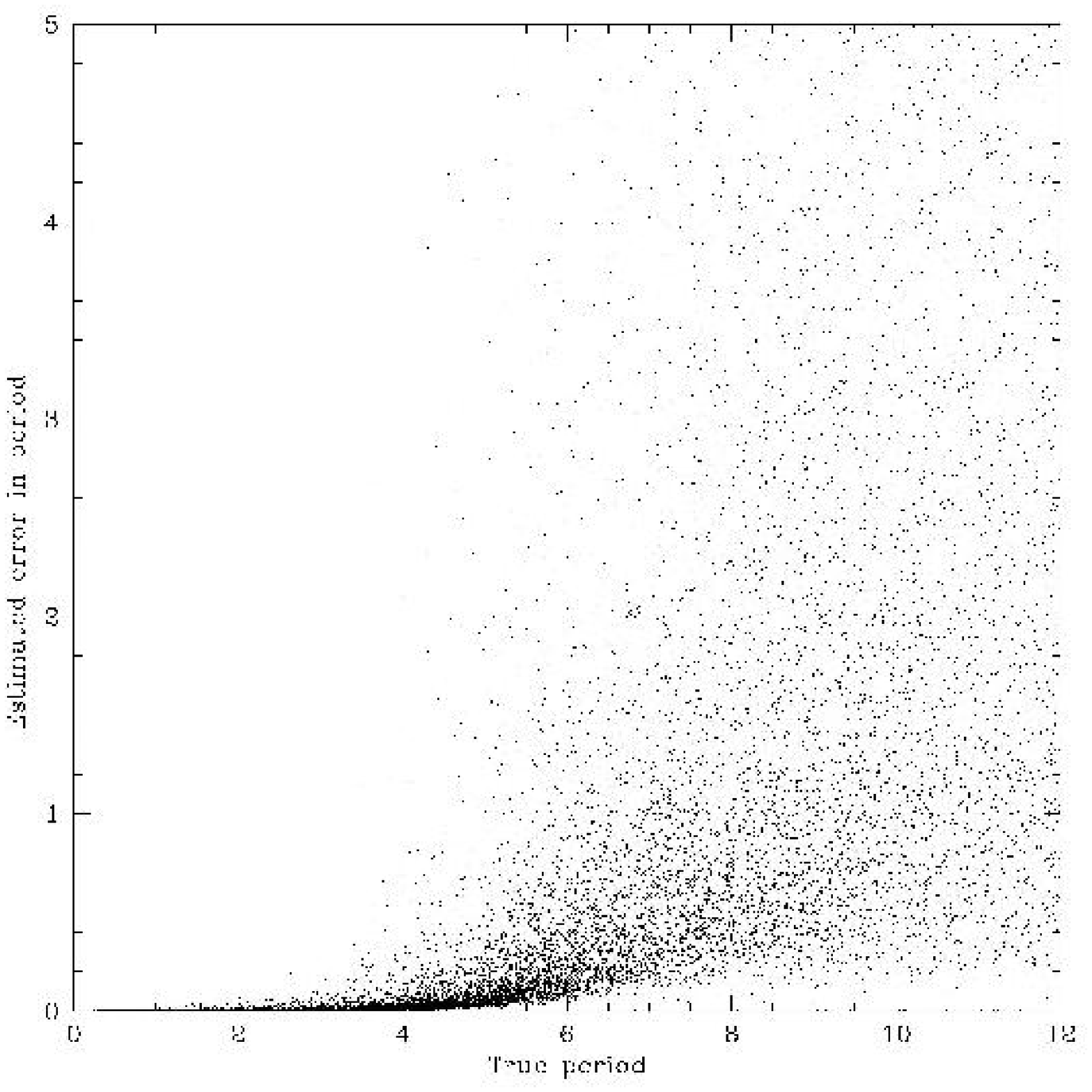} \\
\end{array} $
 \caption{Distribution of estimated error in the period as a function of
true period for Solver A (left) and B (right). }
\label{fig5}
 \end{figure*}

 \begin{figure}[t]
\centering
\includegraphics[width=0.4\textwidth]{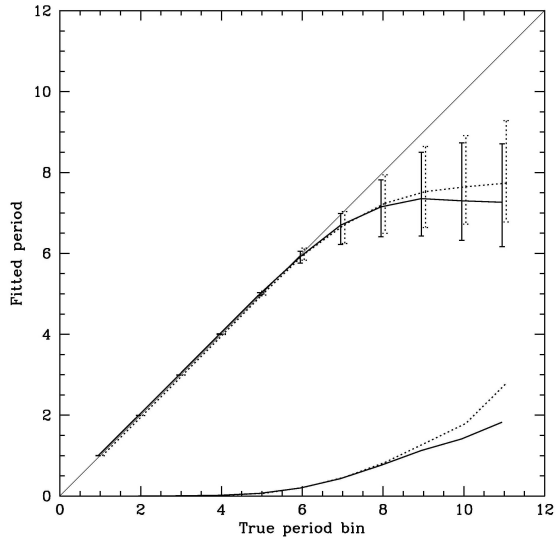}
 \caption{Distribution of estimated periods and their errors for orbits
with signature larger than 0.4 mas as a function of true period.  The
lines with error bars show the median and interquartile range for the
period estimated by Solver A (solid) and B (dashed). The lines without error bars 
represent the median estimated errors from the fitting procedure for 
Solver A (solid) and B (dashed).}
\label{fig6}
 \end{figure}

\begin{figure*}
\centering
\includegraphics[width=1.\textwidth]{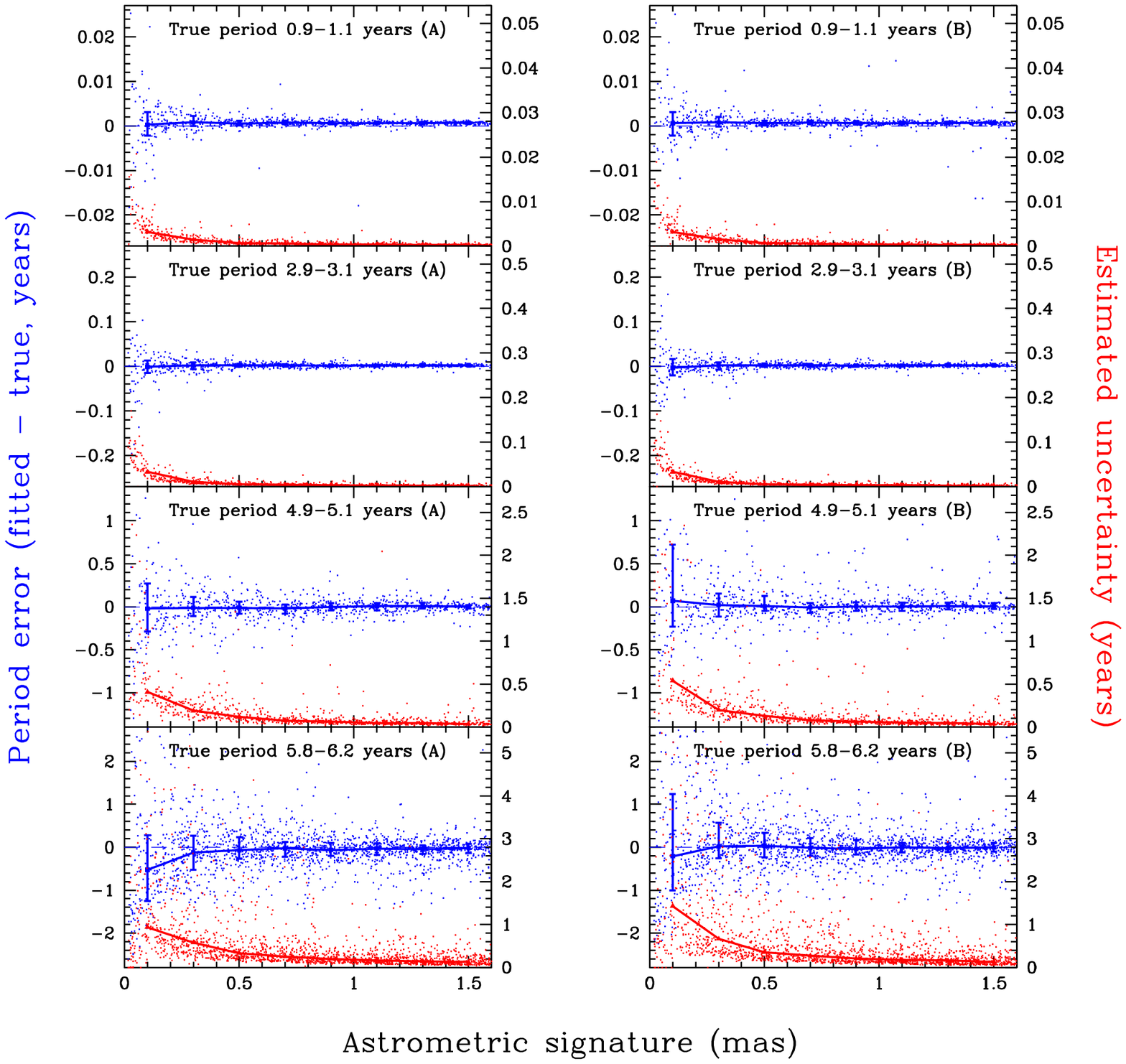}
 \caption {Error in period as a function of astrometric signature for different
  period ranges and for both solvers.  The dots show the difference between fitted
  and true period (blue, left axis) and the estimated uncertainty from the solution 
  (red, right axis).  Shown on the left panels are the solutions by Solver A, and on the right
  those by Solver B.  Heavy dots represent the median values, binned in astrometric signature;
  error bars represent the interquartile range.}
\label{fig7}
  \end{figure*}

The orbital period is perhaps the most important of the orbital
parameters, and generally the most critical in terms of obtaining an
orbital solution that is close to the truth.  Period search is usually a
delicate process, and aliasing, especially for relatively sparsely
sampled orbits, can be a serious concern.  Therefore the evaluation of
the solutions starts with the orbital period. 
In summary, the period is retrieved with very good accuracy and small
bias for true periods ranging from 0.3 to 6 years.  A small fraction of
very short and very long periods are aliased to very different periods;
these cannot be readily identified by simply inspecting the estimated
errors.  Long periods are systematically underestimated; this trend is
predictable on the basis of simulations, and the amount of bias is
comparable to the estimated period error.

Figure~\ref{fig4} shows the quality of the match between the true period and the
solution by Solver B (Solution 1) and by Solver A (Solution 2).  For the 20,411
stars with true period shorter than 5 years, both solvers recover over
98\% with a fractional error in the period of 10\% or
smaller (20,054 for Solver A, 20,158 for Solver A).  This includes a few
cases (45 for Solver A, 27 for Solver B) for which no valid solution is returned. 
Almost all the cases with poor period determination have either very
small signatures or periods shorter than 3 months, for which aliasing
can occur with the relatively sparse sampling of the Gaia scanning law.  Such
cases are rare, no more than 2\% of all short-period planets, but
are not readily identified by the nominal error in the period. 
Short-period solutions will probably need to be looked at more carefully
to eliminate the possibility of aliasing in the solution. 

While fidelity is extremely good for planets with true period ranging from a 
few months to the mission lifetime, the quality of the solution degrades quickly 
for periods longer than the mission duration.  Visually, it is clear that - for given amplitude of
the perturbation - the ability to recover the planet's period with 
modest errors starts degrading at periods of about 6 years. Note also that 
for very long true periods, the fitted period is systematically shorter than 
the truth; at 10 years, the typical recovered period is substantially shorter, 
about 7 years, with a very large dispersion. 
In a small number of cases (418 for Solver A, 150 for Solver B), a very small 
period is fitted to a long period object (resulting in the small cloud of points
near the $P = 0$ axis in both panels of Figure~\ref{fig4}), indicating that the fit has 
aliased into a completely different range. 

Figure~\ref{fig5} shows the error in the period, as estimated by each Solver, 
as a function of true period.  As in the period difference, the estimated
error also increases greatly with increasing period, and in fact 
the estimated uncertainties are comparable with the error in the
fitted period shown in Figure~\ref{fig4}.  

The comparison between error in fitted period and estimated error is
shown in a more quantitative way in Figure~\ref{fig6}.  The curves and
error bars illustrate the median and quartiles of the fitted period
distribution in bins of true period, solid for Solver A and dotted for Solver B; the
thin diagonal dashed line corresponds to exact solutions.  As it can be
clearly seen, the period solution is very good, without indication of
significant bias, up to about 6 years, beyond which the solution
underestimates the period.  The median estimated errors (lower curves)
match the interquartile range reasonably well.

\begin{figure*}[hbt]
\centering
$\begin{array}{cc}
\includegraphics[width=0.4\textwidth]{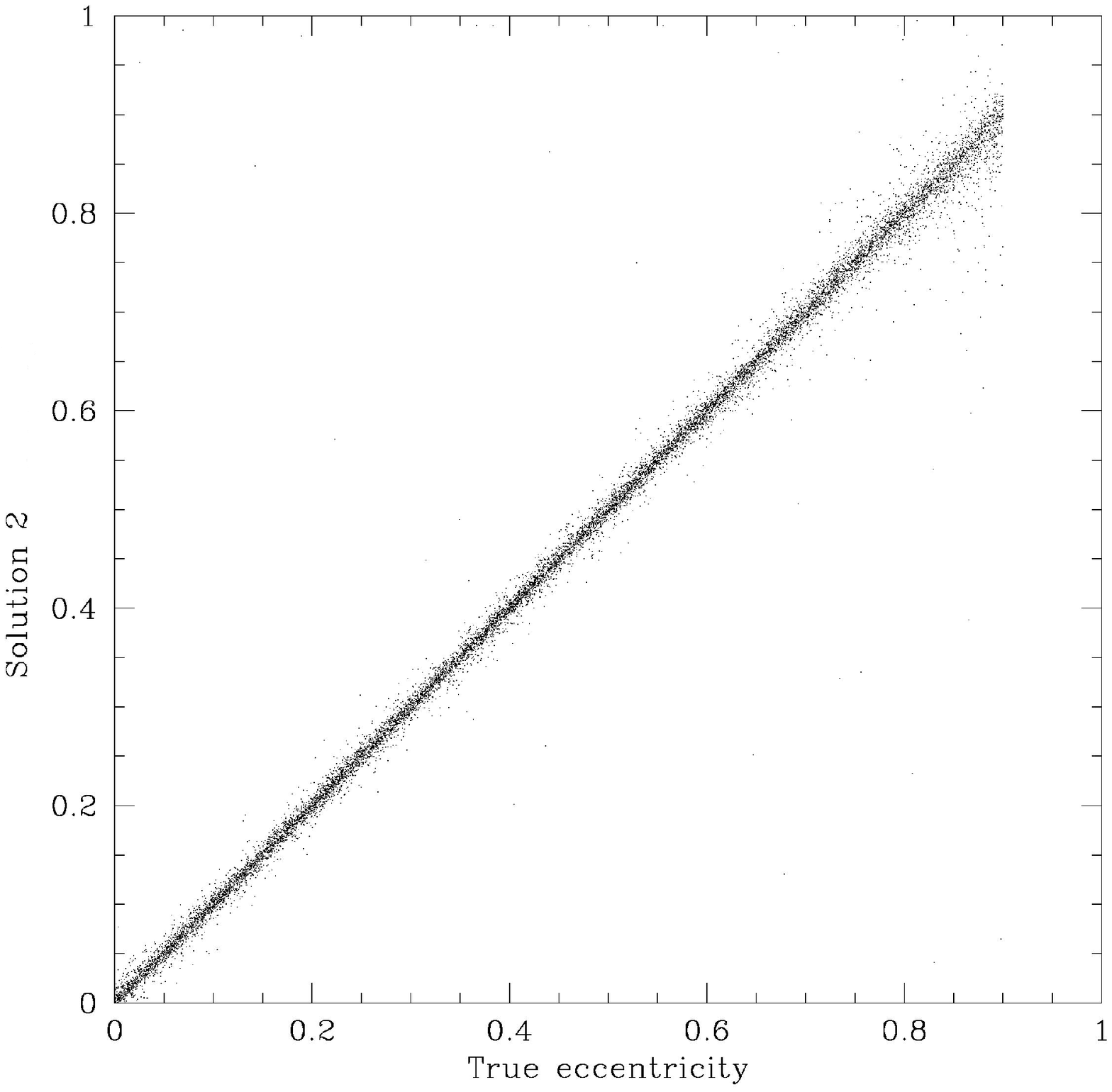} & 
\includegraphics[width=0.4\textwidth]{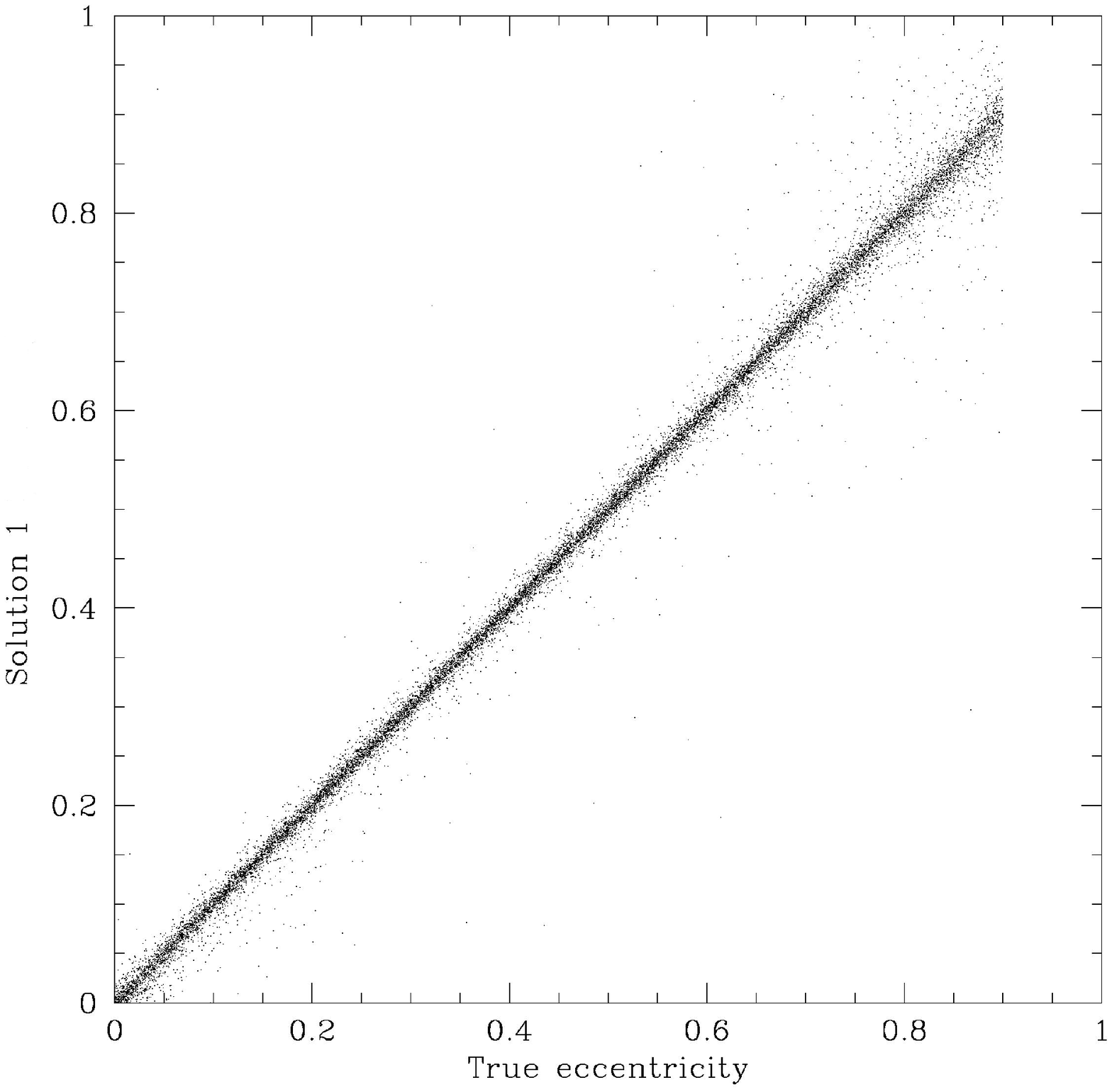} \\
\end{array} $
 \caption {Fitted vs. true orbital eccentricity for Solver A (left) and Solver B (right).
 Included are the orbits with signature larger than 0.4 mas---approximately 75\% of the
 cases studied---and period shorter than 5 years.}
\label{fig8}
  \end {figure*}

\begin{figure*}
\centering
$\begin{array}{cc}
\includegraphics[width=0.4\textwidth]{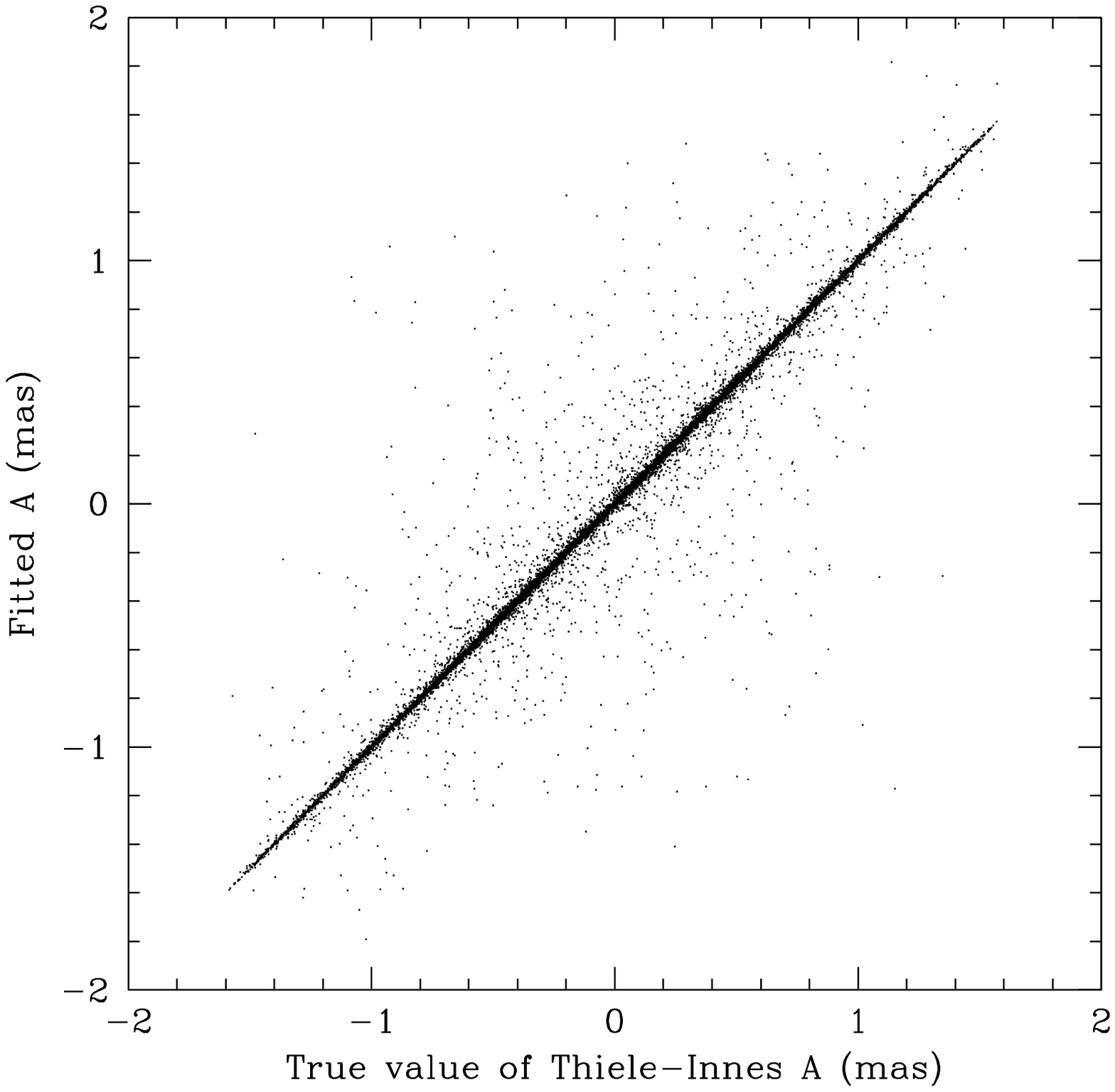} & 
\includegraphics[width=0.4\textwidth]{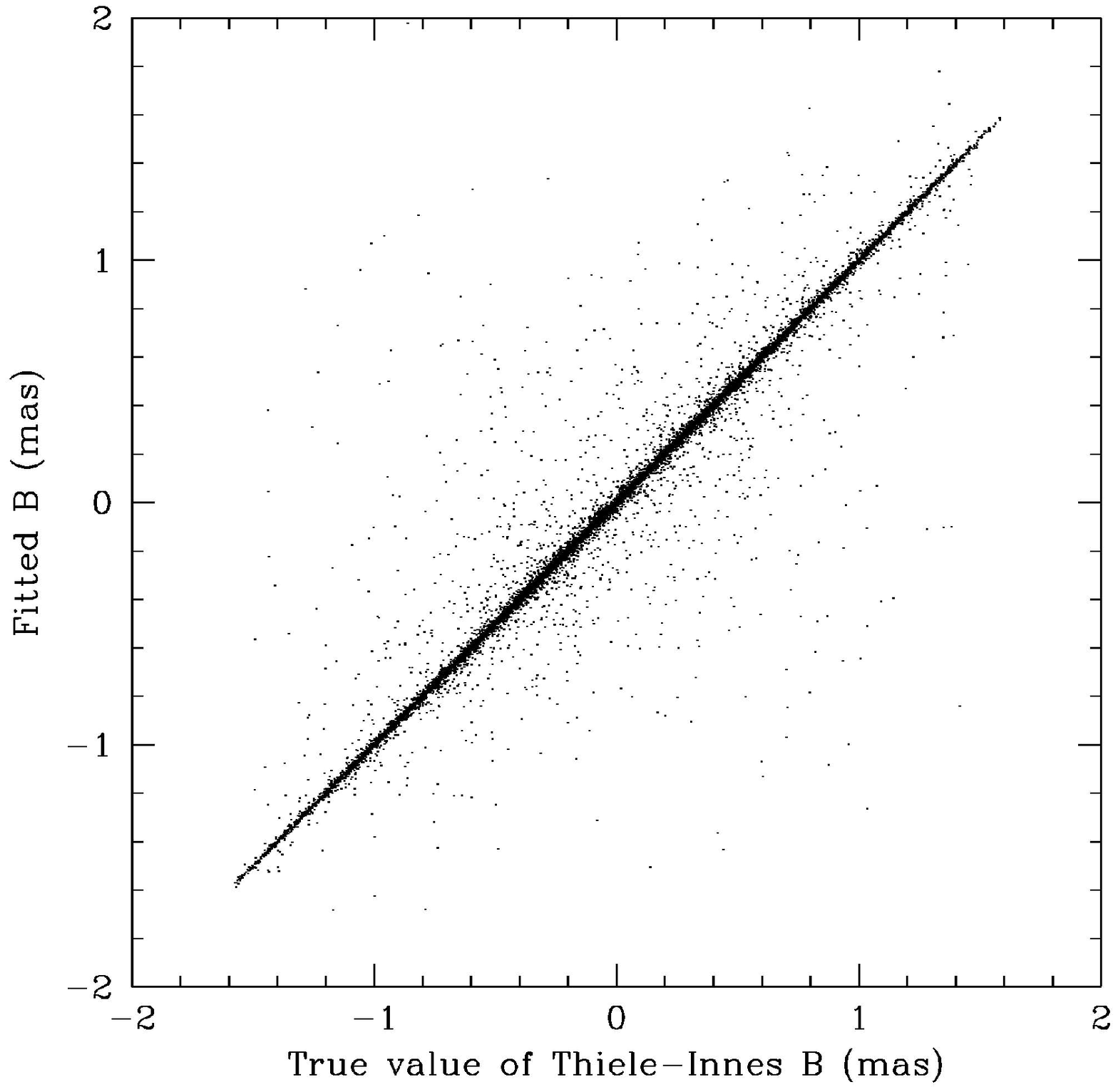} \\
\end{array} $
 \caption {Fitted vs. true values of the Thiele-Innes parameters A and B, according to the
 solution by Solver B.  As in Figure~\ref{fig8}, included are orbits with $\alpha > 0.4$ mas 
 and $P > 5$ yr.}
\label{fig9}
 \end {figure*}

\begin{figure*}
\centering
\includegraphics[width=.96\textwidth]{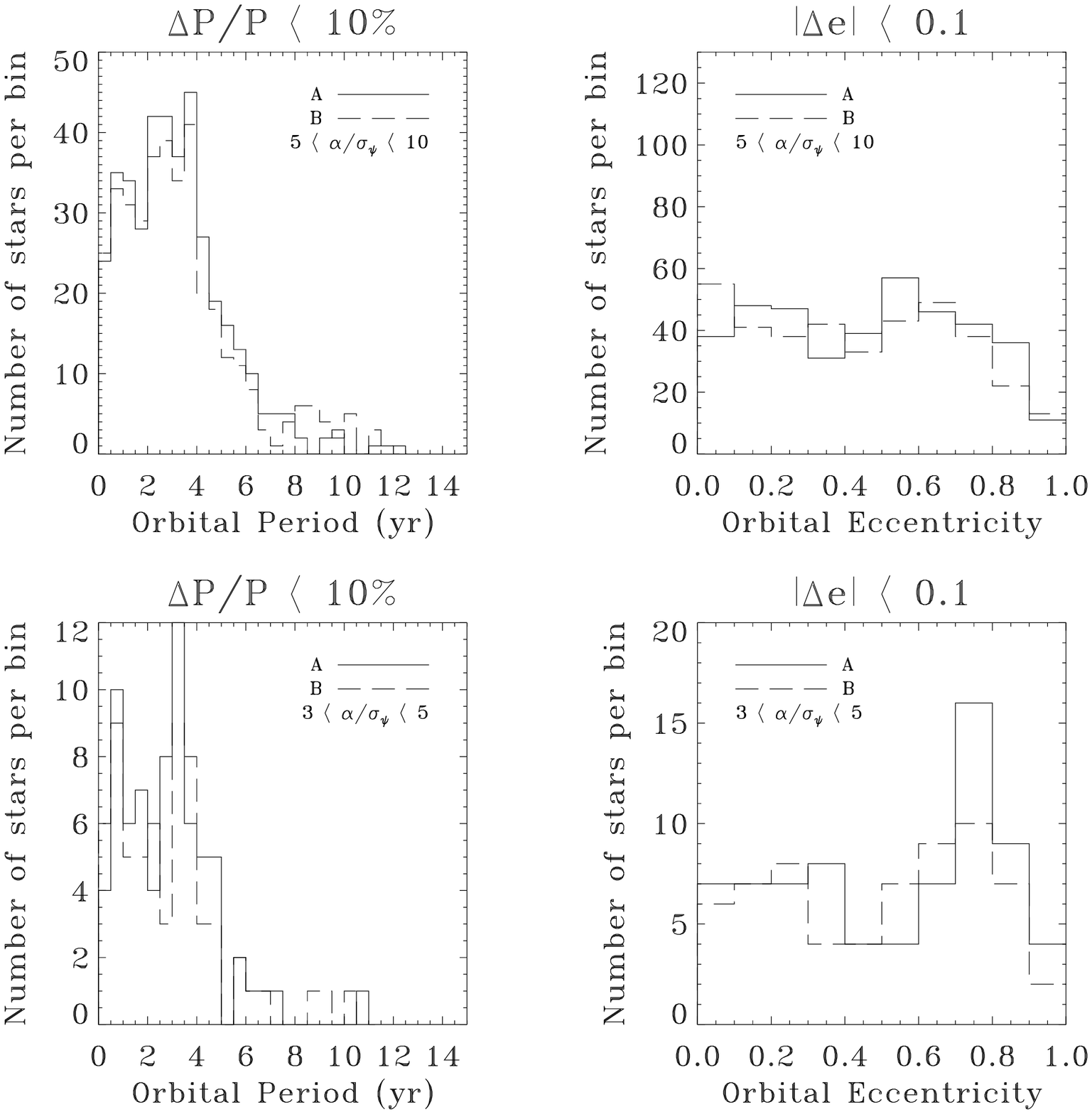}
\caption {Top left and right: distributions of well-measured values of 
$P$ and $e$ for the two Solvers in the case of $5<\alpha/\sigma_\psi<10$. 
Bottom left and right: the same, but for the case of $3<\alpha/\sigma_\psi<5$.}
\label{lowsn}
  \end{figure*}

Figure~\ref{fig7} shows how the period accuracy varies with signature for periods
around 1, 3, 5, and 6 years.  In each case, larger signatures mean a
stronger astrometric signal, and thus better accuracy; the distribution
of errors matches well the estimate from the solution itself.  In each
panel, the blue dots (scale to the left) represent the difference
between fitted and true period as function of true signature in the
stated period range, and the red dots (scale to the right) show the
error as estimated by the solver for that particular orbit.  The solid
lines and points represent the median values for a 0.2 mas bin in
signature; the error bars for the period error show the range between
the first and third quartile in each bin.  Panels to the left refer to
solutions by Solver A, to the right by Solver B.  In each panel and for each
signature bin, the median estimated error (red triangles) is very close
to the difference between median and quartile error for the same set of
solutions, indicating that the estimated errors are a good guide to the
true errors.  The median of the difference between fitted and true
period (blue squares) is generally small, showing that there is very
little bias in the period estimate. 

The other orbital parameters are similarly well estimated for the vast
majority of "good" orbital solutions, excluding those with low signature
and long period.  For example, Figure~\ref{fig8} compares the eccentricity fitted
by the two Solvers with the true value for all stars with period shorter
than 5 years and signature larger than 0.4 mas, which corresponds to the
top 75\% in signature.  Similarly, Figure~\ref{fig9} shows the true and fitted
(by Solver B) values of the Thiele-Innes parameters A and B for the same cases.
Clearly both sets of parameters represent high quality orbital fits.  
Other orbital parameters follow similar patterns.

Finally, it is worth mentioning how subtle differences in the orbital solutions 
carried out by the two solvers can be seen if one focuses on regimes of relatively 
low astrometric signal. We show for example in Figure~\ref{lowsn} the comparison between 
the distributions of fitted $P$ and $e$ for Solver A and Solver B in cases of 
$5<\alpha/\sigma_\psi<10$ and $3<\alpha/\sigma_\psi<5$, and restricting 
ourselves to good solutions for which $P$ is within 10\% of the true value and 
$e$ differs from the true value by no more than 0.1. On the one hand, 
from the Figure it is clear how both solvers identify and measure precisely 
orbital periods for virtually the same stars; a Kolmogorov-Smirnov (K-S) test gives 
a probability that the two distributions are the same of 0.15 and 0.98 for the 
two regimes of signal strength investigated. On the other hand, the distributions 
of well-measured eccentricities are significantly different, with a the K-S test 
giving a probability of the null hypothesis of 0.04 and 0.005, respectively. The 
most obvious feature is the increase in the number of very large eccentricity values 
($e\gtrsim 0.6$) correctly identified by Solver A with respect to Solver B. In particular, 
in the range $3<\alpha/\sigma_\psi<5$ Solver A measures accurately the eccentricity 
for some 23\% more stars than Solver B. A possible explanation for this discrepancy 
maybe found in the different approaches the two solvers adopt to reach the 
configuration of initial starting guesses for the parameters in the orbital fits. 
Both solvers tackle this issue implementing a two-tiered strategy consisting of a 
combined global+local minimization procedure. Solver A uses a methodology similar to 
that described in Konacki et al. (2002), in which a Fourier expansion of the Keplerian 
motion is used to derive initial guesses of the full set of orbital elements, 
subsequently utilized in a local non-linear least-squares analysis. 
Instead, Solver B adopts a scheme in which a guess to 
$P$ is obtained using a period-search technique (e.g. Horne \& Baliunas 1986), 
and then an exploration of the ($e$, $\tau$)-space is carried out to 
derive the linear parameters $A$, $B$, $F$, and $G$ as the unique minimizer of $\chi^2$ 
when $e$, $P$, and $\tau$ are fixed (e.g., Pourbaix 2002). However, for highly 
non-linear fitting procedures with a large number of model parameters 
the statistical properties of the solutions are not at all trivial 
(and significantly differ from those of linear models). A serious study of 
differences in the fitting procedures adopted by the two Solvers would require, 
for example, an in-depth analysis of the relative agreement between a 
variety of statistical indicators of the quality and robustness of the fits. 
Such a study lies beyond the scope of this work, and we leave it for future investigations.

\subsubsection{Estimated and actual errors}

A more quantitative analysis of the fitted parameters can be carried out by
comparing the distribution of differences between true and fitted parameters with
the errors estimated as part of the solution process.  The distribution of 
differences can be used to determine the actual uncertainties in the fit, which
in the ideal case would match the uncertainties estimated by the fit.  In reality,
this is not a perfect process; the estimated uncertainties are based on noisy
data, and therefore tend to be biased towards smaller values when the noise produces
an apparently larger signal.  Nonetheless, a general agreement between estimated
and actual errors is to be expected for a good fitting process.

The results presented in this Section demonstrate that both Solvers are 
not only capable of recovering the expected signal for the overwhelming 
majority of the simulated orbits under the conditions of the T2 test (as 
shown in the previous Section), but also that error estimates
are generally accurate, with the overall distribution of the difference
between fitted and true parameters very close to the solution results. 
Some discrepancies---a bias of up to 2 sigma in estimated period and a
mismatch of up to a factor 2 in estimated errors---do occur under
special circumstances, such as very short and very long periods.  These
discrepancies, small in the economy of this test, can be evaluated and
corrected for by a more thorough understanding of the estimation process
and its error estimates.  An incorrect solution is returned for about
2\% of the planets.  Such cases are not identified from their formal
error estimates, and will need to be addressed by a more aggressive
understanding of possible aliasing in orbital parameter space. 
Simulations and solutions show conclusively that correct solutions with
accurate error estimates can be obtained for about 98\% of the simulated
planets. 

Indeed, the estimated and actual errors do match with good accuracy
under most conditions.  An indication can be seen in Figure~\ref{fig6}, where we
show that the typical difference between true and fitted period, as
estimated from the interquartile range, is very close to the median
estimated uncertainty for diverse values of orbital period and
amplitude. 

A more quantitative---and challenging---test can be carried out by
studying the distribution of differences in the parameters compared with
their predicted errors.  Since predicted errors can in principle depend
on the amplitude of the signature, period, times of observation, and
other orbit details, we define the {\it scaled difference} as the
difference between the fitted and the true value of an orbital
parameter, divided by its estimated uncertainty for that same case.  If
the errors are predicted correctly and follow a Gaussian distribution,
this quantity will also be distributed normally with zero mean and unit
dispersion.  Discrepancies between predicted and actual errors will show
up as distortions in this distribution. 

\begin{figure}
\centering
\includegraphics[width=\columnwidth]{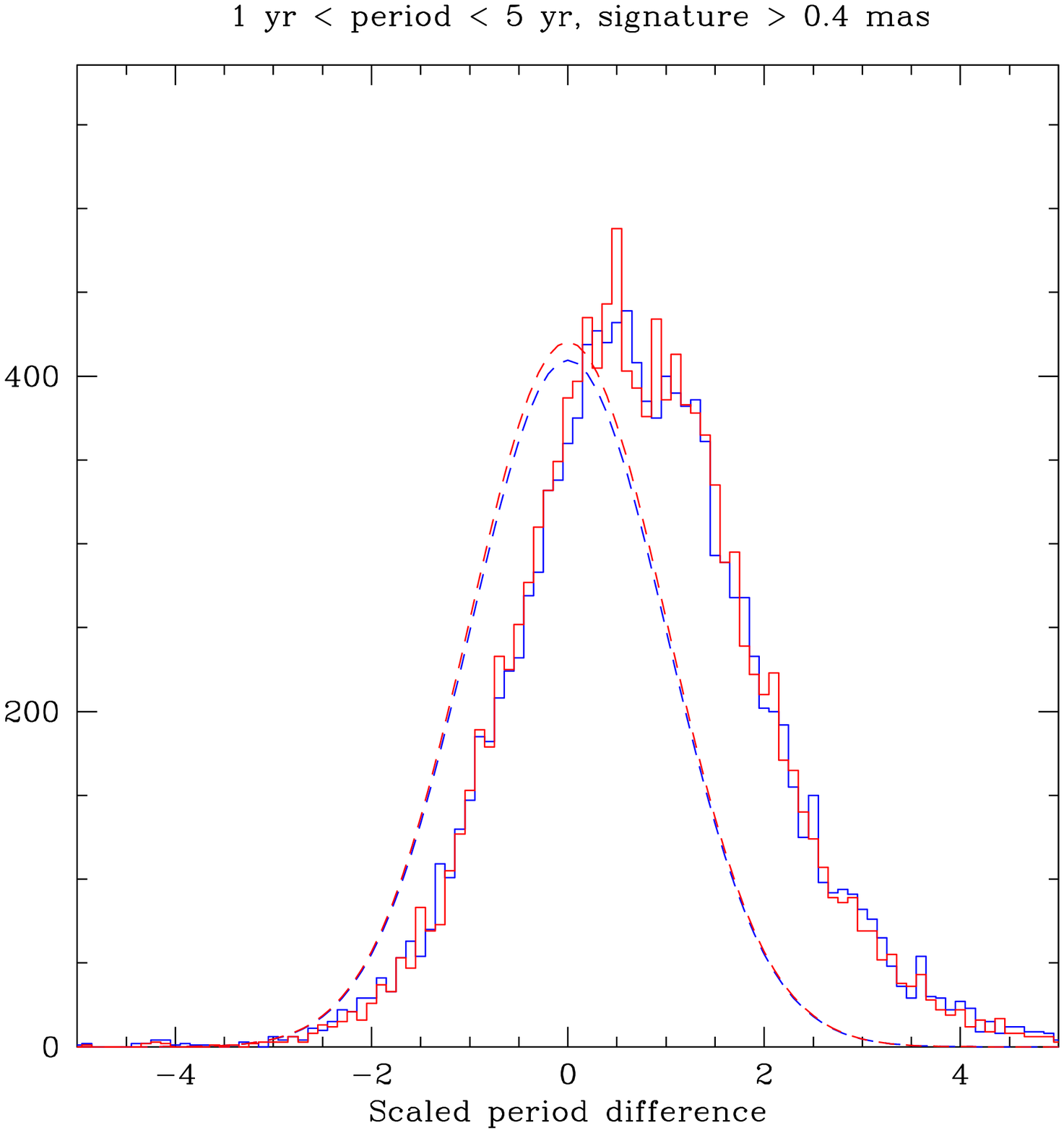}
 \caption{Histogram of scaled period differences for planets with period
between 1 and 5 years and signature larger than 0.4 mas.  The red histogram
is for Solver A, blue for Solver B.  The dashed lines represent a Gaussian distribution
with zero mean and unit dispersion.}
\label{fig10}
 \end{figure}

The expectation of a good error distribution should hold primarily for the
cases with good signal and solid orbit reconstruction, for which the true and
the reconstructed orbits are close.  We therefore focus on planets with 
$P < 5 $ years and $\alpha > 0.4 $ mas, about 20,000 cases.  

Figure~\ref{fig10} shows a definite distortion of the
overall scaled difference in period for both Solver B (blue) and Solver A (red);
the width of the distribution is similar to the predicted value (dashed),
but the peak is shifted towards positive values (i.e., the fitted
value of the period is statistically biased towards positive errors, 
or longer periods). The difference is small, about 0.5-sigma, but it is nonetheless
statistically significant because of the large number of simulations
used. 

 \begin{figure*}
\centering
$\begin{array}{cc}
\includegraphics[width=0.4\textwidth]{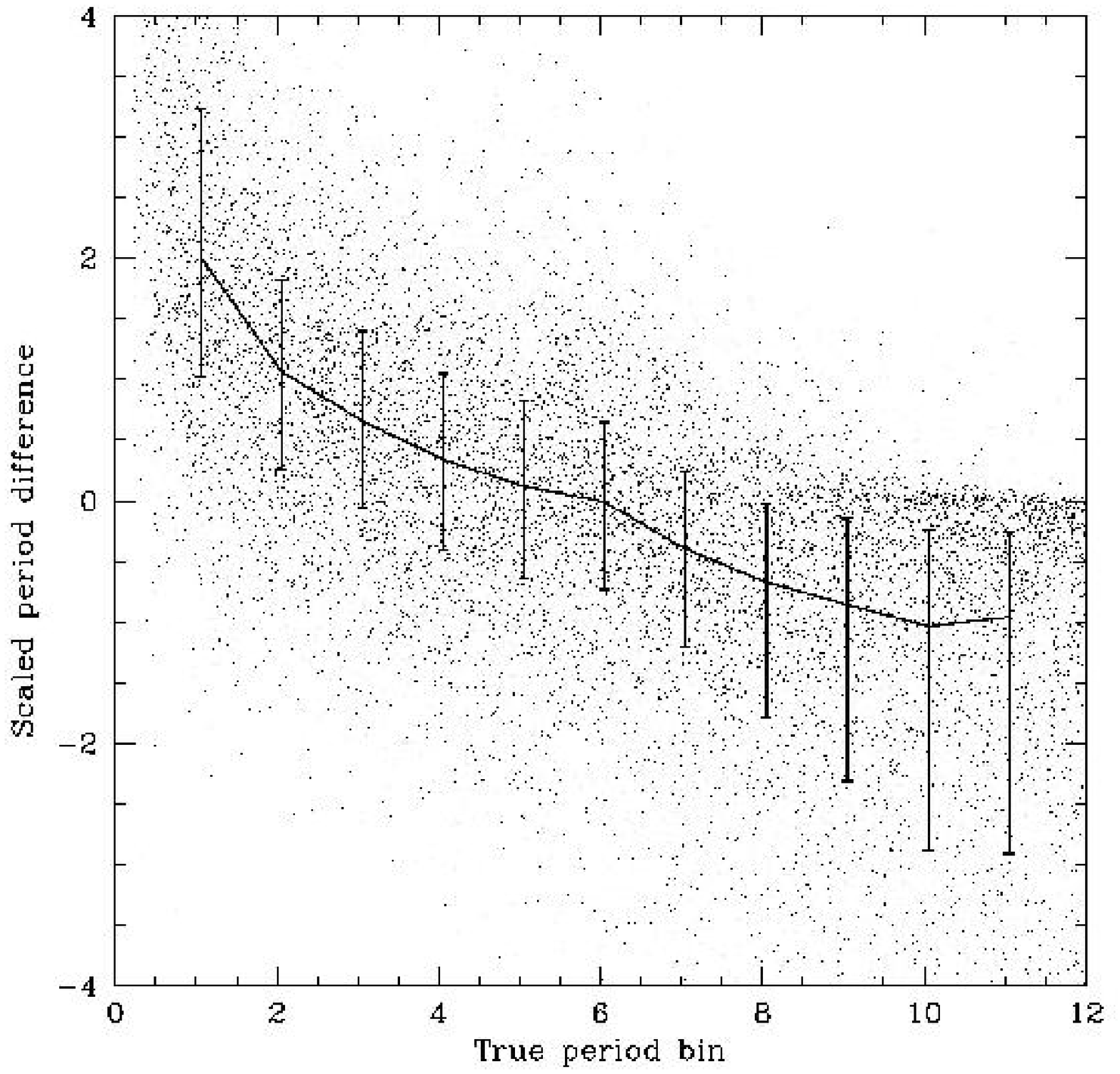} & 
\includegraphics[width=0.4\textwidth]{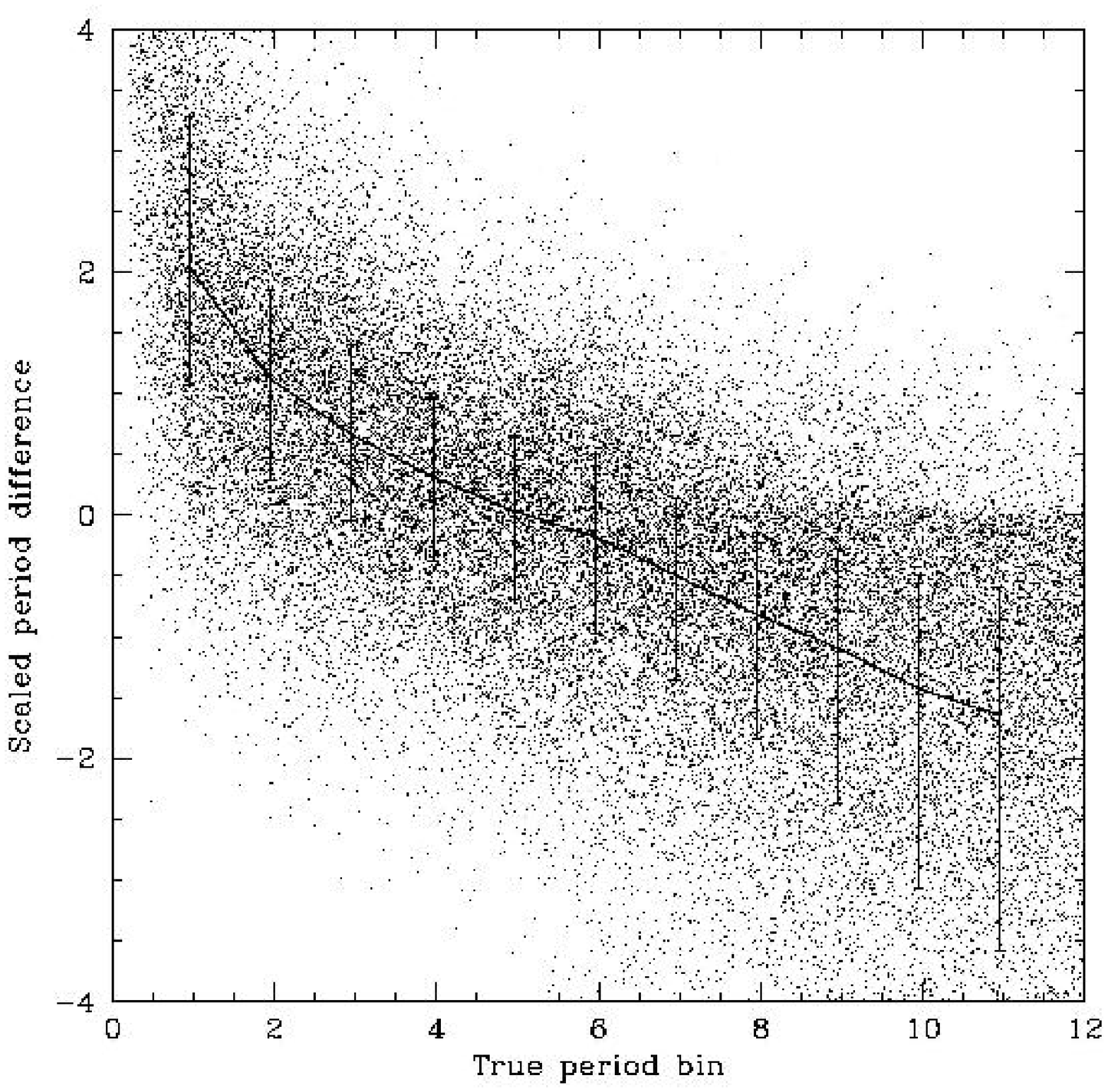} \\
\end{array} $
 \caption {Scaled period differences for Solver A (left) and Solver B (right), for
all orbits with signature larger than 0.4 mas.  The curve and error bar
represent the median and quartiles in 1-year bins in true period.}
\label{fig11}
 \end{figure*}

 \begin{figure*}
\centering
$\begin{array}{cc}
\includegraphics[width=0.4\textwidth]{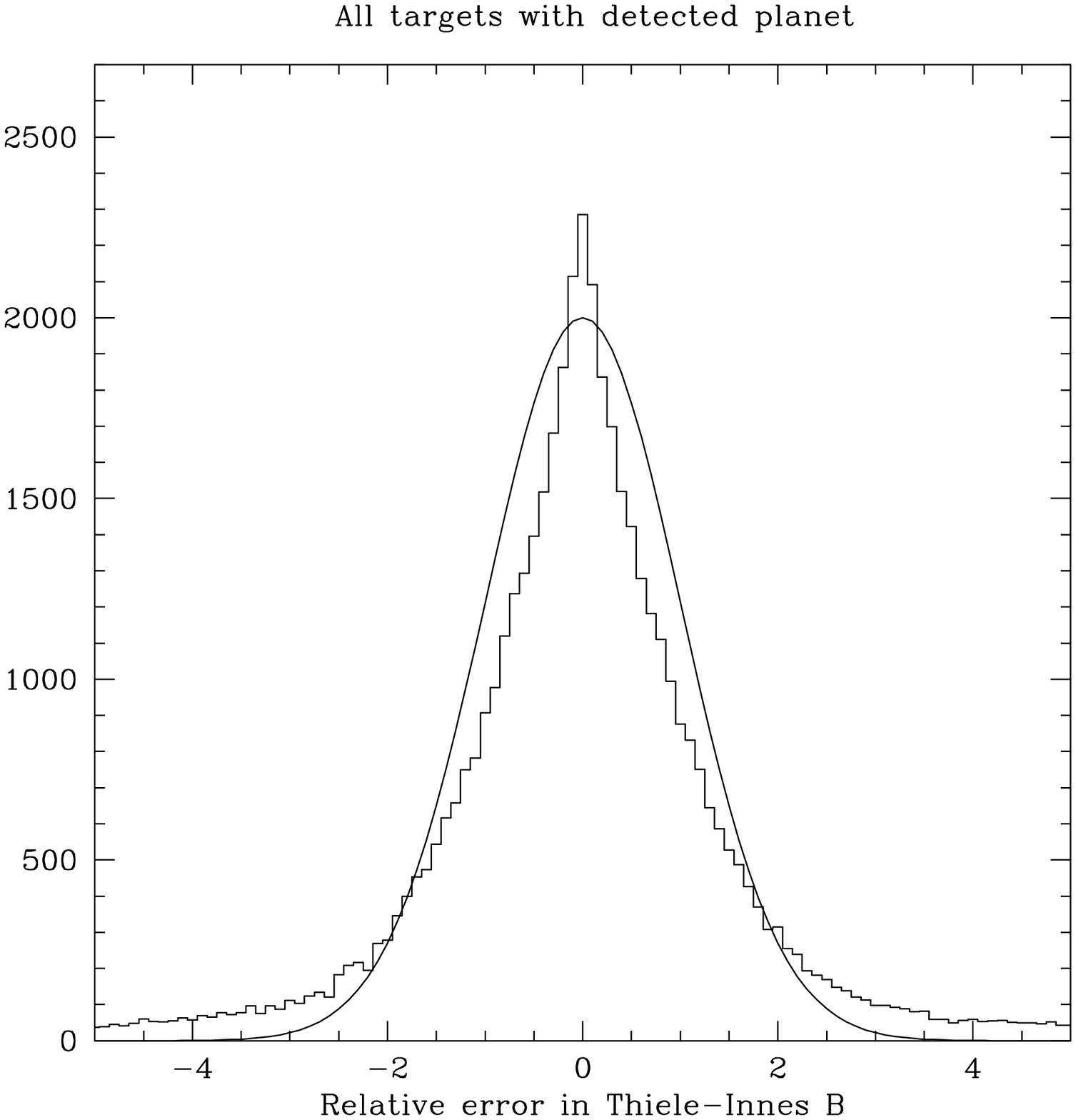} & 
\includegraphics[width=0.4\textwidth]{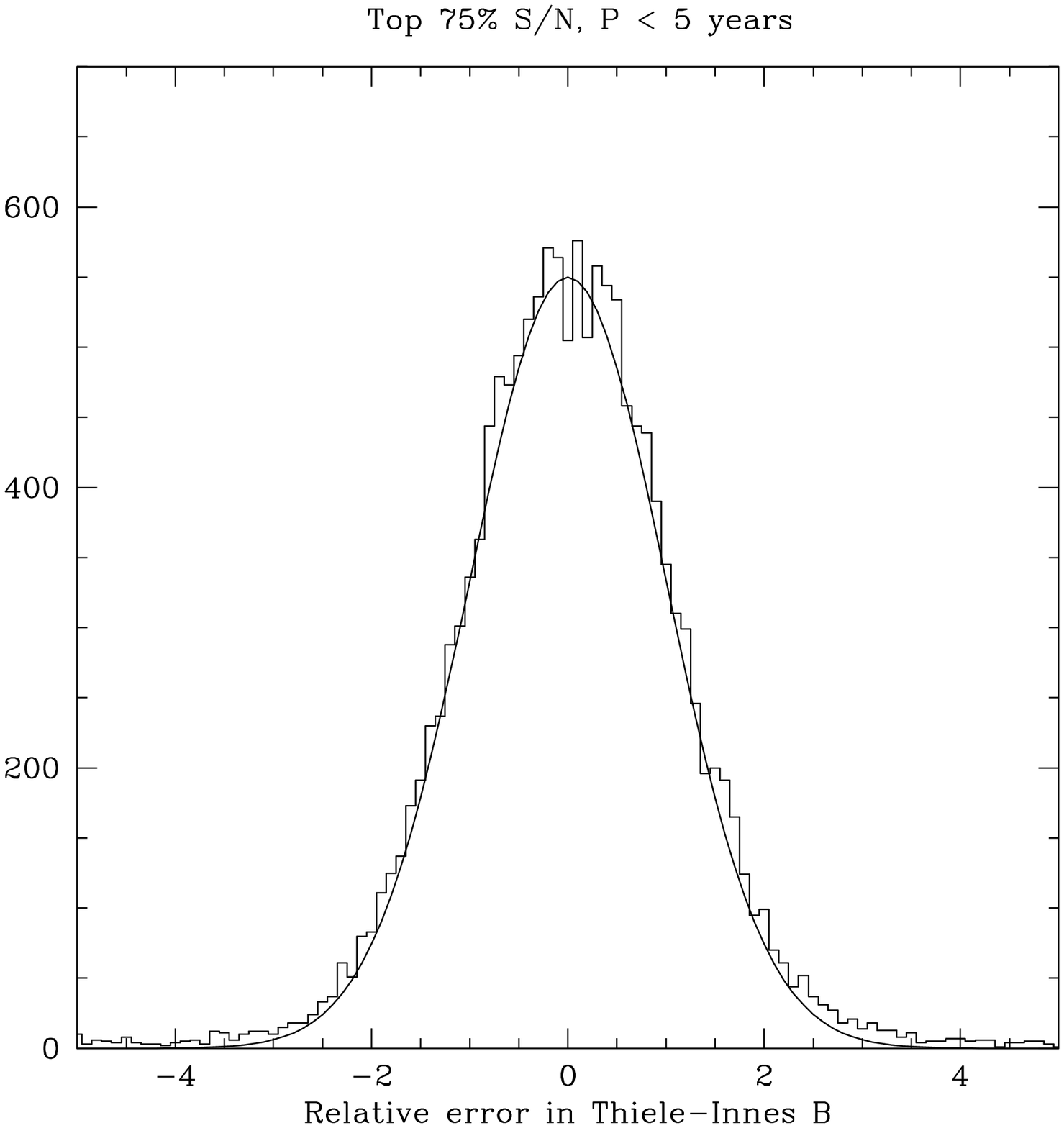} \\
\end{array} $
 \caption{Distribution of scaled difference in the Thiele-Innes
parameter $ B $ for the Solver B solution.  The left panel shows all data
points; the right panel only the planets with signature larger than 0.4
mas and period shorter than 5 years.  The dashed curve in each plot is a
reference Gaussian with zero mean and unit dispersion.}
\label{fig12}
 \end{figure*}

The difference in period appears to be a function of the period itself. 
When considering planets with different periods, it appears that the
period difference decreases for longer periods, and vanishes at $ \sim 5 $ years. 
This appears clearly in Figure~\ref{fig11}, where the median and interquartile
scaled period difference is binned as a function of period for both
solutions.  Periods shorter than 5 years are overestimated, while longer
periods are underestimated.  The difference remains comparable to the
estimated error (one sigma), except for periods around 1 year and
shorter which are overestimated by up to 2 sigma.  We remind the reader
that this is in part a result of the errors being very small; the
typical period error at 1 year is 0.005 years (see Figure~\ref{fig7}), so as a
fraction of the period itself, this bias is typically less than a
percent.  Nonetheless, the fact that the difference is systematic and
present in both solutions suggests that there is a conceptual issue
worth of further analysis.

We next consider the distribution of linear parameters, using the
Thiele-Innes $ B $ parameter in the Solver B solution as an example.  The
overall distribution of scaled errors is, not surprisingly, unbiased in
the mean, and is comparable in width to the expected distribution (Figure~\ref{fig12}, 
left panel). However, the observed distribution does differ from the nominal
Gaussian, both for small and for large errors.  The core of the distribution
appears narrower than the Gaussian, indicating that errors may be overestimated for
part of the distribution; on the other hand, the elevated tails---and the 2\% of
solutions that fall outside the 5-sigma range of the histogram---indicate 
that errors are underestimated for some objects.

A closer analysis shows that the narrow peak is due primarily to planets
with small signatures ($ < 0.4 $ mas) and periods shorter than 5 years,
while the tails are largely due to long-period planets.  Figure~\ref{fig12}, 
right panel, shows
that the distribution of scaled differences for $ B $ for all planets
with signature larger than 0.4 mas and period shorter than 5 years is
very close to Gaussian, although about 2\% of outliers remain.

\subsection{Test T3: Multiple-planet solutions and coplanarity}

The T3 experiment is designed 
primarily to determine how well multiple-planet systems can be identified 
and solved for, as well as how well the mutual inclination angle of 
pairs of planetary orbits can be measured. In addition, the accuracy 
of multiple-planet solutions will be compared with that of single-planet 
solutions for systems with similar properties. 
The noise characteristics of the data are assumed to be fully understood. 

Each solver was asked to carry out a full orbital reconstruction
analysis for each star, beginning from the period search and including
error estimates for each of the orbital parameters.  As for the T1 and 
T2 tests, two solvers, Solver A and Solver B, participated in this test, each with 
their independently developed numerical code.  

Simulated data were prepared for 3,000 stars, in two separate experiments 
(T3a and T3b). 
In the two cases, respectively 310 and 307 objects had one planet, 
while the remaining 2690 and 2693 had 
two planets. In both experiments, astrometric signatures ranged between 
$\alpha=16$ $\mu$as (astrometric signal-to-noise $\alpha/\sigma_\psi\simeq 2$) and 
$\alpha=400$ $\mu$as ($\alpha/\sigma_\psi\simeq 50$). The first planet was 
always generated with a mass M$_p=1$ M$_J$, and with $P$ uniformly, 
randomly distributed between 0.2 and 9 years. The second planet was constrained 
to have $P$ at least a factor 2 shorter or longer than the first planet, 
and its corresponding mass was assigned 
as to produce an astrometric signal falling in the above mentioned range. 
The orbital eccentricity was 
randomly distributed, but limited to the ranges $0.1\leq e\leq 0.6$  
and $0.0\leq e\leq 0.6$ in the T3a and T3b experiments, respectively. 
In the first experiment, no constraints were placed on the value of 
the mutual inclination angle $i_\mathrm{rel}$ 
between pairs of planetary orbits. In the second experiment, it 
was constrained to be $i_\mathrm{rel}\leq 10^{\degr}$. 

Both Solvers run their respective pipelines, consisting of detection,
initial parameter determination, and orbital reconstruction, on each of
the 3,000 simulated time series provided by the Simulators.  They have
no a priori knowledge of the orbital properties of each planet, nor 
they know whether a star has none, one, or more planets. 

In both cases, solvers use the equivalent of a least-squares algorithm
to fit the astrometric data for each planet; they need to solve for the
star's basic astrometric information (position, parallax, proper
motion), for which only low-accuracy catalog parameters are provided, as
well as for the parameters of the reflex motion, for each detected 
companion.  For all planets fitted for, Solver B provides the
results in the form of period $ P $, eccentricity $ e $, epoch of
pericenter passage $ T $, and the Thiele-Innes parameters $ A, B, F, G
$.  He provides also estimated uncertainties for each parameter.  
Solver A also provides period, eccentricity,
and pericenter passage, but instead of the Thiele-Innes parameters, he
returns semi-major axis $ a$, inclination $ i $, position angle of the
ascending node $ \Omega $, and longitude of pericenter $ \omega $.  Like 
Solver B, he computes formal errors for each parameter. 

In summary, the results presented in Sec. 3.3 demonstrate that 
the expected signal can be recovered for 
over 70\% of the simulated orbits under the conditions of the T3 test 
(for every two-planet system, periods shorter than 9 years and 
differing by at least a factor of two, $2\leq\alpha/\sigma_\psi\leq 50$, 
moderate eccentricities). Favorable orbital configurations (both planets with periods 
$\leq 4$ years, both astrometric signals 
at least ten times larger than the nominal single-measurement 
error, and redundancy of over a factor two in the number of 
data points with respect to the number of fitted parameters) 
have periods measured to better than 10\% accuracy $> 90\%$ 
of the time, and comparable results hold for other orbital 
elements. A modest degradation of up to 10\% 
(slightly different for different parameters) is observed 
in the fraction of correct solutions with respect to the 
single-planet solutions of the T2 test. The useful range of 
periods for accurate orbit reconstruction is reduced by about 30\% 
with respect to the single-planet case. The overall 
results are mostly insensitive to the mutual inclination 
of pairs of planetary orbits. Over 80\% of favorable 
configurations have $i_\mathrm{rel}$ measured to better 
than $10^{\degr}$, with only mild dependencies on its 
actual value, or on the inclination angle with respect to 
the line of sight of the planets. 
Error estimates are generally accurate, particularly for 
fitted parameters such as the orbital period, while (propagated) 
formal uncertainties on the mutual inclination angle seem to often 
underestimate the true errors. Finally, it is worth mentioning how, 
as already shown by radial-velocity surveys, long-term astrometric 
monitoring, even with lower per-measurement precision, would be very 
beneficial for improving on the determination of multiple-planet system 
orbits and mutual alignment, thanks to the increasingly higher redundancy 
in the number of observations with respect to the number of estimated 
model parameters in the solutions.

\subsubsection{Overall quality of the solutions}

For both experiments, Solver A reports solutions for all stars. Solver A
 initially carries out an orbital 
solution for a single planet orbiting each star. He then performs 
a $\chi^2$-test on the post-fit observation residuals, at the $99.73\%$ 
confidence level. This allows one to provide an initial 
assessment of the detectability of the signal of a second planet 
in the system, as a function of its properties. 

For the first experiment, 
a total of 509 objects have $P(\chi^2) \geq 0.0027$, 
thus are classified as systems with only one planet. Of these, 
289 out of 310 simulated ones are truly star+planet systems. Of the 
remaining 220 objects orbited by 2 planets but for which a single 
planet solution appears satisfactory, the overwhelming majority of 
the cases (93\%) are constituted by systems in which at least one 
planet has $P$ exceeding the time-span of the observations ($T=5$ yr), 
and often times the inner planet has $P\simeq T$. In virtually 
all cases, the fitted value of the period is close to that of 
the inner planet, or it's intermediate between that of the 
inner and that of the outer planet. 
In the 7\% of cases in which both planets have $P\leq 5$ yr, one 
of the astrometric signatures is always close to the detection limit 
($\alpha/\sigma_\psi\leq 3$). Essentially identical results hold for 
the second experiment. A more thorough investigation of the 
behavior of false detections and of the realm of degradation of 
detection efficiency in presence of a second planet is beyond the 
scope of this report, and it will require much larger sample sizes.
Finally, Solver A performs a two-planet solution 
on all stars. In both experiments, essentially the same fraction of 
systems with two planets ($\approx 73\%$) passes the 
$\chi^2$-test on the post-fit residuals, at the 99.73\% level. 
For the remaining 27\% of cases in which a two-planet solution 
is not satisfactory within the predefined statistical tolerances, 
Solver A does not attempt to fit for a three-planet configuration. 

From the results reported by Solver B for the T3b experiment, 24 
stars have no solution (in 85\% of the cases objects with less than 25 
observations). For the remaining 2976 objects, Solver B fits at least 
two planets, and a 3-planet orbital solution 
is reported for 43\% of the sample. Overall, $\sim 56\%$ of the 
systems are correctly identified by Solver B as having only two planets, 
with post-fit $P(\chi^2) \geq 0.05$. The T3a experiment yielded very similar 
results. 

Overall, only $\sim 40\%$ of the two-planet systems simulated have a good solution according to 
both Solvers. Simply based on the post-fit 
$\chi^2$ test, the two fitting algorithms thus perform differently 
in a measurable fashion, unlike the T2 test case, in which the 
performance of the two codes for single-planet orbital fits 
was essentially identical.

The next steps are to focus on good ($P(\chi^2) \geq 0.0027$ for 
Solver A, $P(\chi^2) \geq 0.05$ for Solver B) two-planet solutions reported 
by the Solvers when the simulated systems are truly composed of 
two planets, and investigate a) how well solvers actually recover the orbital 
parameters of the planets, b) how the quality of multiple-planet 
solutions compares with that of single-planet fits for planets with 
comparable properties, and c) how accurately the actual value of the 
mutual inclination angle $i_\mathrm{rel}$ is recovered in the case 
of quasi-coplanar and randomly oriented pairs of planetary orbits.
 
\subsubsection{Multiple-Keplerian orbit reconstruction}

\begin{figure*}
\centering
\includegraphics[width=1.\textwidth]{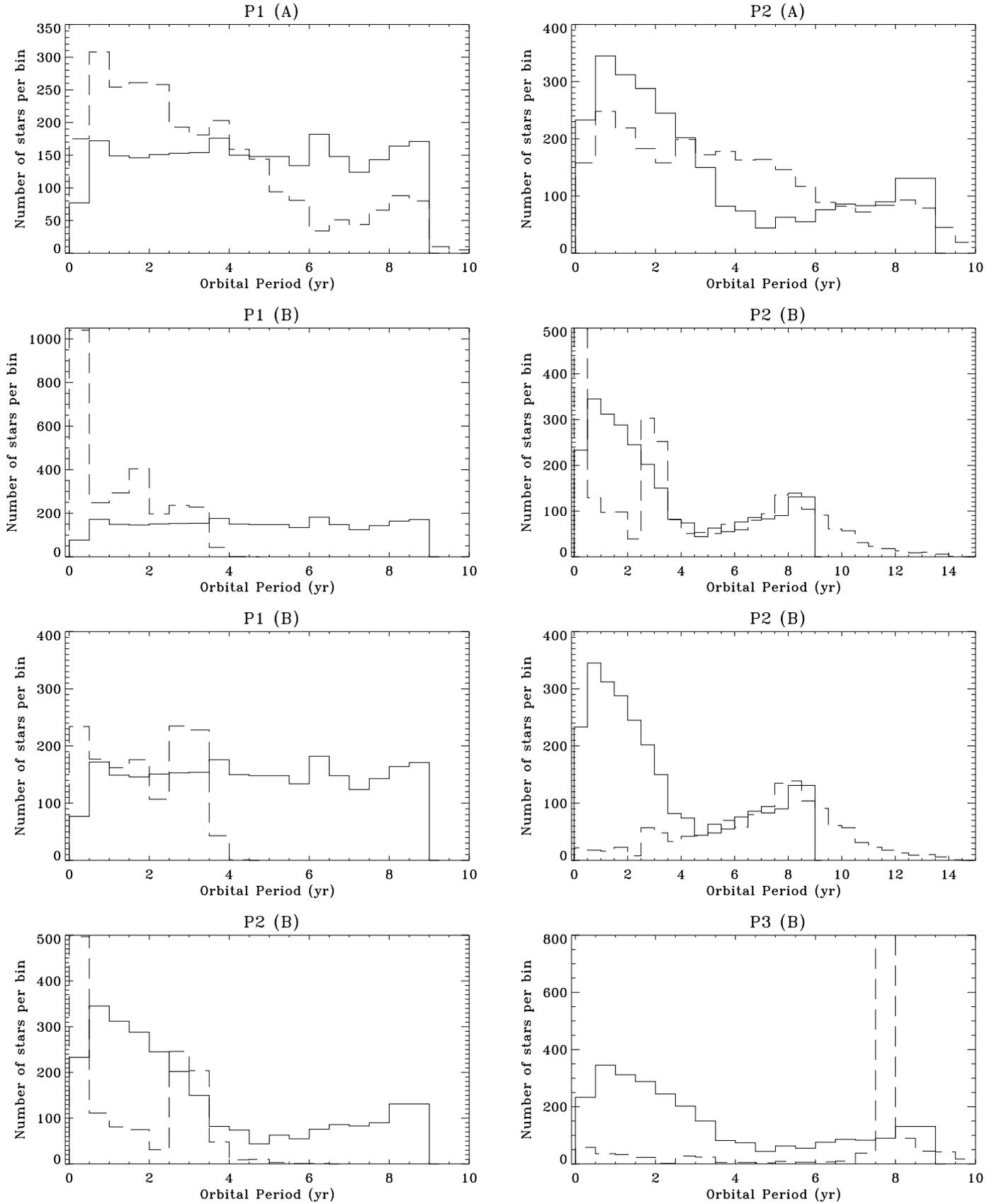}  
\caption{\small Distribution of orbital periods in the multiple-planet 
solutions (dashed and dashed-dotted lines), compared with the 
true underlying distributions (solid lines). 
Top two panels: results for planet 1 and 2 obtained by 
Solver A (all stars). Panels 3 and 4: the same for Solver B, including 
stars with both two and three planets found. Panels 5 and 6: 
the same for Solver B, but excluding stars with three planets fitted. 
Bottom two panels: the true distribution of the second planet 
compared with the same distributions for planet two and 
three obtained by Solver B in the sample of three-planet orbital fits.}
\label{pcomp1}
\end{figure*}

\begin{figure*}
\centering
\includegraphics[width=1.\textwidth]{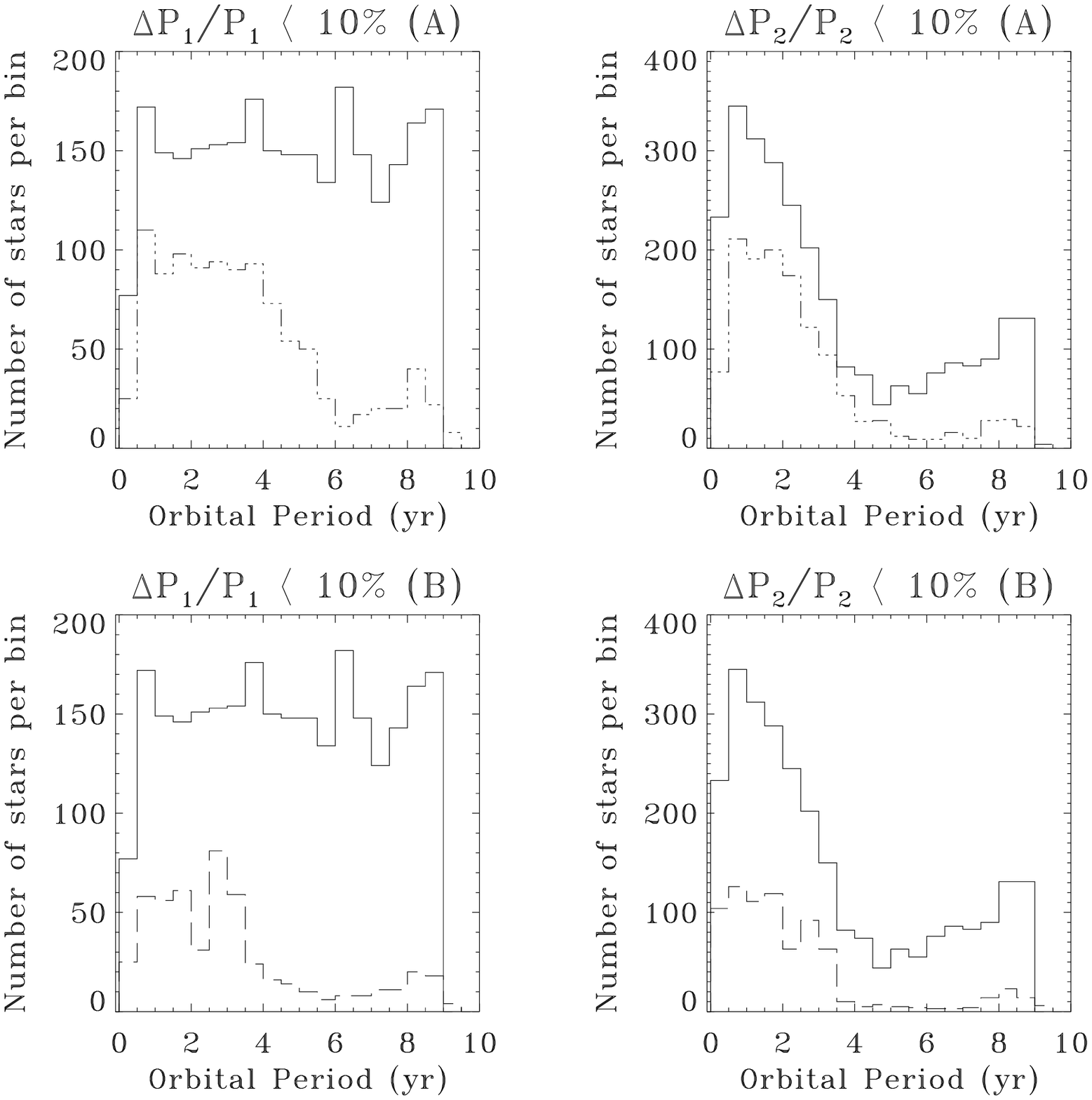}  
\caption{True distributions for planet 1 and 2 (solid histogram) 
compared 
with the same distributions derived by Solver A (top two panels) 
and Solver B (bottom two panels) when the fitted values of the 
periods lie within 10\% of the simulated values.}
\label{pcomp2}
\end{figure*}

The relative performances of Solver A's and B's algorithms in accurately recovering 
the orbital parameters in the case of two-planet systems are quantified using the results for 
the orbital period of the two planets. This is the most important of the orbital parameters, and 
the most critical in terms of obtaining an orbital solution that is close to the truth. 
As already noted above, we find that the overall performance in multiple-planet orbit 
reconstruction does not depend significantly on the relative alignment of the 
orbits, so that we present here results from the T3b experiment, i.e. the quasi-coplanar 
orbits case.

The first noticeable result are the large differences in the 
distributions of orbital parameters for the two Solvers. Figure~\ref{pcomp1} 
shows, compared to the true simulated ones (solid histograms), 
the recovered distributions 
of orbital periods of the first and second planet. In the upper four 
panels, the results for all stars (excluding objects with only one planet, 
but for Solver B including those for which three planets are fitted) 
are presented for both Solvers. In panels five and six, Solver B's 
results are shown only for stars with good two-planet solutions, 
while in the last two panels Solver B's distributions of periods of 
the second and third planet are presented, for the sample of 
stars with three-planet orbital solutions. 

On the one hand, for Solver A's solutions (panels 1 and 2) the most 
obvious feature observed is the fact that in a significant number of 
orbital solutions the periods are swapped (roughly 
30\% of the cases, averaging over all periods), i.e. the 
first planet identified in the data is the second generated in 
the simulations, and vice-versa. This result is easily understood, 
as, given the simulation setup, the dominant signal (identified 
by, for example, a better sampled period, or a larger 
astrometric signature) is not necessarily the one of the first 
planet generated. Otherwise, Solver A's solutions appear to recover 
reasonably well the true underlying distributions.

On the other hand, for Solver B's solutions no obvious pattern 
of this kind can be found. Instead, over 1/3 of the periods 
identified as dominant is within 0.5 years, and no periods 
greater than 5 years are identified (panel 3). The former feature is 
in common to the solutions for the second planet (panel 4). 
When only two-planet solutions (with good post-fit $P(\chi^2)$) 
are considered (panels 5 and 6), the recovered distributions still 
look largely different from those obtained by Solver A and from the 
true ones. Finally, as it appears clear by comparing 
panels 7 and 8, and 5 and 6, the vast majority of short-period 
period orbits fitted for the second planet ($\sim 90\%$), 
and $\sim 50\%$ of those for the first planet, seems to be 
the undesired consequence of three-planet fits, with a correspondingly 
very large number of long periods found for the third planet.  

Such differences translate in a lower percentage of correctly 
identified two-planet systems by Solver B (even when the post-fit 
$\chi^2$-test is satisfactory). In fact, 
in Figure~\ref{pcomp2} we show the distributions of true periods 
for the first and second planet compared to the fitted 
distributions when the value of the period falls within 10\% 
of the simulated one. In order to compare results between the 
two Solvers in almost identical conditions, for Solver A only stars 
with post-fit $P(\chi^2) \geq 0.05$ are included, while for Solver B only 
two-planet solutions are considered (all having $P(\chi^2) \geq 0.05$). 
Overall, Solver B's algorithm performs about 40\% worse than Solver A's 
(for the first and second planet respectively, 554 and 807 stars satisfy 
the above constraints for Solver B, while for 
Solver A the equivalent numbers are 993 and 1223). 
This difference increases to over a factor of two 
if Solver A's $P(\chi^2) \geq 0.0027$ criterion is adopted. 
The number of stars with both 
periods simultaneously satisfying the above conditions is also 
lower for Solver B, by some 15\%. It is true that about 10\% 
of the stars for which Solver B performs three-planet fits actually 
have the orbital period of the first and second planet fitted 
falling within the above-mentioned criteria, thus helping to 
somewhat reduce the observed discrepancy in performance. However, 
we will only focus on Solver A's 
$\sim 70\%$ of good two-planet orbital solutions (at the 99.73\% 
confidence level), a total of 1912 and 1900 stars for the 
T3a and T3b tests, respectively. Focusing on Solver A's cleaner, and larger, 
sample of good orbital solutions allows one to effectively 
undertake the comparison between the T2 and T3 tests, by using 
stellar samples for which orbital solutions have comparable 
quality. 

\begin{figure*}
\centering
$\begin{array}{cc}
\includegraphics[width=0.4\textwidth]{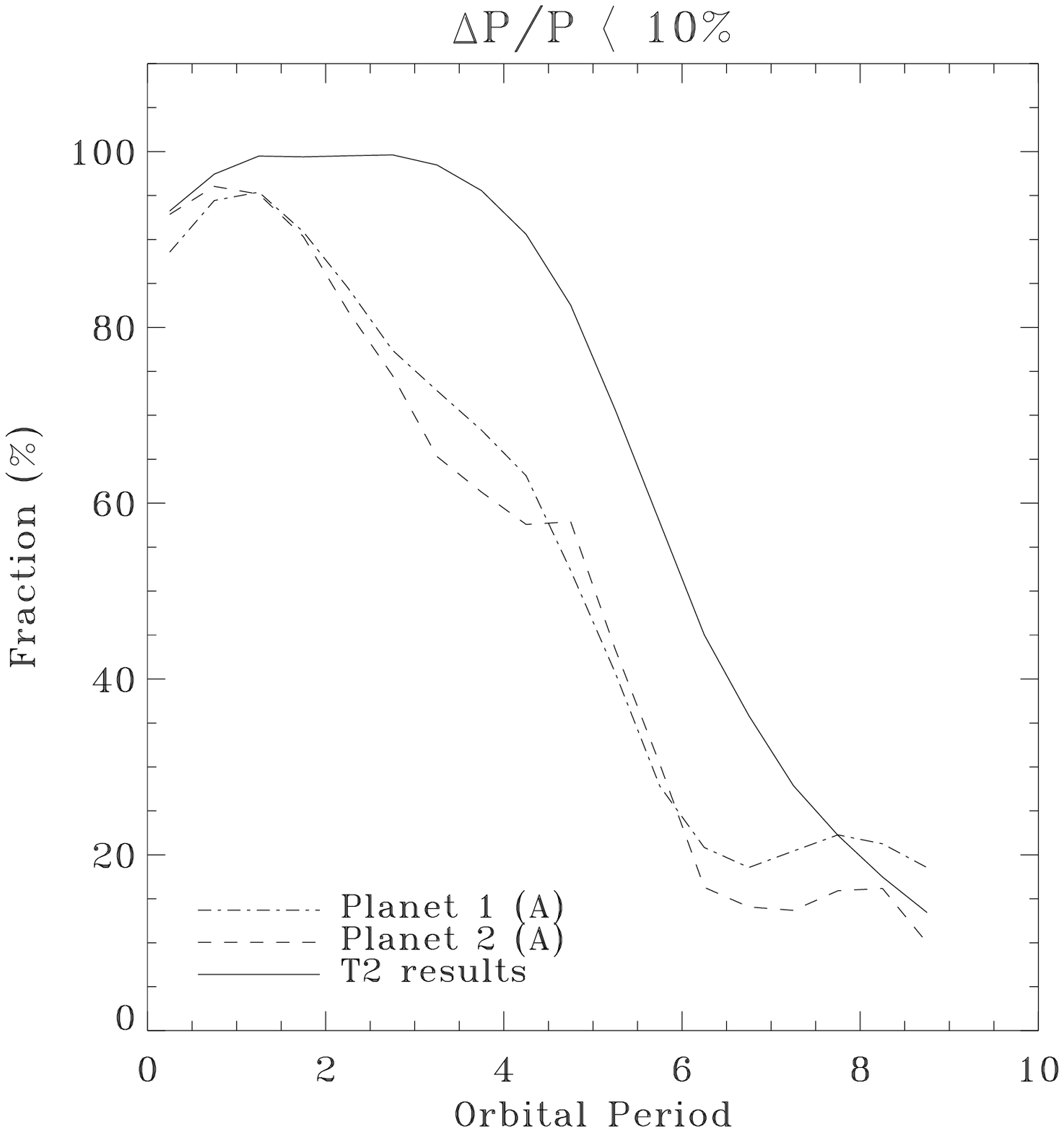} & 
\includegraphics[width=0.4\textwidth]{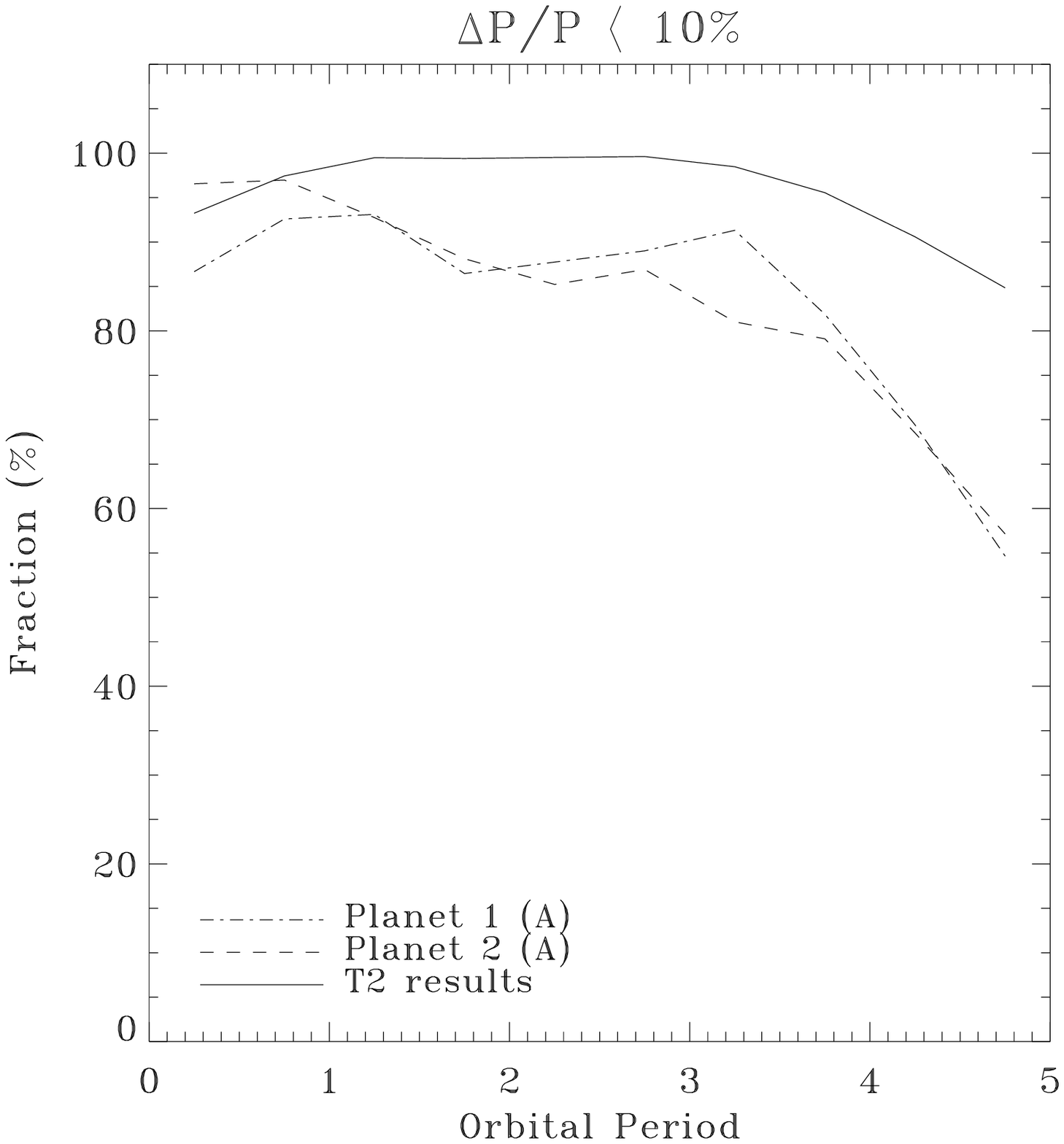} \\
\includegraphics[width=0.4\textwidth]{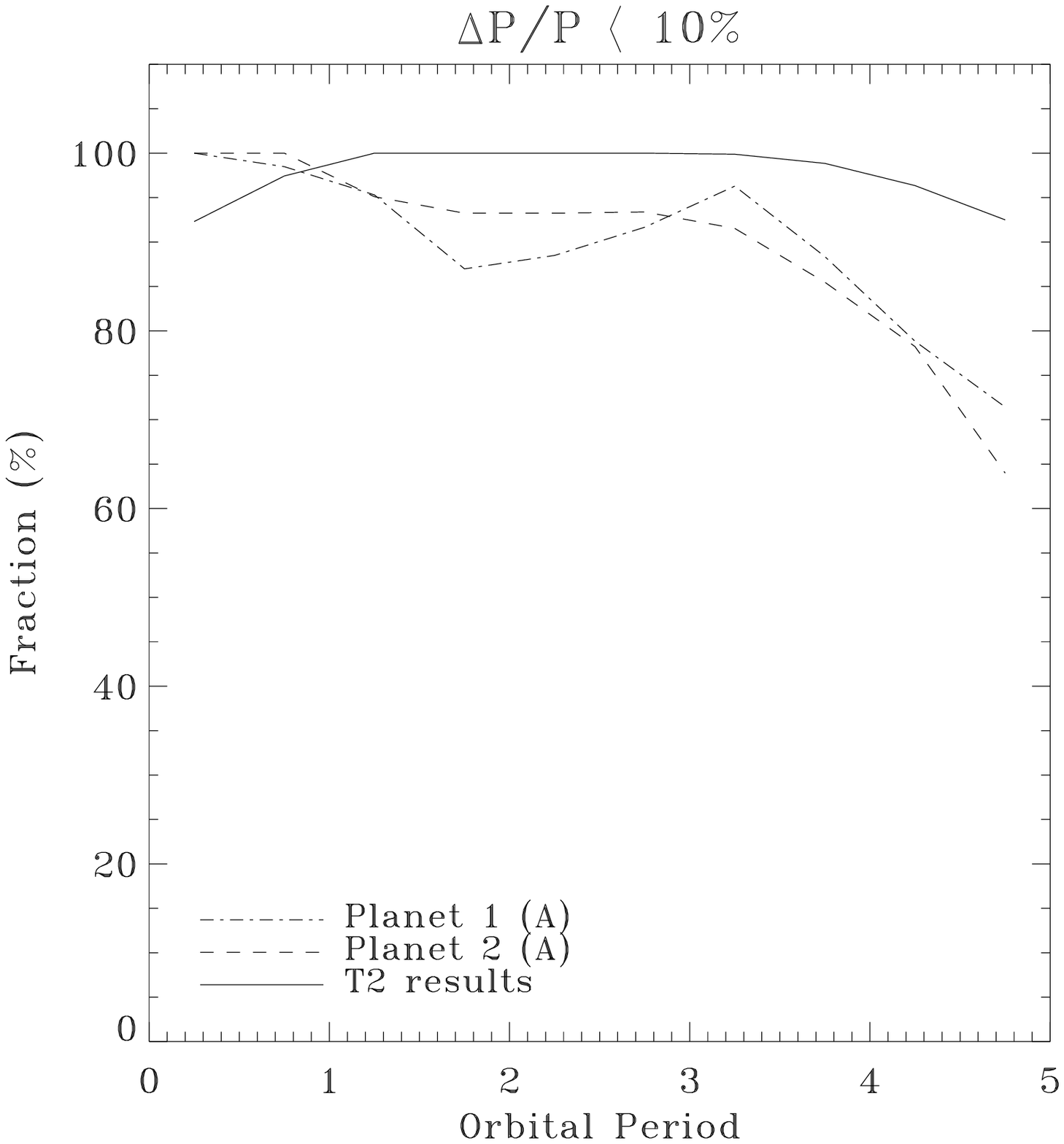} &
\includegraphics[width=0.4\textwidth]{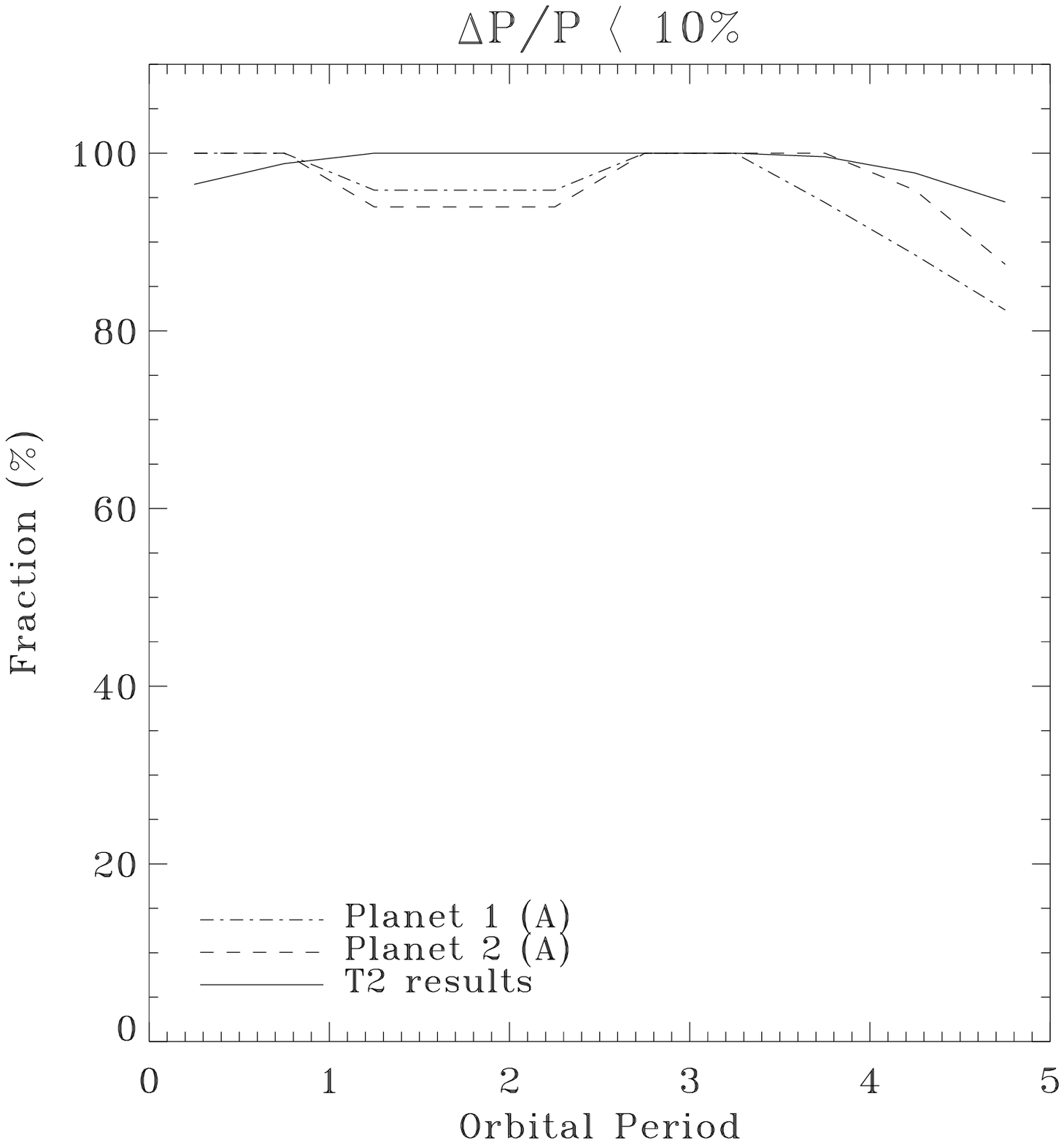} \\
\end{array} $
\caption{Fraction of systems with good orbital solutions 
($P(\chi^2)\geq 0.0027$) in the T3b experiment for 
which both orbital periods are recovered 
by Solver A with a fractional uncertainty $\leq 10\%$, as a function of period 
(0.5-yr bins). For comparison, the same results are displayed for the T2 test. 
Top left: all stars. Top right: systems with both periods $\leq 5$ yr. 
Bottom left: systems with both periods $\leq 5$ yr, 
and with $\alpha/\sigma_\psi \geq 10$. 
Bottom right: systems with both periods $\leq 5$ yr, $\alpha/\sigma_\psi \geq 10$, 
and with $N_\mathrm{oss}\geq 45$.}
\label{pdist}
\end{figure*}

%\begin{figure}[!th]
%\centering
%$\begin{array}{cc}
%\includegraphics[width=0.4\textwidth]{err_dist_p1.eps} %&
%\includegraphics[width=0.4\textwidth]{err_dist_p2.eps} \\
%\end{array} $
%\caption{Distribution of estimated error in the period 
%of planet 1 as a function of period for AS 
%(T3b experiment). All stars with $P(\chi^2)\geq 0.0027$ are included. 
%The quickly decreasing fraction of accurately determined periods 
%(to better than 10\%) demonstrated in the top left panel of 
%Figure~\ref{pdist} is reflected in the nominal period error, which 
%increases to a comparable extent.}
%\label{errdist}
%\end{figure}

\begin{figure}[!t]
\centering
\includegraphics[width=0.9\columnwidth]{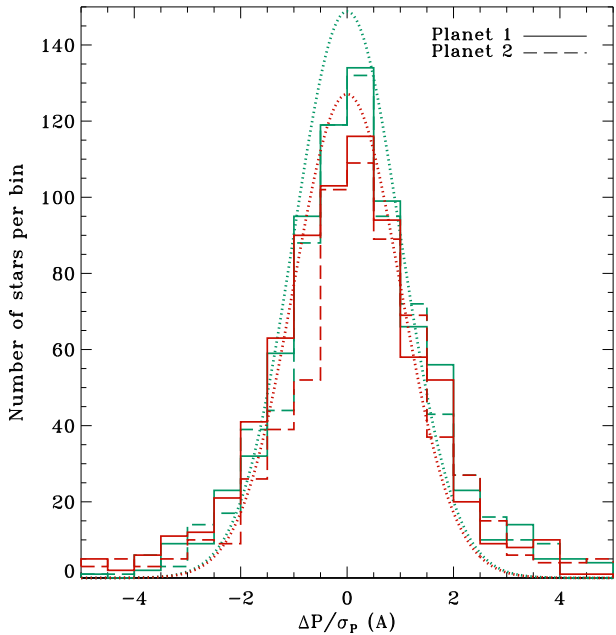} 
%$\begin{array}{cc}
%\includegraphics[width=0.4\textwidth]{scaled_p_diff_rand.eps} &
%\includegraphics[width=0.4\textwidth]{scaled_p_diff.eps} \\
%\end{array} $
\caption{Histogram of scaled period differences $\Delta P/\sigma_P$ for 
good two-planet fits ($P(\chi^2)\geq 0.0027$) with periods 
accurate to better than 10\% for the T3a (green solid and dashed lines) 
and T3b (red solid and dashed lines) experiments. The dotted curves are 
reference Gaussians with zero mean and unit dispersion.}
\label{scaled}
\end{figure}

\begin{figure*}[!t]
\centering
$\begin{array}{cc}
\includegraphics[width=0.4\textwidth]{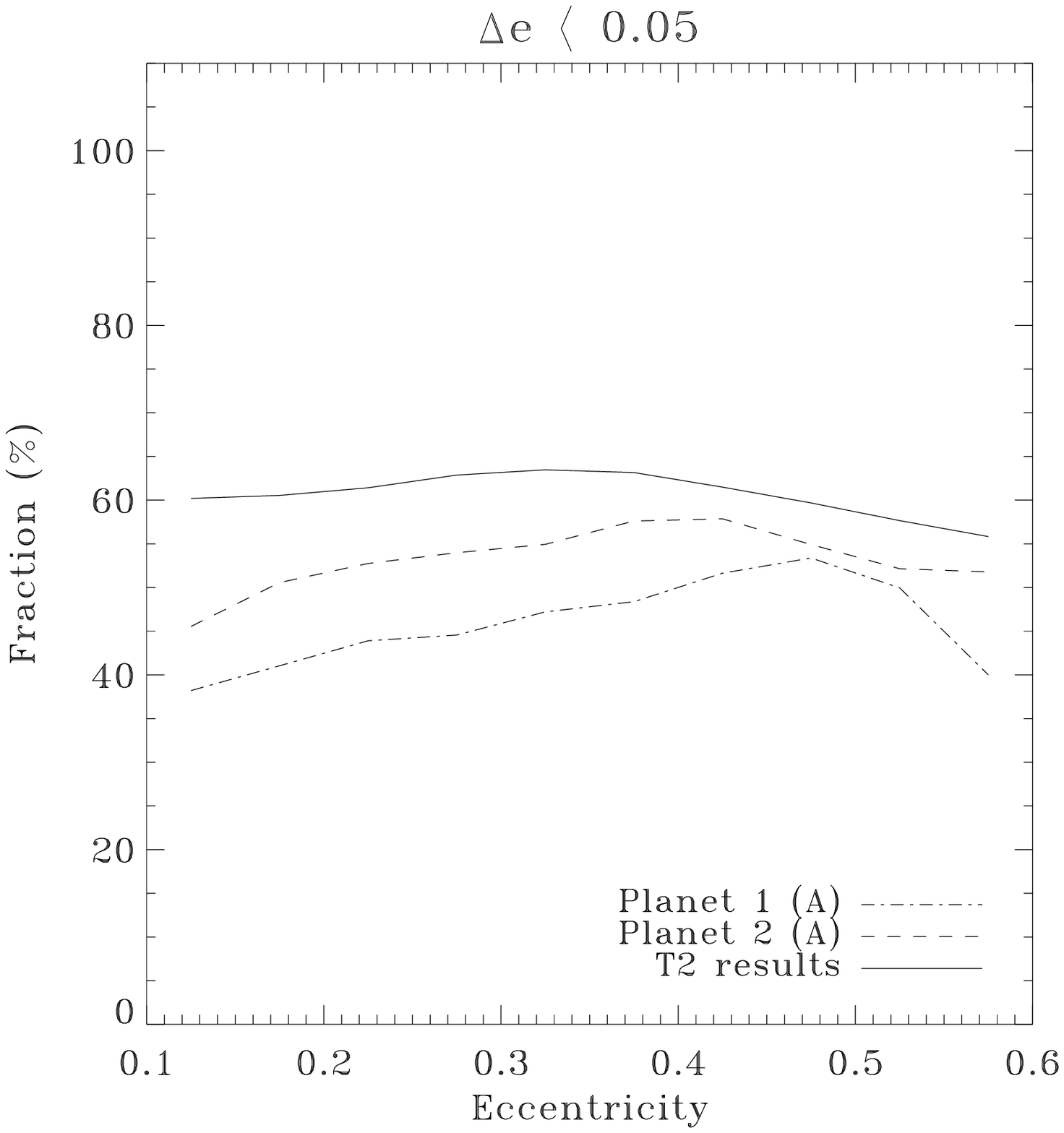} &
\includegraphics[width=0.4\textwidth]{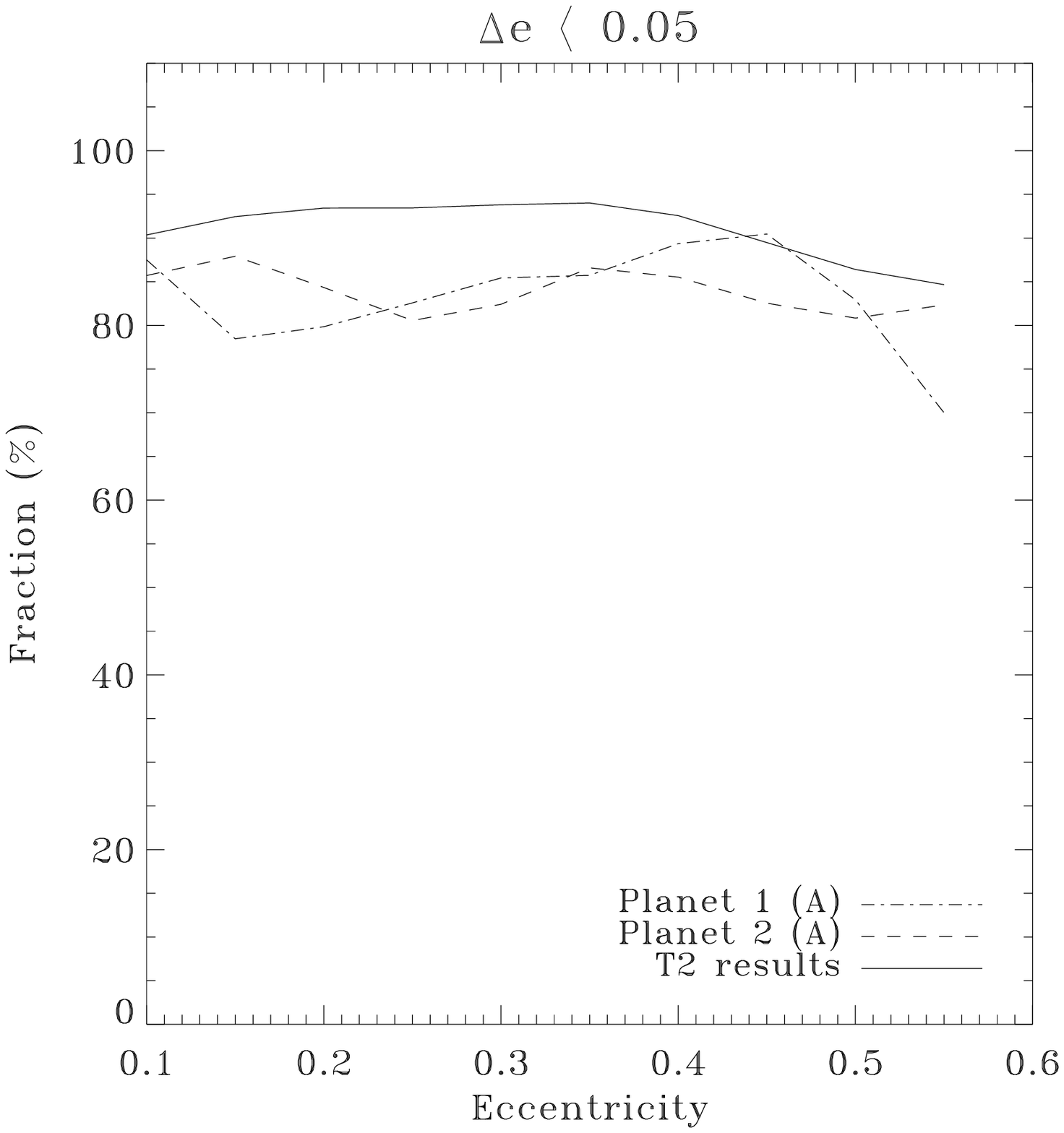} \\
\end{array} $
\caption{Left: Fraction of systems with good orbital solutions 
($P(\chi^2)\geq 0.0027$) in the T3b experiment for 
which both eccentricities are determined 
within 0.05 of the true values. For comparison, analogous results for the 
T2 test sample are displayed. Left: all stars. 
Right: systems with both periods $\leq 5$ yr, with $\alpha/\sigma_\psi \geq 10$, 
and with $N_\mathrm{oss}\geq 45$.}
\label{edist}
\end{figure*}

\subsubsection{Comparison with test T2}

We use orbital period and eccentricity as proxies to understand the behavior 
of the two-planet orbital solutions, and compare them with analogous 
results obtained in the T2 experiment. The properties of good two-planet solutions should 
thus be easier to understand.

For the T3b case (quasi-coplanar orbits), 
the four panels of Figure~\ref{pdist} show, as a function of the value of the 
true orbital period, the fraction of stars with good orbital 
solutions for which the periods 
of both planets are recovered by Solver A with a fractional uncertainty 
$\Delta P/P\leq 10\%$ (where $\Delta P$ is the difference between fitted and 
true period value). For comparison, analogous results from the 
T2 experiment are over-plotted, after constraining orbital periods, eccentricities, 
and astrometric signals to lie in the same ranges of the T3b experiment 
($P\leq 9$ yr, $e\leq 0.6$, and $\alpha\leq 400$ $\mu$as). 

Overall, the quality of the solutions degrades quickly already for periods 
$\geq 2$ years, and the fraction of systems with both orbital periods 
recovered to within 10\% of the true value is at least 5\%-10\% lower than the 
single-planet case. For configurations in which both planets have 
$P\leq 5$ yr, $\alpha/\sigma_\psi\geq 10$, and for which a number 
$N_\mathrm{oss}\geq 45$ of observations are 
carried out over the 5-yr simulated mission lifetime 
(bottom right panel), 
the situation improves significantly. Over 90\% of all orbital configurations 
have both periods measured to better than 10\%, 
and the 5\%-10\% deficit with respect to the T2 experiment applies 
for periods in the range $0.2\leq P\leq 4$ yr, for both planets in the systems. 
A very similar behavior is observed (but not shown) in the T3a experiment, 
in which no constraints are put on the mutual inclination angles. 

Formal errors from the fitting procedure appear to match the actual errors reasonably well. 
To determine more quantitatively how good an approximation the estimated errors are 
for the true ones, we utilize the same metric adopted in the T2 experiment, 
i.e. the scaled difference $\Delta\mathrm{P_j}/\sigma_\mathrm{P_j}$ ($j=1,2$) 
defined as the ratio between the fitted and the true value of the 
orbital period of the $j$-th planet and its corresponding formal 
uncertainty. We limit ourselves to the sample of stars for which 
Solver A obtains good solutions (99.73\% confidence level), and for which 
orbital periods are recovered to within 10\% accuracy. Figure~\ref{scaled} 
shows that, for both planets, and in both the T3a and T3b experiment, 
the distributions of scaled period differences 
are quite close to the predicted value (a Gaussian with zero mean and unit 
dispersion). A small shift in the peak of the $\Delta\mathrm{P}/\sigma_\mathrm{P}$ 
distribution for the second planet in the T3b test might be present, but its 
statistical significance is low. Elevated tails, however, indicate that 
a non-negligible fraction of objects have underestimated periods (7\% of the 
objects lie above the 3-$\sigma$, and 2\% above the 5-$\sigma$ threshold 
out of the scale of the plot in Figure~\ref{scaled}).

Finally, the two panels of Figure~\ref{edist} show results for the eccentricities 
of both planets in the systems. Displayed are the fractions of systems 
with good orbital solutions for 
which the fitted values of $e$ are within 0.05 of the true value, 
the left panel displaying results from the full sample with good 
orbital solutions, and 
the right after applying the above-mentioned constraints on periods, 
astrometric signal, and number of observations. Overall, 
for both planets a degradation of $\sim 20\%$ between the single-planet and the 
two-planet solutions is observed, independently of the actual value of 
$e$. Favorable configurations have $e$ determined within 0.05 
of the true value about 80\% of the time, with a degradation of $\sim 10\%$ 
with respect to the single-planet solutions of the T2 test, in line 
with what is found for the orbital periods. The modest 
degradation of $\sim 5\%-10\%$ in the fraction of well-measured periods and 
eccentricities with respect to the result of the T2 test is likely due to the increased  
number of parameters in the two-planet fits (19 vs. 12 in the single-planet 
solutions), given the same number of observations. Other orbital parameters 
follow similar patterns. And again, essentially 
identical results are obtained for the T3a test, demonstrating that 
the relative alignment between pairs of planetary orbits does not 
seem to play a significant role in terms of the ability of Solver A's 
algorithm to reconstruct with good accuracy the orbits of both planets, 
under favorable conditions.

\subsubsection{Coplanarity measurements}

 \begin{figure*}[!t]
\centering
$\begin{array}{cc}
\includegraphics[width=0.4\textwidth]{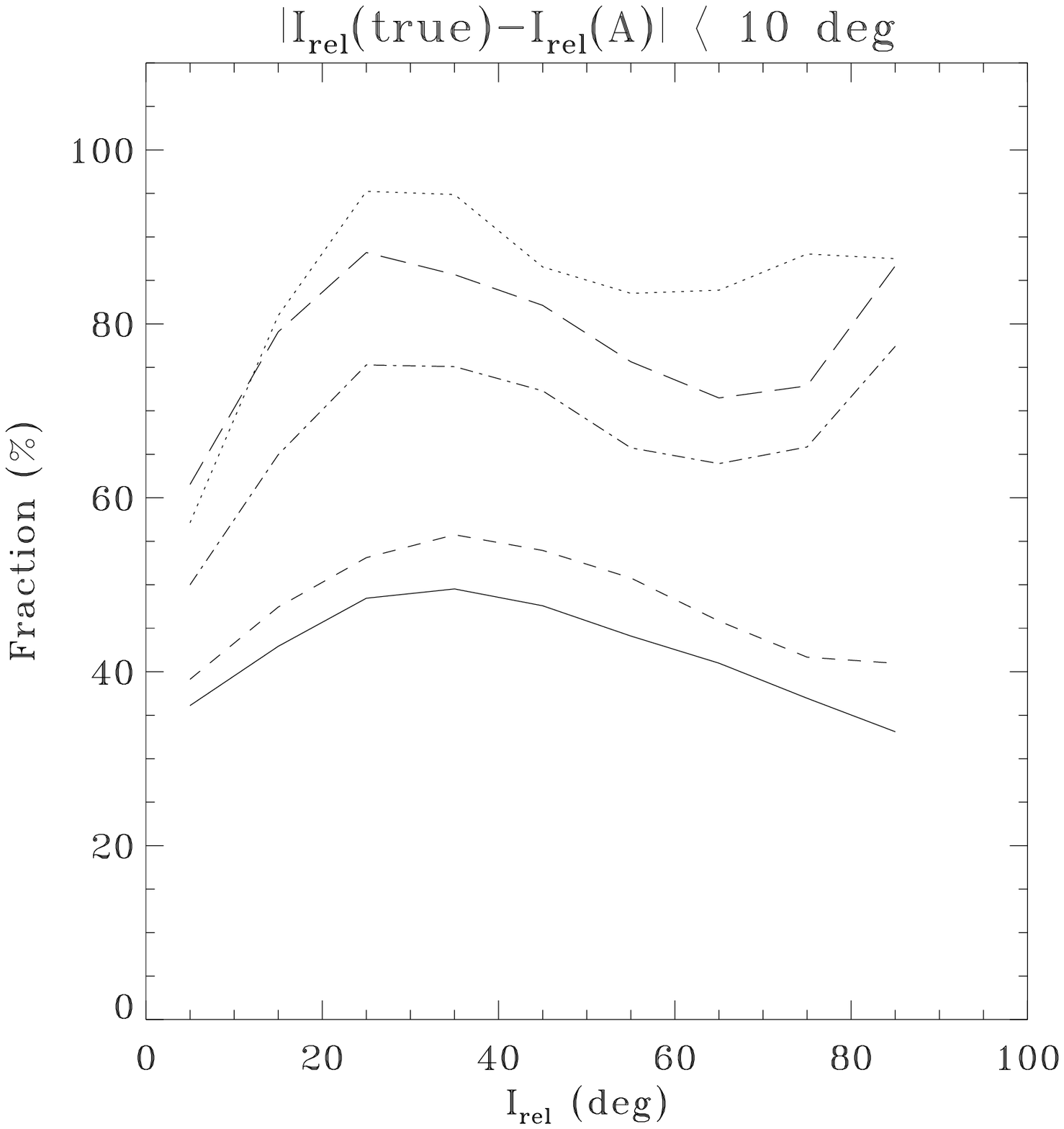} &
\includegraphics[width=0.4\textwidth]{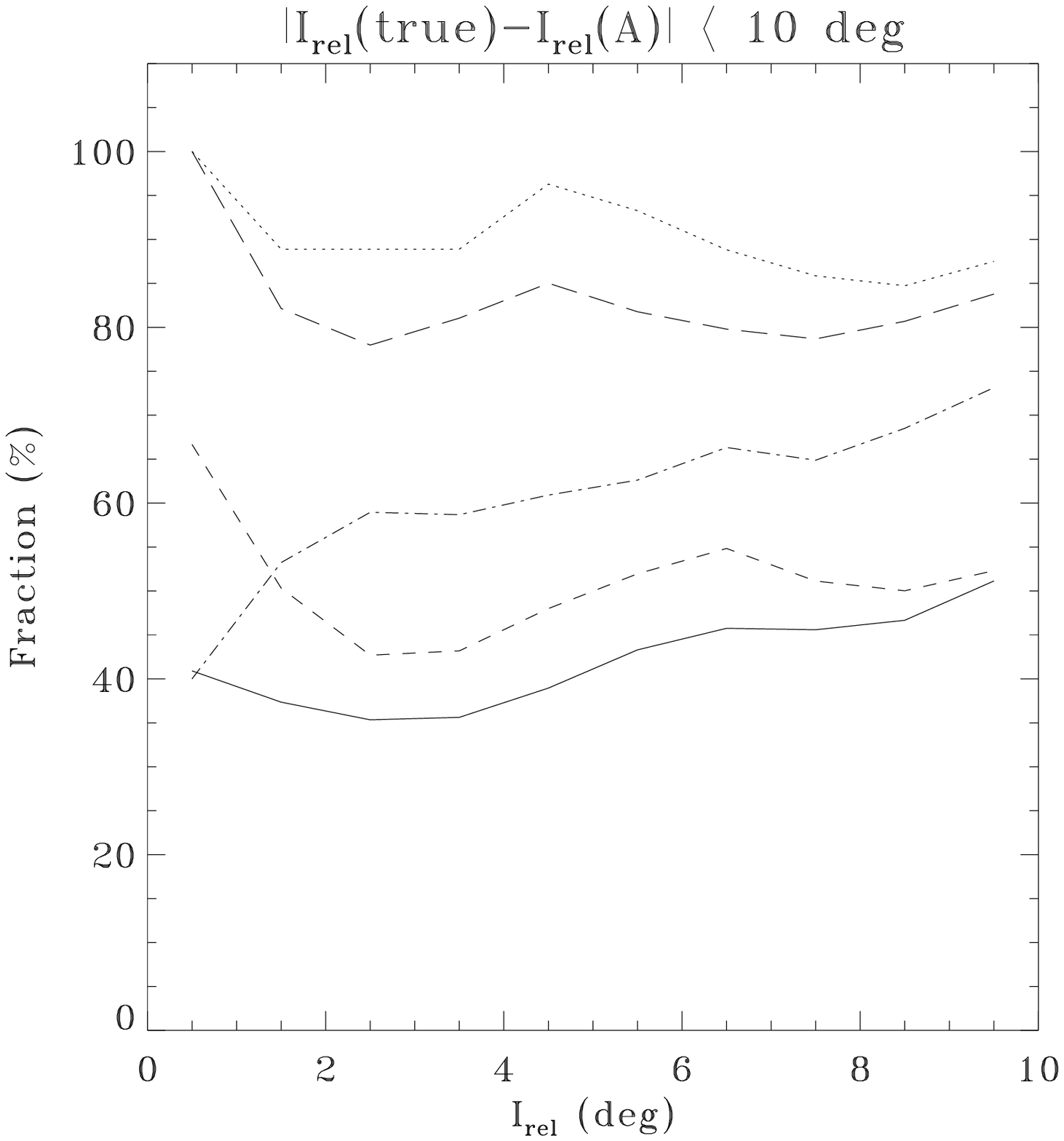} \\
\includegraphics[width=0.4\textwidth]{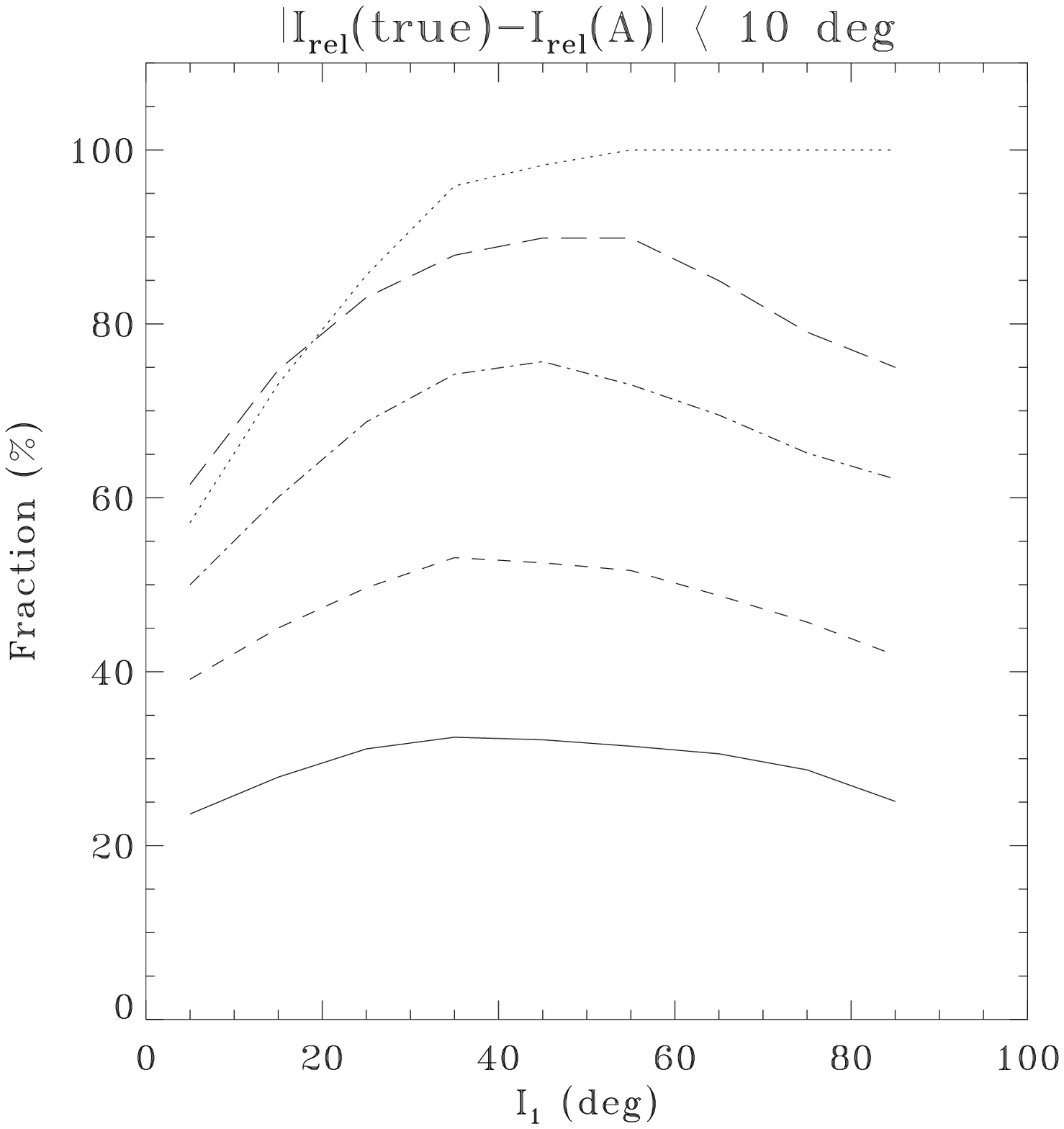} &
\includegraphics[width=0.4\textwidth]{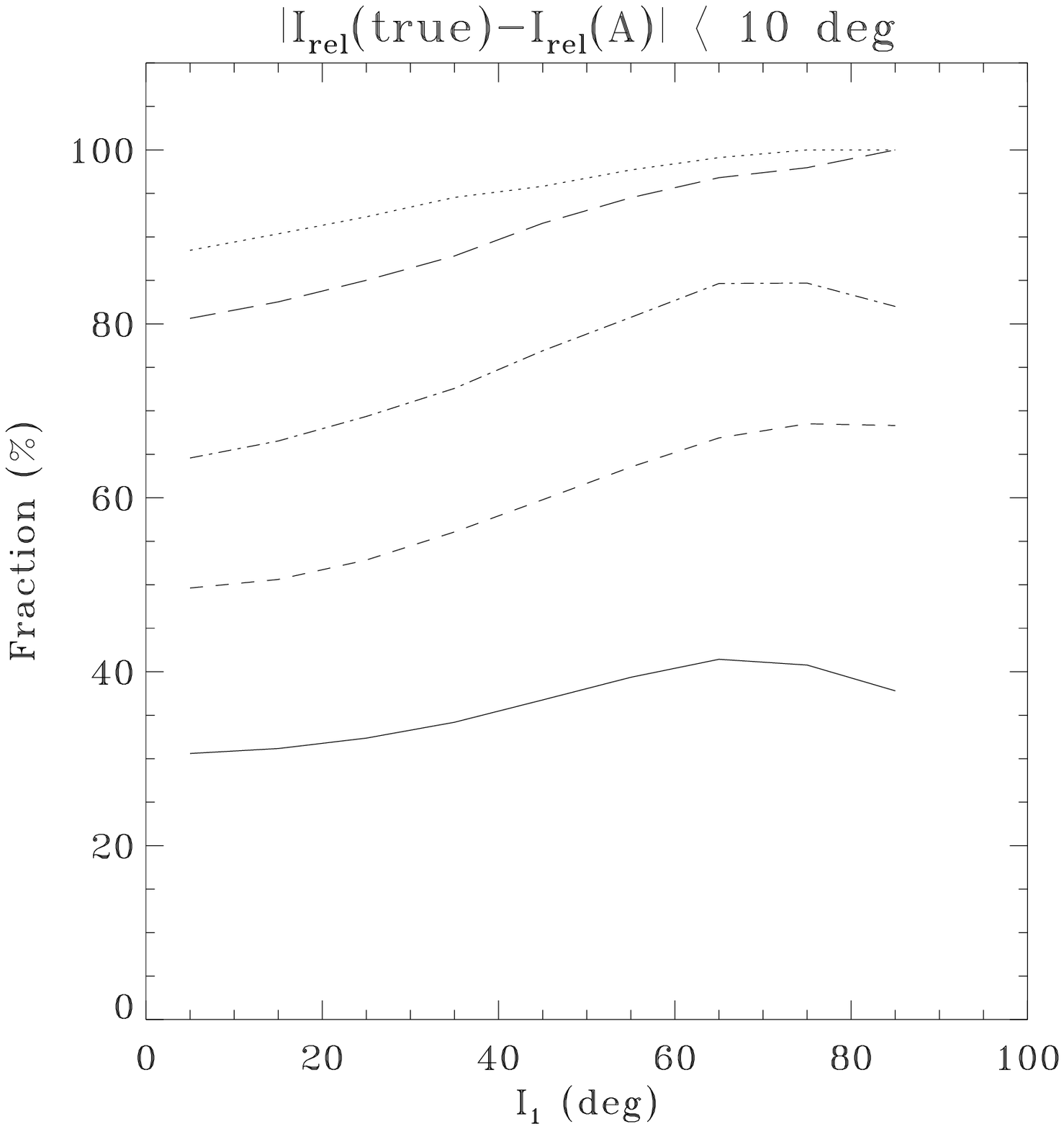} \\
\end{array} $
 \caption{Top left: Fraction of systems in the T3a experiment 
 with satisfactory goodness-of-fit ($P(\chi^2)\geq 0.0027$) 
 for which $i_\mathrm{rel}$ is determined to within $10^{\degr}$ 
 of the true value, as a function of $i_\mathrm{rel}$ itself 
 (10 deg bins).  Top right: the same for the T3b test 
 (1 deg bins in $i_\mathrm{rel}$). Solid lines: all stars; 
 dashed line: both orbital periods $\leq 5$ years; dashed-dotted 
 line: $\alpha/\sigma_\psi\geq 10$; long-dashed line: both 
 orbital periods $\leq 5$ years and $\alpha/\sigma_\psi\geq 10$; 
 dotted line: both 
 orbital periods $\leq 5$ years, $\alpha/\sigma_\psi\geq 10$, 
 and $N_\mathrm{oss}\geq 45$. Bottom left 
 and right: Same as the top two panels, this time as a function of 
 the inclination angle of one of the two planets.}
\label{irel}
 \end{figure*}

 The mutual inclination $i_\mathrm{rel}$ of two orbits is 
defined as the angle between the two orbital planes, and 
is given by the formula: 
 \begin{equation}\label{inclrel}
\cos i_\mathrm{rel} = \cos i_\mathrm{in}\cos i_\mathrm{out}+ \sin
i_\mathrm{in}\sin i_\mathrm{out} \cos(\Omega_\mathrm{out}-
\Omega_\mathrm{in}),
\end{equation}

where $i_\mathrm{in}$ and $i_\mathrm{out}$, $\Omega_\mathrm{in}$
and $\Omega_\mathrm{out}$ are the inclinations and lines of nodes
of the inner and outer planet, respectively.
The value of $i_\mathrm{rel}$ is thus a trigonometric function of $i$ and $\Omega$ of 
both planets, and the latter two are in turn derived as non-linear 
combinations of the four Thiele-Innes elements, which are the actual 
parameters fitted for in the orbital solutions. It is thus conceivable 
that any uncertainties in the determination of the linear parameters 
in the two-planet solutions might propagate in a non-trivial manner 
onto the derived value of $i_\mathrm{rel}$, and consequently a 
value of mutual inclination angle close to the truth might be more 
difficult to obtain.  

In the top two panels of Figure~\ref{irel} we show the fraction of 
stars with good orbital solutions in the T3a and T3b experiments 
for which the derived value of the mutual inclination angle 
$i_\mathrm{rel}$ is determined within $10^{\degr}$ of the true one by Solver A. 
The results are expressed as a function of $i_\mathrm{rel}$ itself. 
Overall, for Solver A both experiments give similar results, 
showing that his fitting algorithm is only mildly sensitive to the 
mutual inclination of pairs of planetary orbits. 

In both cases, Solver A globally recovers $\sim 40\%$ of the $i_\mathrm{rel}$ values to within 
$10^{\degr}$ uncertainty, independently of the value of mutual 
inclination. The fraction of systems for which the actual value of $i_\mathrm{rel}$ 
is determined within the above tolerance increases when the constraints 
on well-sampled, high signal-to-noise orbits, with a sufficient number of 
observations, are set, up to ~90\%. In the top left panel of Figure~\ref{irel}, 
both ends of the upper three curves are not significant, 
due to very low number statistics considerations. 
Actually, the results shown in the top right panel can be mapped in the top left panel 
(at least for $2^{\degr}\leq i_\mathrm{rel}\leq 10^{\degr}$), thus highlighting that the 
apparent quick degradation in the fraction of systems with 
$i_\mathrm{rel}$ accurately determined is not real. It does nevertheless appear 
that, for random mutual orientation of the orbits, values of $i_\mathrm{rel}$ 
between $30^{\degr}$ and $40^{\degr}$ are slightly more likely to be identified 
correctly (by some 20\%) than quasi-coplanar cases or cases 
with $i_\mathrm{rel}$ close to $90^{\degr}$. 
For the quasi-coplanar case, perfectly coplanar orbits are slightly 
less likely to be correctly identified. 

\begin{figure*}
\centering
$\begin{array}{cc}
\includegraphics[width=0.4\textwidth]{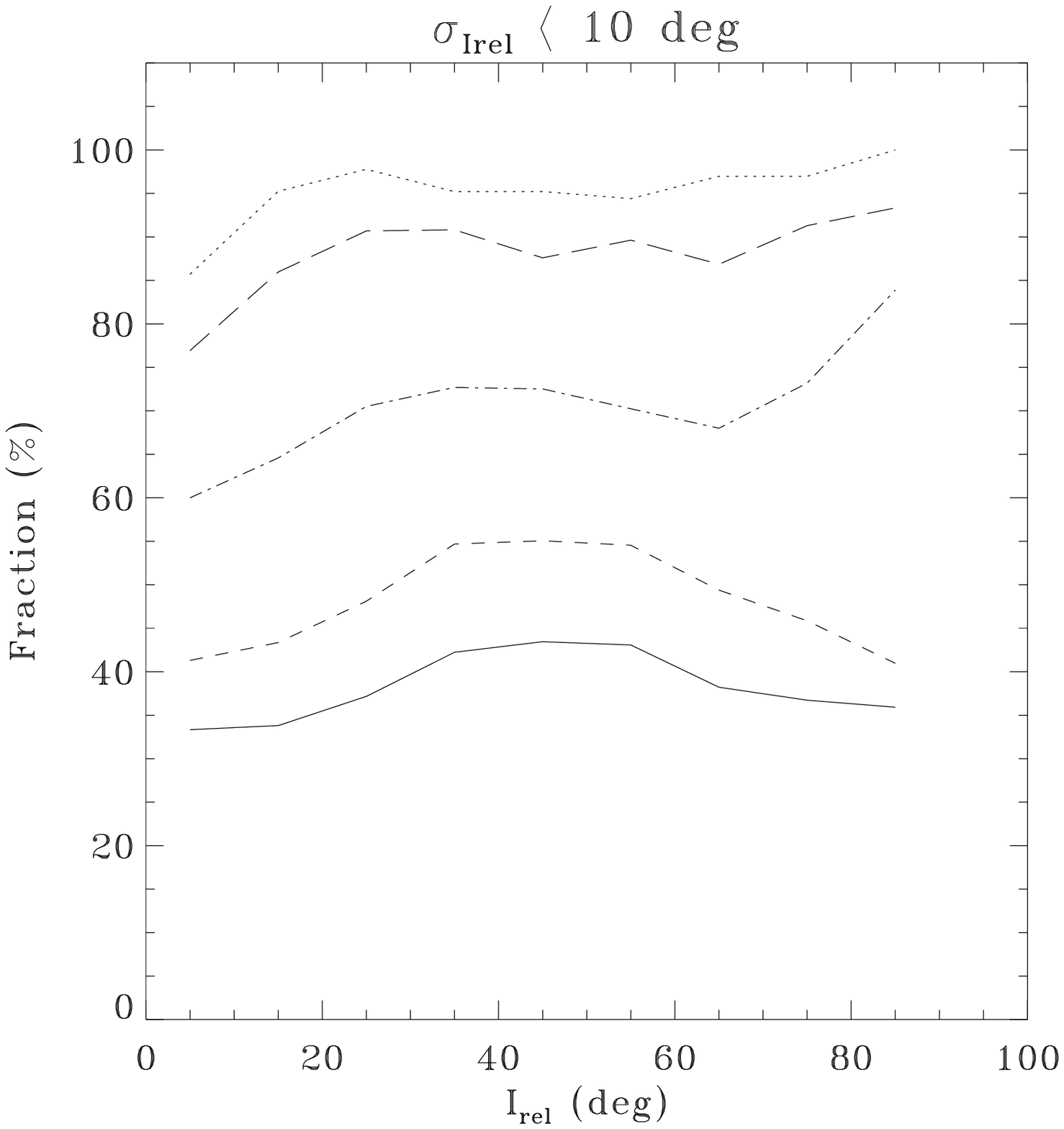} & 
\includegraphics[width=0.4\textwidth]{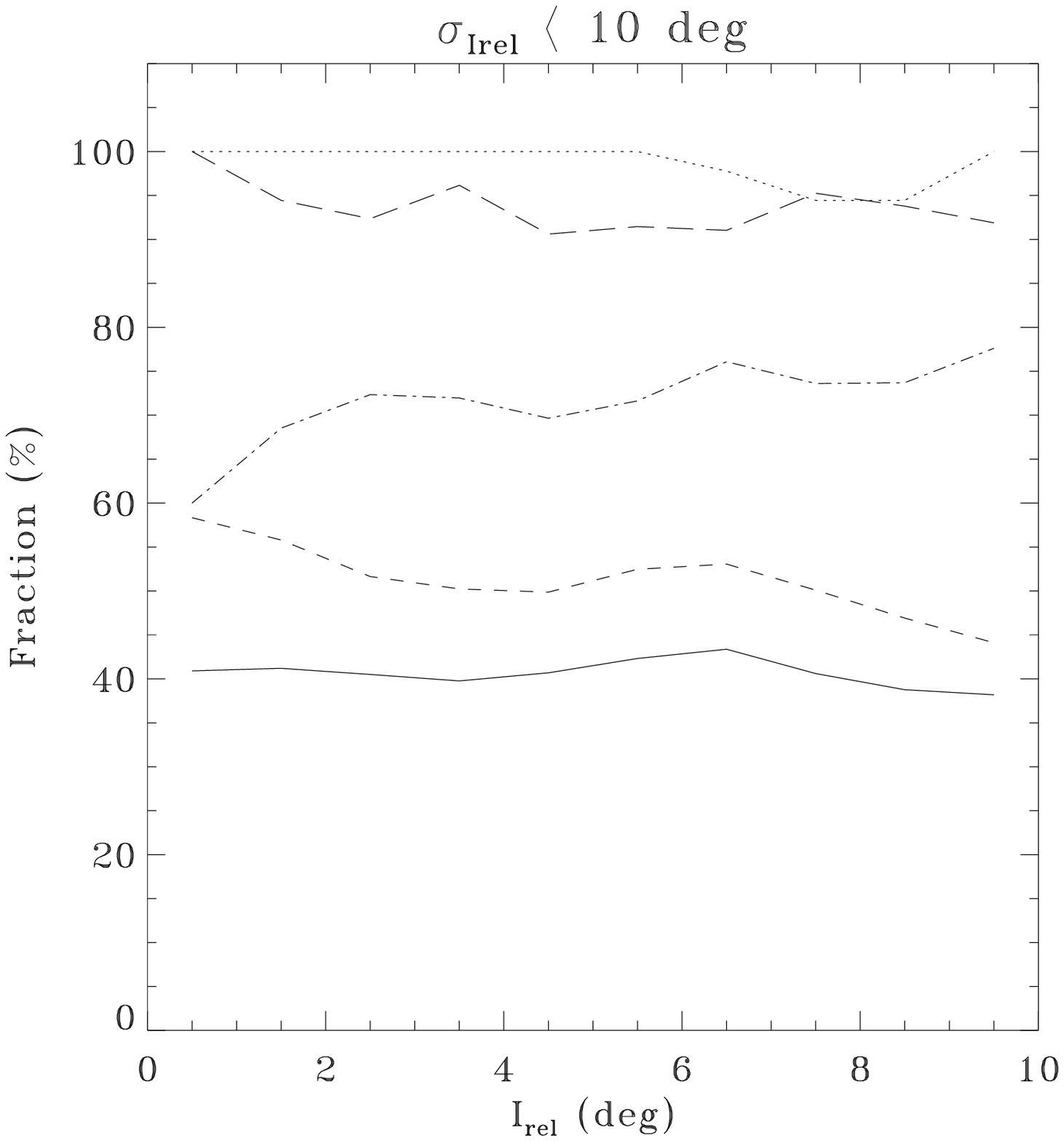} \\
\includegraphics[width=0.4\textwidth]{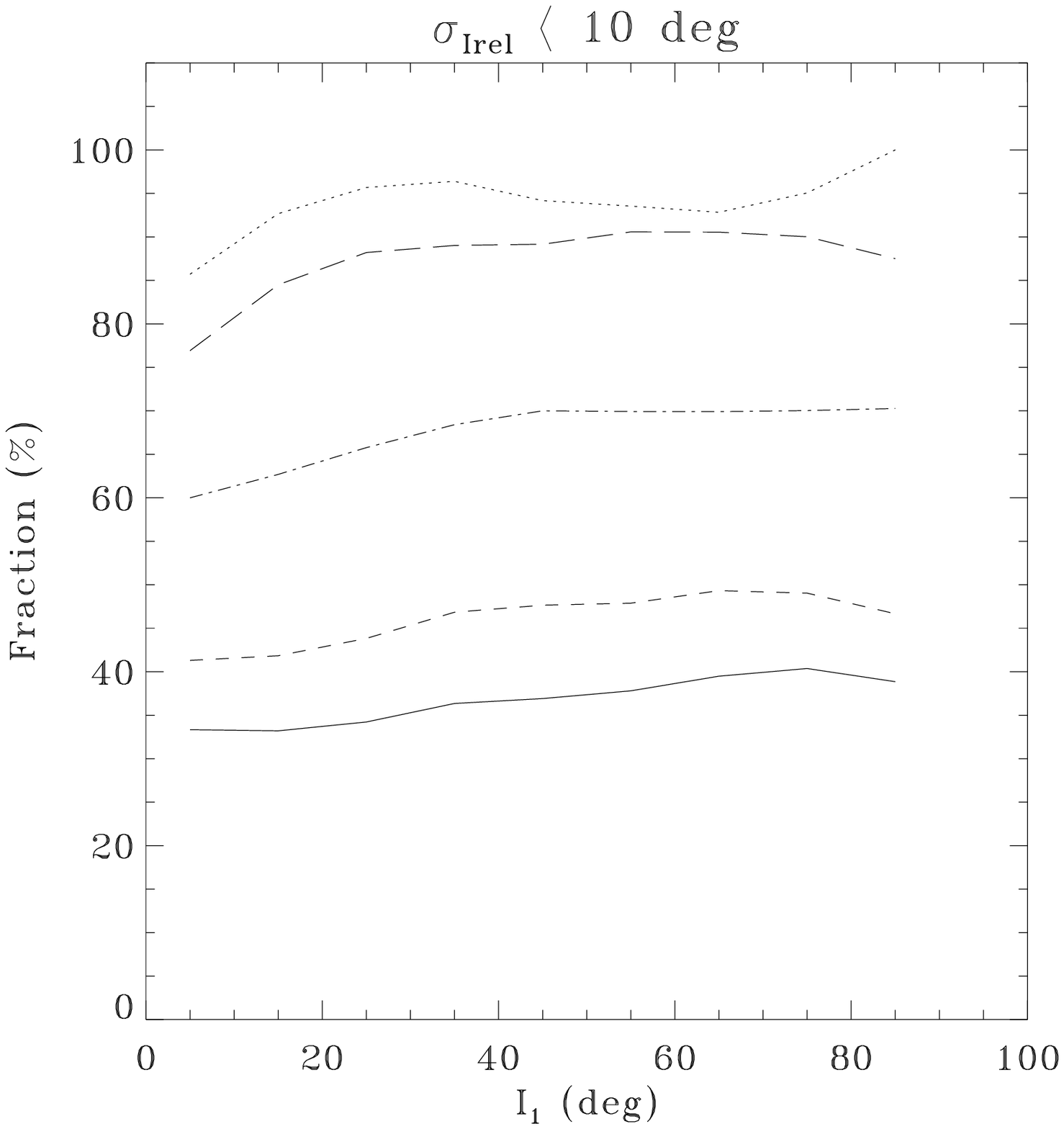} &
\includegraphics[width=0.4\textwidth]{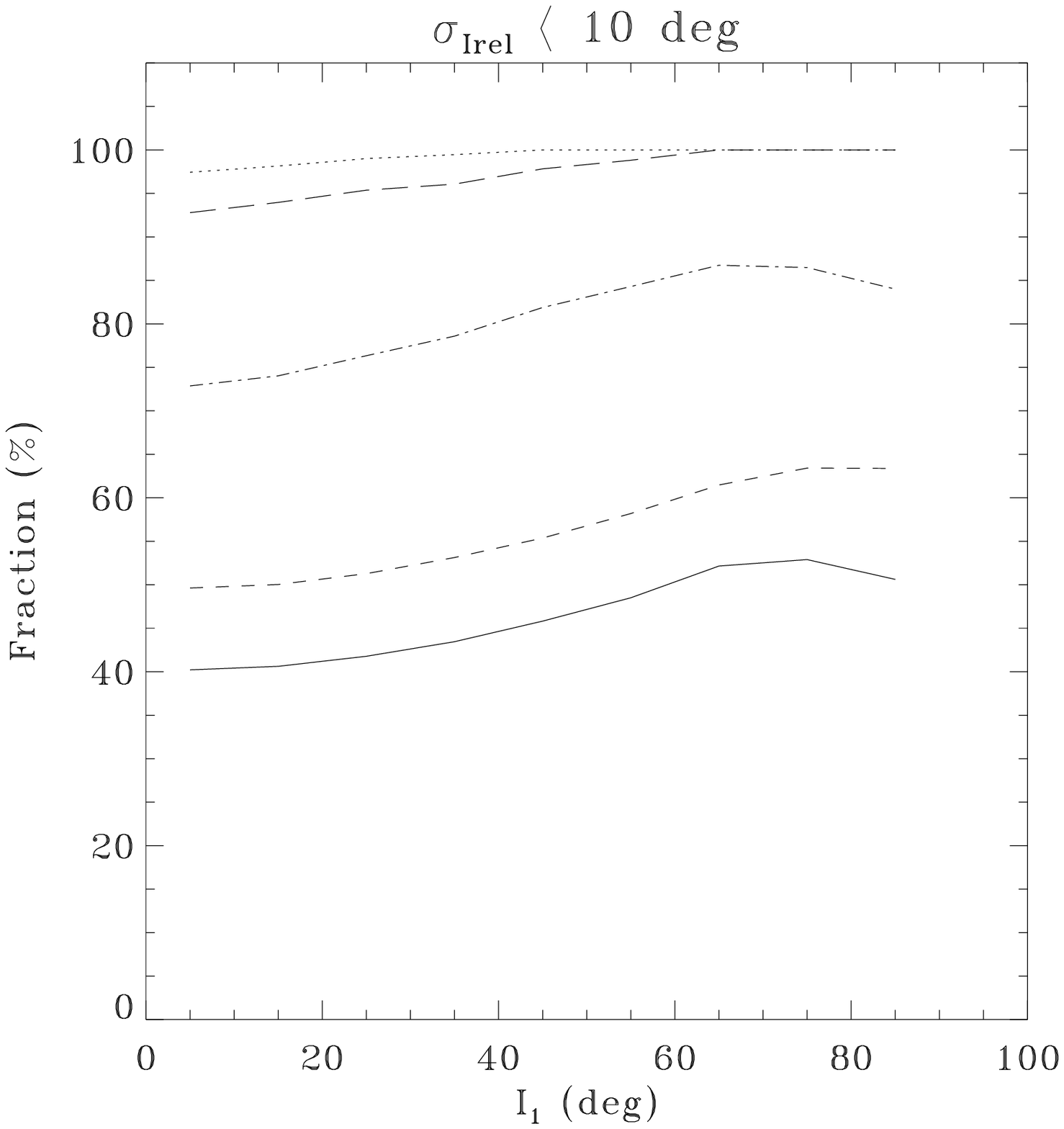} \\
\end{array} $
\caption{Top left and right: Same as the two upper 
panels of Figure~\ref{irel}, but for 
the formal uncertainties on $i_\mathrm{rel}$ as calculated 
by propagating the formal errors on the Thiele-Innes elements 
from the covariance matrix of the solutions. Bottom left and right: 
Same as the two lower panels of Figure~\ref{irel}, but for the formal 
uncertainties on $i_\mathrm{rel}$ calculated as for the top two panels.}
\label{sigirel}
\end{figure*}

The two lower panels of Figure~\ref{irel} show similar results, but this time expressed 
as a function of the inclination angle of one of the two planets. 
Again, Solver A's results for the T3a and T3b sample are similar in terms 
of fractions of systems with $i_\mathrm{rel}$ correctly identified 
within $10^{\degr}$ of the true value, when the various constraints are 
applied. However, the fraction of quasi-coplanar orbits 
correctly identified seems to be systematically 
higher, by up to 10\%, than those with random values of 
$i_\mathrm{rel}$, except for the region with inclination angles 
in the intermediate range $30^{\degr}$-$50^{\degr}$, in which random values of 
$i_\mathrm{rel}$, away from face-on or edge-on configurations, 
appear to be somewhat favored (by up to 20\% more). Configurations 
in the T3a experiment in which one of the two planets is seen almost face-on appear 
unfavorable particularly when high signal-to-noise, well-sampled 
orbits are considered. A similar, but less significant (differences 
up to 10\%), trend is seen for the case of the T3b experiment 
(orbits viewed close to face-on are less likely to have 
$i_\mathrm{rel}$ measured accurately than quasi edge-on 
configurations). The responsible for such an effect is not, 
however, small-number statistics. That determining precisely 
the value of $i_\mathrm{rel}$ for almost face-on orbits is 
somewhat more difficult should not in fact come as a surprise, as 
this result had already been discussed in our previous papers 
on Gaia and SIM multiple-planet detection and orbit determination (Sozzetti et al. 2001, 2003b). 
When $i\rightarrow 0^{\degr}$, the uncertainty on the position angle 
of the line of nodes grows, as eventually $\Omega$ becomes 
undefined for $i = 0^{\degr}$. If one of the two planetary orbits 
is close to face-on, but $i_\mathrm{rel}$ is large, then 
the incorrect identification of $\Omega$ is reflected in 
a poorer determination of $i_\mathrm{rel}$. The effect 
is less severe if the two orbits are quasi-coplanar, because 
in this case, as $i\rightarrow 0^{\degr}$ for both planets, 
the term depending on $\Omega$ in equation~\ref{inclrel} 
becomes very small, and ultimately an accurate knowledge 
of $\Omega$ is not required.

Finally, in Figure~\ref{sigirel} we show the behavior of the 
nominal uncertainties on $i_\mathrm{rel}$ obtained by propagating 
the formal errors on the Thiele-Innes elements from the covariance 
matrix of the solutions. The results are plotted as a function of 
$i_\mathrm{rel}$ (upper panels) and $i$ of one of the two planets 
(lower panels). The nominal uncertainties appear to follow rather closely 
the actual errors. We note, however, that in several cases formal 
errors seem to underestimate the real ones. This effect is 
highlighted by systematically higher fractions of 
objects with low values of the nominal errors with respect to 
the real ones. This mild trend is observed for all values of 
$i_\mathrm{rel}$ and $i$, and in both experiments.

\subsection{Directions for future work}

Several complex issues have been left aside in the preliminary 
analyses carried out for all experiments of the double-blind tests 
program, such as correlations between orbital parameters and 
their errors, more thorough investigations of how well 
formal errors map the real ones, or in-depth studies of 
the conditions in which two-planet orbital fits are more 
likely to fail (e.g., due to covariance between proper motion 
solutions and long-period orbits). These topics will 
require rather sophisticated approaches and a more 
aggressive understanding of correlations and aliasing 
in orbital parameter space, and significantly larger 
sample sizes. 

Another area of potential improvement concerns the possibility 
to explore alternative methods for orbit fitting to improve 
on the interpretaton of the observations and ultimately the 
inferences concerning the overall population of planets. One 
possible venue could be the evaluation of the applicability of 
Bayesian model selection, based on Markov chain Monte Carlo 
algorithms (e.g., Ford \& Gregory 2007), to simulated Gaia data, 
in order to gauge their potential for accurate characterization 
of orbital parameters and their uncertainties.

The understanding of the technical specifications of the Gaia 
satellite and its astrometric instrument will develop further 
with time, therefore some of the simplifying assumptions in 
our simulations will be progressively relaxed and a more 
realistic error model (e.g., including zero-point uncertainties, 
calibrations errors, chromaticity effects, attitude error) 
and a realistic error distribution for 
$\psi$, including bias and magnitude terms, adopted.

Finally, there is margin for adding more realism to our 
reference model of planetary systems, by considering actual 
distributions of orbital parameters and masses, and up-to-date 
values of planetary frequencies. We will include some degree 
of mutual dynamical interactions in representative cases of 
planetary systems, and evaluate in detail the impact of 
possible sources of astrometric noise that might pollute and/or 
mimic planetary signatures (e.g., binarity of the parent star, 
stellar spots, and protoplanetary disks, whose impact can be 
seen in terms of additional dynamical perturbations as well as 
contamination by scattered light). 

\section{Discussion: Gaia in context}

The striking properties revealed by the observational data on extrasolar planets 
(for a review, see e.g. Udry et al. 2007) reflect the complexities 
inherent in the processes of planet formation and evolution. 
The comparison between theory 
and observation has shown that several difficult problems are 
limiting at present our ability to elucidate in a unified manner 
all the various phases. Rather, one often resorts to attempt 
to investigate separately limited aspects of the physics of planet 
formation and evolution using a `compartmentalized' approach. 

However, improvements are being made toward the definition of more robust 
theories capable of simultaneously explaining a large range of 
the observed properties of extrasolar planets, as well as of making new, 
testable predictions. To this end, help from future data obtained 
with a variety of techniques will prove invaluable. 
In light of the results of the double-blind tests campaign presented 
in the previous sections, we focus here 
on the potential of high-precision global astrometry 
with Gaia, as compared to other planet detection methods, to 
help answer several outstanding questions in 
the science of planetary systems.

\subsection{Gaia discovery space}

\begin{figure*}
\centering
\includegraphics[width=0.8\textwidth]{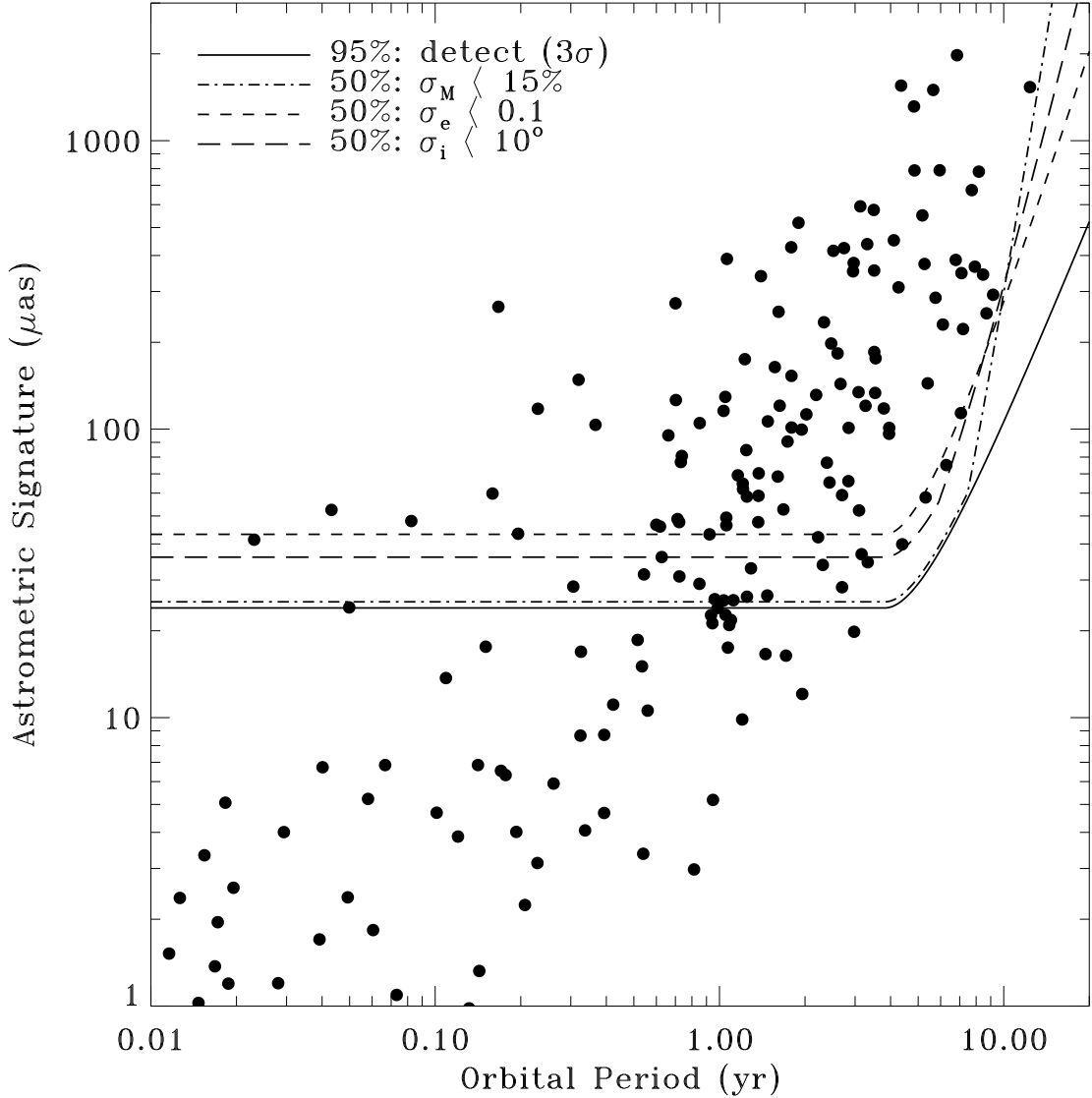}
\caption{Boundaries of secure ($\simeq 3\sigma$, for $\sigma = 8$ $\mu$as) 
detection and accurate mass and orbital 
parameters determination with Gaia compared to the known extrasolar planets 
(data from http://exoplanet.eu), which are plotted for the minimum case: orbit viewed edge-on, 
true mass equals radial velocity minimum mass, and astrometric signature minimum. 
Lines of different shape represent the minimum astrometric signature for 95\% 
probability of a 3$\sigma$ detection (solid line), 
the minimum astrometric signature needed to determine at least 50\% of 
the time the mass of a planet with better than 15\% accuracy (dash-dotted line), 
the eccentricity with uncertainties $< 0.1$ (short-dashed line), and the inclination angle 
with uncertainties $< 10^{\degr}$ (long-dashed line), respectively. The true astrometric signature, which is 
proportional to the true mass, will be generally higher, much higher in some 
cases, with the effect that more reliable detections and orbital fits will be possible.}
\label{sign_per}
\end{figure*}

We show in Figure~\ref{sign_per} a summary of the results presented 
in the previous sections, in terms of the minimum astrometric signature 
required for detection and measurement of orbital parameters and masses 
with Gaia, as a function of the orbital period of the companion, 
and averaging over all other orbital parameters. 

The curves in Figure~\ref{sign_per} correspond, respectively, to 
iso-probability contours for 95\% efficiency (virtual completeness) in 
detection at the 99.73\% confidence level, 50\% probability of measuring the companion 
mass to better than 15\% accuracy, and for the same likelihood of measuring 
eccentricities with uncertainties lower than 0.1 and the inclination 
angle of the orbital plane to better than 10$^{\degr}$ accuracy. 
All curves are polynomial fits to the actual iso-probability curves, 
with extrapolations for values of $P < 0.2$ yr and $P > 12$ yr, i.e. out 
of the period range covered by our simulations. For comparison, the 
minimum astrometric signatures (assuming $\sin i = 1$) and orbital periods of the 
present-day planet sample are overplotted. The plot, which closely 
resembles those presented in our earlier works (Lattanzi et al. 2000a; 
Sozzetti et al. 2002) indicates that Gaia would detect $\sim 55\%$ of 
the extrasolar planets presently known (the exact fraction depending on 
the actual value of $\sin i$), and for $> 50\%$ of these it would be 
capable of accurately measuring orbital parameters and actual masses. 

However, ongoing and planned 
surveys for planets with a variety of techniques are being designed to 
embrace the three-fold goal of 1) following-up and improving on the 
characterization of the presently known extrasolar planet sample, 
2) targeting more carefully defined and selected stellar samples, and 3) 
covering new areas of the planet discovery space, with the ultimate 
expectation of eventually reaching the capability to discover Earth-sized 
planets in the Habitable Zone (e.g., Kasting et al. 1993) of nearby stars. 
Indeed, by the time Gaia flies various other observatories will be 
operational, gathering additional information on the already known 
extrasolar planets sample and producing a wealth of new discoveries. 
For example, both ground-based as well 
as space-borne instrumentation for astrometric planet searches is 
being developed, such as VLTI/PRIMA (Delplancke et al. 2006) 
and SIM PlanetQuest, 
with targeted single-measurement precision comparable to, if not higher than, 
Gaia's. Then, the most effective way to proceed in order to gauge the relative 
importance of the Gaia global astrometric survey is not by looking at its 
discovery potential {\it per se}, but rather 
in connection with outstanding questions to be
addressed and answered in the science of planetary systems, 
thus helping to discriminate between proposed models of 
planet formation and evolution. 

\begin{figure*}
\centering
%$\begin{array}{cc}
\includegraphics[width=0.8\textwidth]{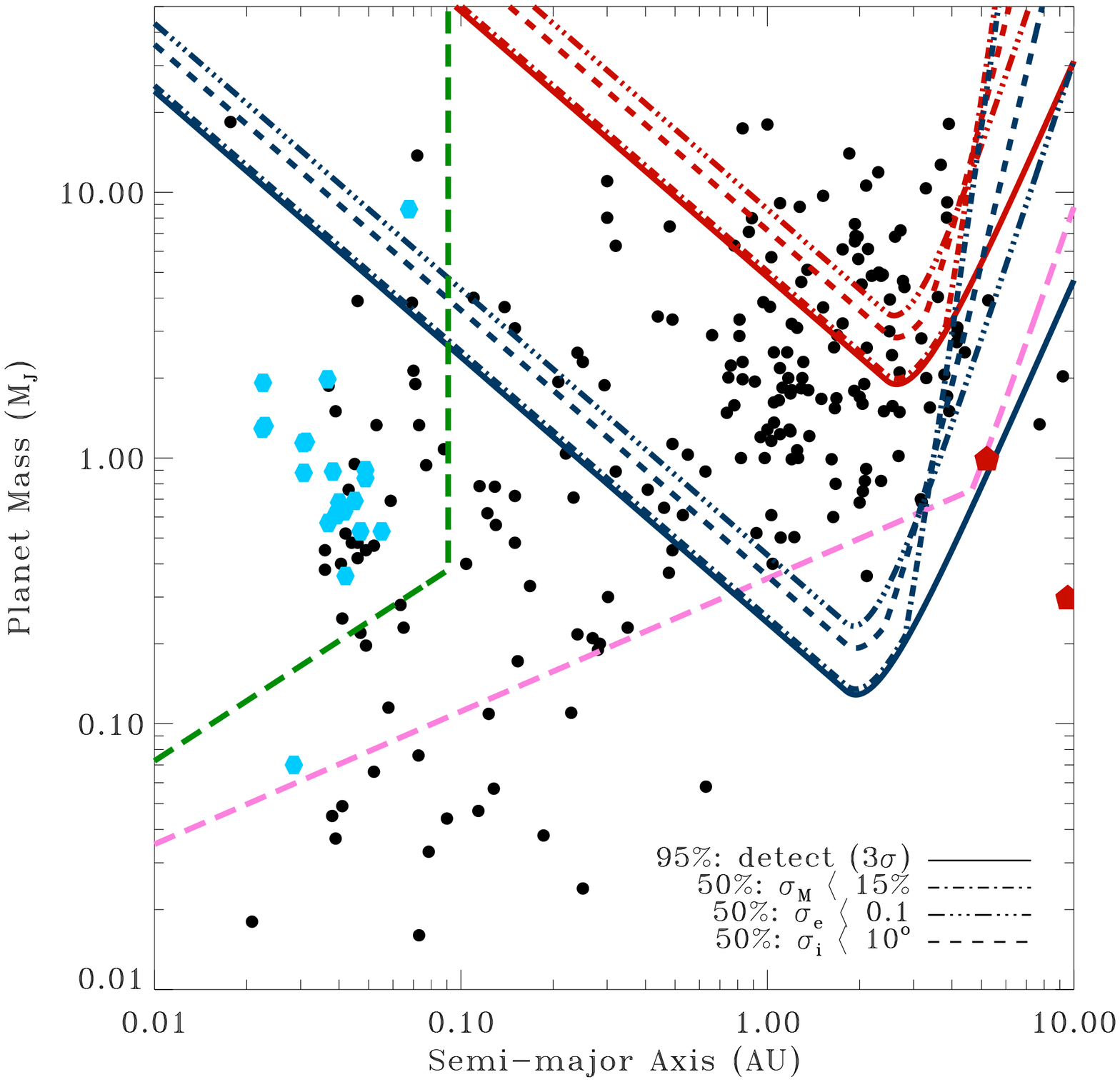}
%\includegraphics[width=0.4\textwidth]{det_meas_200pc_gdwarf.eps} &
%\includegraphics[width=0.41\textwidth]{det_meas_25pc_mdwarf.eps} \\
%\end{array} $
\caption{Gaia discovery space for planets of given mass and orbital radius compared 
to the present-day sensitivity of other indirect detection methods, namely Doppler 
spectroscopy and transit photometry. Red curves of different styles have the same 
meaning as in Figure~\ref{sign_per} assuming a 1-$M_\odot$ G dwarf primary at 200 pc, 
while the blue curves are for a 0.5-$M_\odot$ M dwarf at 25 pc. The radial velocity 
curve (pink line) is for detection at the $3\sigma_\mathrm{RV}$ level, assuming 
$\sigma_\mathrm{RV} = 3$ m s$^{-1}$, $M_\star = 1 M_\odot$, and 10-yr survey duration. 
For transit photometry (green curve), the assumptions of Gaudi et al. (2005) are 
used, i.e. $\sigma_V = 5$ milli-mag, $S/N = 9$, $M_\star=1$ $M_\odot$, $R_\star = 1$ $R_\odot$, 
uniform and dense ($> 1000$ datapoints) sampling. 
Black dots indicate the inventory of exoplanets as of September 2007. Transiting systems 
are shown as light-blue filled pentagons. Jupiter and Saturn are also shown as red pentagons.}
\label{detmeas}
\end{figure*}

By doing so, one immediately realizes that Gaia's most unique contribution will likely reside in
the unbiased and complete magnitude limited census of stars of all ages, spectral types, 
and metallicity in the solar neighborhood that could be screened for 
{\it new} planets, rather than on the 
additional insight its measurements might give on {\it already discovered} 
planets. In order to quantify our statement, we convert the results in 
Figure~\ref{sign_per} in the equivalent range of companion masses and 
semi-major axes that could be detected and measured orbiting a star of given mass 
and at a distance from the Sun. For illustration, we show in 
Figure~\ref{detmeas} Gaia's discovery space in the $M_p-a$ plane 
for 3$\sigma$ detection (with 95\% probability) and for accurately 
measuring $> 50\%$ of the time orbital elements and masses of planets orbiting 
a 1-$M_\odot$ star at 200 pc, and a 0.5-$M_\odot$ M dwarf at 25 pc 
(objects with $V < 13$, for which Gaia's highest astrometric precision 
can be achieved). From the Figure, one would then conclude that Gaia 
could discover and measure massive giant planets ($M_p \gtrsim 2-3$ $M_\mathrm{J}$) 
with $1<a<4$ AU orbiting solar-type stars as far as the nearest star-forming regions, 
as well as explore the domain of Saturn-mass planets with similar 
orbital semi-major axes around late-type stars within 30-40 pc. Particularly 
for the latter case, the Gaia sensitivity nicely complements at wider separations 
the area of the discovery space covered by ground-based transit photometry and 
decade-long Doppler surveys (see caption for details). 

\subsubsection{How many planets will Gaia find?}

\begin{table}[tbh]
\begin{center}
   \caption{Number of giant planets detected and measured by Gaia.}
   \renewcommand{\arraystretch}{1.4}
   \setlength\tabcolsep{5pt}
   \begin{tabular}{cccccc}
       \hline\noalign{\smallskip}
       $\Delta d$ & $N_\star$  & $\Delta a$ & $\Delta M_p$ &  $N_{\rm d}$ &  $N_{\rm m}$ \\
       (pc) & & (AU) & ($M_J$) & & \\
              \noalign{\smallskip}
       \hline
       \noalign{\smallskip}
0-50 & $\sim$10\,000 & 1.0 - 4.0 & 1.0 - 13.0 & $\sim 1400$ & $\sim 700$\\ 
50-100 & $\sim$51\,000 & 1.0 - 4.0 & 1.5 - 13.0 & $\sim 2500$ & $\sim 1750$\\ 
100-150 & $\sim$114\,000 & 1.5 - 3.8 & 2.0 - 13.0& $\sim 2600$ & $\sim 1300$\\
150-200 & $\sim$295\,000 & 1.4 - 3.4 & 3.0 - 13.0& $\sim 2150$ & $\sim 1050$\\

      \hline\noalign{\smallskip}
   \end{tabular}
\label{nplan}
\end{center}
\end{table}

To better gauge the Gaia potential for planet discovery, we update the early results 
of Lattanzi et al. (2000b), and re-compute the number of possible 
planetary systems within Gaia's grasp using 
estimates of the stellar content in the solar neighborhood 
and our present-day understanding of the giant planet frequency 
distribution $f_p$. For the former, we use the 
Besancon model of stellar population synthesis (Bienaym\'e et al. 1987; 
Robin \& Cr\'ez\'e 1986), constrained to 
$V < 13$, and for spectral types earlier than K5. According to this Galaxy 
model, we should expect $N_\star\sim 15\,000$, $\sim 61\,000$, $\sim 175\,000$, 
and  $\sim 470\,000$ stars within radii of 50 pc, 100 pc, 150 pc, and 200 pc, 
respectively (see Figure~\ref{totcounts}). For $f_p$, 
we take the Tabachnik \& Tremaine (2002) approach, and use a power-law 
functional form to integrate a differential fraction within an arbitrary 
range of $M_p$ and $P$:

\begin{equation}
\mathrm{d}f_p = C M_p^\beta P^\gamma \mathrm{d}M_p\mathrm{d}P
\end{equation}

We find the normalization $C$ by using the Tabachnik \& Tremaine (2002) 
values for the exponents ($\beta=-1.1$, $\gamma=-0.73$), which still provide 
a good description for the observed mass and period distributions of 
exoplanets (see for example Butler et al. 2006), and by imposing that 
the fraction of planets with $1\leq M_p\leq 15\,M_J$ and $2\leq P\leq 3000$ d 
equals the observed 7\% for F-G-K normal stars with $-0.5\leq$[Fe/H]$\leq 0.5$ 
(Marcy et al. 2005).

An estimate of the number of giant planets at a given distance $d$ 
(in pc) whose astrometric signal could be detected by Gaia with 3$\sigma$ 
confidence 95\% of the time is then given by $N_d\sim 0.95\times f_p\times N_\star$,
where $N_\star$ is computed within a sphere of radius $d$ 
centered on the Sun for given limiting magnitude and spectral type, 
while the value of $f_p$ is calculated integrating over a specific 
range of masses and periods. The number of planets for which, say, 
masses will be determined at least 50\% of the time with an accuracy 
of better than 15\% will instead be: $N_m\sim 0.50\times N_d$. The 
results are summarized in Table~\ref{nplan}. One then realizes that, 
based on our present knowledge of giant planets frequencies ($M_p > 1-3 M_J$), 
integrated over a wide range of spectral types and metallicities, 
Gaia could then find $\sim 8\,000$ such objects, and accurately 
measure masses and orbital parameters for $\sim 4000$ of them.

\subsubsection{How many multiple-planet systems will Gaia find?}

As of December 2007, 24 planet-bearing stars are orbited by more than one 
planet, corresponding to $\sim 12\%$ of the total sample of RV-detected systems
\footnote{Johnson et al. (2007) and Setiawan et al. (2008) report possible 
multiple companions around GJ 317 and HD 47536. We elect not to include them 
as their orbits are either only loosely constrained or not yet statistically 
very significant}. However, 
many systems known to host one exoplanet show more distant, long-period, sub-stellar companions
with highly significant but incomplete orbits (with inferred semi-major axis typically 
beyond 5 AU). Recent analyses of these long-term trends (Wright et al. 2007) indicate 
that $\sim 30\%$ of known exoplanet systems show significant evidence of multiplicity. 
Considering that the mass distribution of planets increases steeply toward lower masses 
(e.g., Marcy et al. 2005), incompleteness must be considerable between 1.0 and 0.1 
Jupiter-masses. Thus, the actual occurrence of multiple planets among stars having one 
known planet is likely considerably greater than 30\%. 

We report in Table~\ref{multsign} the relevant parameters of the multiple-planet systems with well-measured 
orbits known to-date, ordered by increasing distance of the system from the Sun. The expected values of 
the astrometric signature ($\alpha_\mathrm{min}$) are computed assuming perfectly edge-on, 
coplanar configurations ($\sin i_j = 1$, for $j = 1,\dots,n_p$). The single-measurement
precision is $\sigma_\psi = 8$ $\mu$as for all stars. 
Of these systems, $\sim 50\%$ have more than one component with $\alpha_\mathrm{min} > 3\sigma_\psi$, 
$\sim 40\%$ have components with $\alpha_\mathrm{min} > 3\sigma_\psi$ as well as $P < 5-6$ yr, 
and some $16\%$ have both $\alpha_\mathrm{min} > 10\sigma_\psi$ as well as $P < 5-6$ yr. 
Extrapolating from the numbers obtained in the previous Section and the ones above, one then infers that 
of the $\sim 8000$ new planetary systems discovered by Gaia, $\sim 1000$ would have multiplicity 
greater than one, and $\sim 400-500$ could have orbital parameters and masses measured to better 
than $15\%-20\%$ accuracy. 

%\begin{table*}[tbh]
\begin{longtable}{lccccc}
%\scriptsize
%\begin{center}
%   \renewcommand{\arraystretch}{1.4}
%   \setlength\tabcolsep{7pt}
   \caption{\label{multsign} List of relevant parameters for known planetary systems.} \\
\hline\hline\noalign{\smallskip}
%   \begin{tabular}{lccccc}
%       \hline\hline\noalign{\smallskip}
              Planet & $d$ & $M_\star$ & $M_p\sin i$  & $a$ & $\alpha$  \\
               & (pc) & ($M_\odot$) &  ($M_\mathrm{J}$) & (AU) & ($\mu$as) \\

                            \noalign{\smallskip}
       \hline%\noalign{\smallskip}
\endfirsthead
%\caption{continued.}\\
%\hline\hline\noalign{\smallskip}
%              Planet & $d$ & $M_\star$ & $M_p\sin i$  & P & $\alpha$  \\
%               & (pc) & ($M_\odot$) &  ($M_\mathrm{J}$) & (years) & ($\mu$as) \\
%\noalign{\smallskip}
%\hline\noalign{\smallskip}
%\endhead
%\hline\noalign{\smallskip}
%\endfoot
       \noalign{\smallskip}
         GJ 876b  &    4.72  & 0.32   &    1.93   &    0.21  &    265.6 \\
         GJ 876c  &          &        &    0.56   &    0.13  &     48.0 \\
         GJ 876d  &          &        &    0.02   &    0.02  &      0.2 \\
         GJ 581b  &    6.26  & 0.31   &    0.05   &    0.04  &      1.0 \\
         GJ 581c  &          &        &    0.02   &    0.07  &      0.6 \\
         GJ 581d  &          &        &    0.02   &    0.25  &      3.1 \\
       HD 69830b  &   12.60  & 0.86   &    0.03   &    0.08  &      0.2 \\
       HD 69830c  &          &        &    0.04   &    0.19  &      0.7 \\
       HD 69830d  &          &        &    0.06   &    0.63  &      3.4 \\
         55 Cncb  &   13.40  & 1.03   &    0.78   &    0.11  &      6.7 \\
         55 Cncc  &          &        &    0.22   &    0.24  &      3.9 \\
         55 Cncd  &          &        &    3.92   &    5.26  &   1534.0 \\
         55 Cnce  &          &        &    0.05   &    0.04  &      0.1 \\
 $\upsilon$ Andb  &   13.47  & 1.27   &    0.69   &    0.06  &      2.4 \\
 $\upsilon$ Andc  &          &        &    1.98   &    0.83  &     95.1 \\
 $\upsilon$ Andd  &          &        &    3.95   &    2.51  &    575.3 \\
         47 Umab  &   13.97  & 1.03   &    2.60   &    2.11  &    376.7 \\
         47 Umac  &          &        &    1.34   &    7.73  &    347.5 \\
      HD 160691b  &   15.30  & 1.08   &    1.67   &    1.50  &    153.0 \\
      HD 160691c  &          &        &    3.10   &    4.17  &    781.1 \\
      HD 160691d  &          &        &    0.04   &    0.09  &      0.2 \\
      HD 160691e  &          &        &    0.52   &    0.92  &     29.1 \\
      HD 190360c  &   15.89  & 1.04   &    0.06   &    0.13  &      0.5 \\
      HD 190360b  &          &        &    1.50   &    3.92  &    365.7 \\
      HD 128311b  &   16.60  & 0.80   &    2.18   &    1.10  &    174.8 \\
      HD 128311c  &          &        &    3.21   &    1.76  &    415.1 \\
       HD 82943b  &   27.46  & 1.18   &    1.75   &    1.19  &     64.7 \\
       HD 82943c  &          &        &    2.01   &    0.75  &     46.6 \\
       HD 37124c  &   33.00  & 0.91   &    0.68   &    3.19  &     75.0 \\
       HD 37124b  &          &        &    0.61   &    0.53  &     11.1 \\
       HD 37124d  &          &        &    0.60   &    1.64  &     33.8 \\
       HD 11964b  &   33.98  &1.13    &    0.11   &    0.23  &      0.7 \\
       HD 11964c  &          &        &    0.70   &    3.17  &     58.0 \\
      HD 169830b  &   36.32  & 1.40   &    2.88   &    0.81  &     46.0 \\
      HD 169830c  &          &        &    4.04   &    3.60  &    285.4 \\
      HD 217107b  &   37.00  & 1.02   &    1.33   &    0.07  &      2.6 \\
      HD 217107c  &          &        &    2.50   &    4.41  &    292.3 \\
       HD 12661b  &   37.16  &1.07    &    2.30   &    0.83  &     47.6 \\
       HD 12661c  &          &        &    1.57   &    2.56  &    101.0 \\
      HD 168443b  &   37.88  & 1.06   &    8.02   &    0.30  &     59.8 \\
      HD 168443c  &          &        &   18.10   &    3.91  &   1314.1 \\
       HD 38529b  &   42.43  & 1.39   &    0.78   &    0.13  &      1.7 \\
       HD 38529c  &          &        &   12.70   &    3.68  &    789.4 \\
      HD 155358b  &   42.70  & 0.87   &    0.89   &    0.63  &     15.1 \\
      HD 155358c  &          &        &    0.50   &    1.22  &     16.6 \\
      HD 202206b  &   46.34  & 1.13   &   17.40   &    0.83  &    273.0 \\
      HD 202206c  &          &        &    2.44   &    2.55  &    117.9 \\
      HIP 14810b  &   52.90  & 0.99   &    3.84   &    0.07  &      5.1 \\
      HIP 14810c  &          &        &    0.76   &    0.41  &      5.9 \\
      HD 74156b   &   64.56  & 1.05   &    1.88   &    0.29  &      8.0 \\  
      HD 74156d   &          &        &     0.4   &    1.04  &      6.1 \\ 
      HD 74156c   & 	     &        &    8.03   &    3.85  &    456.1 \\ 
      HD 108874c  &   68.50  & 1.00   &    1.02   &    2.68  &     39.9 \\
      HD 108874b  &          &        &    1.36   &    1.05  &     20.9 \\
       HD 73526b  &   99.00  & 1.02   &    2.90   &    0.66  &     18.6 \\
       HD 73526c  &          &        &    2.50   &    1.05  &     25.5 \\
%       HD 47536b  &  121.36  &     0.94   &    5.00   &    1.18  &     47.9 \\
%       HD 47536c  &     &        &    7.00   &    6.84  &    216.7 \\

      \hline\noalign{\smallskip}
%        \end{tabular}
%\end{center}
%\normalsize
\end{longtable}
%\end{table*}
\twocolumn

\begin{table}[tbh]
\begin{center}
   \caption{Number of multiple-planet systems detected and measured by Gaia.}
   \renewcommand{\arraystretch}{1.4}
   \setlength\tabcolsep{7pt}
   \begin{tabular}{lc}
       \hline\hline\noalign{\smallskip}
        Case & Number of Systems \\
              \noalign{\smallskip}
       \hline
       \noalign{\smallskip}
1) Detection & $\sim 1000$\\ 
2) Orbits and masses to  & \\ 
better than $15\%-20\%$ accuracy & $\sim 400-500$ \\
3) Successful  & \\
coplanarity tests & $\sim 150$\\
      \hline\noalign{\smallskip}
   \end{tabular}
\label{nmult}
\end{center}
\end{table}

For some 150 systems with very favorable configurations, and enough 
redundancy in the number of observations, coplanarity tests could be performed, with expected 
uncertainties on the mutual inclination angle of $\sim 10^{\degr}$, or smaller. In terms of 
systems for which the Gaia data alone could provide reasonably good orbital solutions, this is about 
a twenty-fold improvement with respect to the present-day number of systems with well-determined 
orbits, and even the number of potential systems for which coplanarity analysis could be 
successfully carried out compares favorably to today's sample, presently populated 
by zero objects. These numbers are summarized in Table~\ref{nmult}. 
Again, these results should be considered as lower limits, given the increasingly convincing evidence 
for a frequency of multiple-planet systems at least a factor of 2-3 greater than the value 
used here for the extrapolation.

\subsection{The Gaia legacy}

It is easy to realize how the statistical value of such large 
samples of newly detected giant planets and planetary systems 
would be instrumental for critical 
testing of planet formation and evolution models. To illustrate 
more clearly the wealth of information potentially contained in the 
data collected by Gaia, let us ask four fundamental questions for 
the astrophysics of planetary systems, and see how, based on the results 
presented in this paper, Gaia could help address them 
(complementing other datasets obtained with a variety of techniques). 

\subsubsection{How do planet properties and frequencies depend upon 
the characteristics of the parent stars?}

\begin{figure}[tbh]
\centering
\includegraphics[width=.5\textwidth]{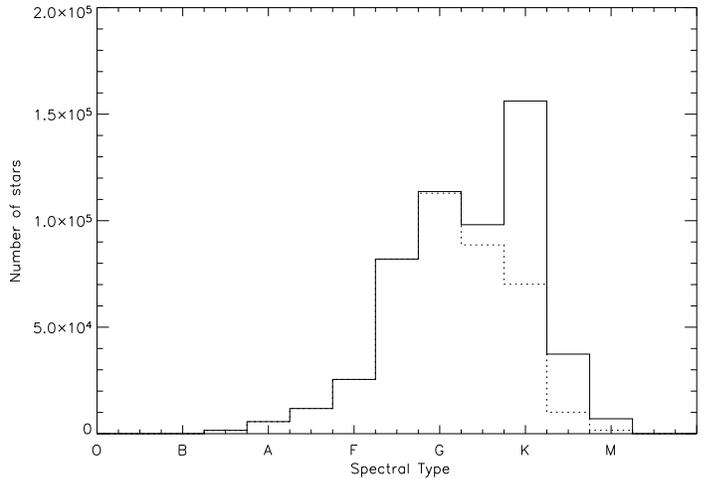}
\caption{Stellar content to $d<200$ pc, 
as function of the spectral type, for $V<13$ ({\it solid line}) and $V<12$ 
(dotted line).}
\label{totcounts}
\end{figure}

Twelve years after the first discovery announcement (Mayor \& Queloz 1995), 
the observational data on extrasolar planets are providing growing evidence 
that planetary systems properties (orbital elements and mass distributions, and correlations
amongst them) and frequencies appear to depend upon the characteristics of
the parent stars (spectral type, age, metallicity, binarity/multiplicity). 
Doppler surveys have begun in the recent 
past to put such trends on firmer statistical grounds. For example, dedicated 
surveys of metal-rich (Fischer et al. 2005; Bouchy et al. 2005) and 
metal-poor dwarfs (Sozzetti et al. 2006; Mayor et al. 2003
\footnote{http://www.eso.org/observing/proposals/gto79/harps/4.txt}) are currently 
providing data to improve the statistical significance of the strong correlation 
between planet occurrence rates and stellar metallicity (e.g., Gonzalez 1997; Santos et al. 2004; 
Fischer \& Valenti 2005). Similarly, other groups have been monitoring samples of 
bright M dwarfs (Butler et al. 2004; Bonfils et al. 2005; Endl et al. 2006; Johnson et al. 2007b, 
and references therein), 
Hertzsprung gap sub-giants (Johnson et al. 2006, 2007a), heavily evolved stars belonging to the red-giant branch 
and clump regions of the H-R diagram (Frink et al. 2002; 
Setiawan et al. 2005; Sato et al. 2003; Hatzes et al. 2005; Lovis \& Mayor 2007; 
Niedzielski et al. 2007, and references therein), early-type dwarfs (Galland et al. 2005), 
and relatively young stars (Setiawan et al. 2007a), 
in order to probe the possible dependence of $f_p$ on stellar mass and age. 
However, the typical sample sizes of these surveys are of order of a few hundred 
objects, sufficient to test only the most outstanding difference between the 
various populations. It is thus desirable to be able to provide as large 
a database as possible of stars screened for planets. 

As we have seen, the size of the stellar sample available for planet detection and 
measurement to the Gaia all-sky astrometric survey will be approximately a few hundred thousand 
relatively bright ($V < 13$) stars with a wide range of spectral types, 
metallicities, and ages out to $\sim 200$ pc. The sample-size
is thus comparable to that of planned space-borne transit surveys,
such as CoRot and Kepler. The expected number of giant planets 
detected and measured (see Table~\ref{nplan}) could be several 
thousands, depending on actual giant planet frequencies as a function of 
spectral type and orbital distance. This number is comparable to the 
size of the combined target lists of present-day ground-based Doppler surveys 
and of future astrometric projects such as VLTI/PRIMA and SIM. 
The Gaia unbiased and complete magnitude limited census of stars screened  for new planets 
will allow, for example, to test the fine structure of giant planet parameters 
distributions and frequencies, and to investigate their possible changes 
as a function of stellar mass with unprecedented resolution. From 
Figure~\ref{totcounts}, of order of tens of thousands of normal stars in 
0.1 $M_\odot$ bins would become available for such investigations. 
Furthermore, the ranges of orbital parameters and giant planet host characteristics
probed by the Gaia survey would crucially complement both
transit observations (which strongly favor short orbital periods and are
subject to stringent requisites on favorable orbital alignment), and
radial-velocity measurements (which can be less effectively carried out for
stars covering a wide range of spectral types, metallicities, and ages and do not allow
to determine either the true planet mass or the full three-dimensional
orbital geometry). 

Thus, the ability to simultaneously and systematically determine 
planetary frequency and distribution of orbital parameters for the stellar 
mix in the solar neighborhood without any potential biases induced by the 
choice of specific selection criteria for target lists, stems out as a 
fundamental contribution that Gaia will uniquely provide, 
the only limitations being those intrinsic to the mission, i.e., to the actual
sensitivity of the Gaia measurements to planetary perturbations, which in 
this paper we have quantitatively gauged.

\subsubsection{What is the preferred method of gas giant planet formation?}

\begin{figure}[tbh]
\centering
\includegraphics[width=.5\textwidth]{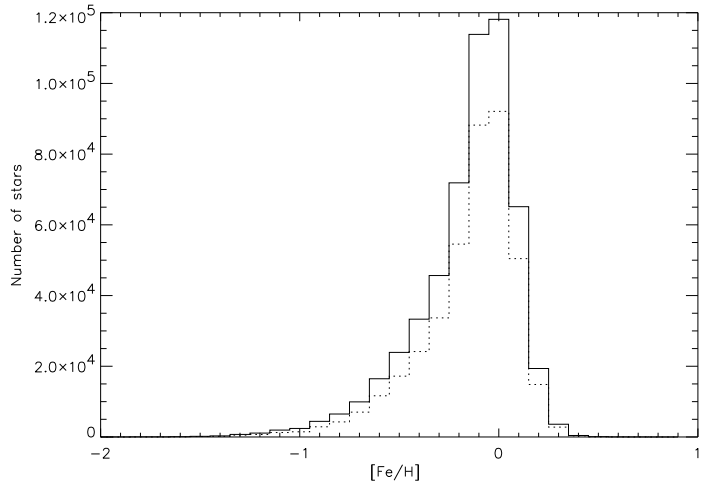}
\caption{Stellar distribution in the solar neighborhood ($d<200$ pc) 
as function metallicity, for $V<13$ (solid line) and $V<12$ 
(dotted line).}
\label{fehcounts}
\end{figure}

The two main competing models of giant-planet formation by 
{\it core accretion} (e.g., Pollack et al. 1996. For a review see Lissauer \& Stevenson 2007) 
and {\it disk instability} (e.g., Boss 2001. For a review see Durisen et al. 2007)
make very different predictions regarding formation time-scales (Mayer et al. 2002; 
Alibert et al. 2005; Boss 2006), planet properties (Armitage et al. 2002; Kornet \& Wolf 2006; 
Ida \& Lin 2004a, 2005, 2008; Rice et al. 2003b), and frequencies as a function 
of host star characteristics (Laughlin et al. 2004; Ida \& Lin 2004b, 2005, 2008; Kornet 
et al. 2005, 2006; Rice et al. 2003b; Boss 2000, 2002, 2006). Furthermore, correlations 
between orbital elements and masses, and possibly between the former and some of the host star
characteristics (metallicity, mass) might reflect the outcome of a variety
of migration processes and their possible dependence on environment (Livio \& Pringle 2003; 
Ida \& Lin 2004a, 2008; Boss 2005; Burkert \& Ida 2007). 
Some of these predictions could be tested on firm 
statistical grounds by extending planet surveys to large samples of 
stars that are not readily accessible to Doppler surveys.

For example, Galaxy models (Bienaym\'e et al. 1987; Robin \& Cr\'ez\'e 1986) 
predict $\sim4000$~F--G--K dwarfs and sub-dwarfs to
200~pc, brighter than $V=13$~mag, and with metallicity [Fe/H] 
$<-1.0$ (see Figure~\ref{fehcounts}). The entire population will be 
screened by Gaia for giant planets on wide orbits
thus complementing the shorter-period
ground-based spectroscopic surveys (e.g., Sozzetti et al. 2006),
which are also limited in the sample sizes due to the intrinsic faintness and
weakness of the spectral lines of the targets. These data combined
would allow for improved understanding of the behavior of the
probability of planet formation in the low-metallicity regime, by
direct comparison between large samples of metal-poor and metal-rich
stars, in turn putting stringent constraints on the proposed planet formation models 
and helping to better the role of stellar metallicity 
in the migration scenarios for gas giant planets.

\begin{table}[tbh]
\begin{center}
   \caption{The closest ($\leq 200$ pc) star forming regions and
young stellar kinematic groups.}
\vskip 0.1cm
   \setlength\tabcolsep{7pt}
   \begin{tabular}{lll}
       \hline\hline\noalign{\smallskip}
        Name               &   Distance (pc) & Age (Myr) \\
       \noalign{\smallskip}
       \hline
       \noalign{\smallskip}
	   Hercules-Lyra    &  15-40 & 100 \\
	   AB Doradus       &  20-50 & 30-50 \\
	   Subgroup B4      &  20-50 & 80-100 \\
	   $\beta$ Pictoris &  30-50 & 8-15 \\
	   Tucana-Horologium & 50-60 & 8-50 \\
       TW Hya           &  50 & 3-50\\
	   MBM 12           &  60-110 & 3-10    \\
       $\eta$ Chamaeleontis  &  90-150 & 8-10\\
       $\eta$ Carin\ae  &  100 & 8\\
	   MBM 20           &  110-160 & 3-10 \\
       Pleaides         &  125 & 75-100\\
       $\varrho$ Ophiuchi &  125-150 & 1-2\\
       Taurus-Auriga    &  135 & 1-2\\
       Corona Austrina  &  140 & 1-2\\
       Lupus            &  140 & 1-2\\
       $o$ Velorum      &  160 & 30\\
       $\vartheta$ Carin\ae &  160 & 30\\
       Scorpio-Centaurus&  160-180 & 2-20\\
       $\alpha$ Persei  &  175 & 85\\
       Serpens          &  200 & 5-10\\
%       Coma Berenices      &  100~pc & 400\\
%       Ursa Major      &  30~pc & 500\\
%       Hyades          &  45~pc & 600\\
%       Praesepe            &  180-200~pc & 600\\
	                      \hline\noalign{\smallskip}
   \end{tabular}
\label{sfr}
\end{center}
%\tablecomments{}
\end{table}

Furthermore, within the useful (for Gaia) distance horizon of $\sim 200$ pc, 
hundreds of relatively bright ($V < 13-14$) young stars can be found 
in some twenty or so nearby star-forming regions and young associations 
(see Table~\ref{sfr} for a list of young associations, open clusters, and moving groups 
in the age range $\sim1-100$ Myr in the solar neighborhood, with ages 
in the approximate range 1-100 Myr. The data, ordered by increasing distance from the 
Sun, are from Zuckermann \& Song (2004, and references therein) and 
L\'opez-Santiago et al. (2006, and references therein)). All these stars will be observed 
by Gaia with enough astrometric precision to detect the presence of 
massive giant planets ($M_p \gtrsim 2$ $M_J$) orbiting at 2-4 AU. 
The possibility to determine the epoch
of giant planet formation in the protoplanetary disk would
provide the definitive observational test to distinguish
between the proposed theoretical models.
These data would uniquely complement near- and
mid-infrared imaging surveys (e.g., Burrows 2005, and references
therein) for direct detection of young, bright, wide-separation
($a > 30-100$ AU) giant planets, such as JWST.

\subsubsection{How do dynamical interactions affect the architecture of 
planetary systems?} 

The highly eccentric orbits of planetary systems have been explained 
so far calling into question a variety of dynamical mechanisms, 
such as interactions between a planet and the gaseous disk, 
planet-planet resonant interactions, close encounters between planets, 
secular interactions with a companion star (see for example Ford \& Rasio 2007, 
and references therein). Some of these eccentricity 
excitation mechanisms can give rise to very different orbital 
architectures, including significantly non-coplanar orbits (Thommes \& Lissauer 2003). 
An effective way to understand their relative roles would
involve measuring the mutual inclination angle between pairs
of planetary orbits. Studies addressing the long-term dynamical stability
issue for multiple-planet systems (presently divided in three broad 
classes of hierarchical, secularly interacting and resonantly interacting 
systems. See for example Kiseleva-Eggleton et
al. 2002; Ji et al. 2003 and references therein; Correia et al.
2005; Barnes \& Quinn 2004; Goz´dziewski \& Konacki 2004
and references therein), as well as the possibility of formation
and survival of terrestrial planets in the Habitable Zone of the
parent star (Menou \& Tabachnik 2003; Jones et al. 2005 and references therein), 
would also greatly benefit from knowledge of 
the mutual inclination angle between planetary orbits.

The only way to provide meaningful estimates of the full
three-dimensional geometry of {\it any} planetary system (without
restrictions on the orbital alignment with respect to the line of sight) is
through direct estimates of the mutual inclinations angles using
high-precision astrometry. We have shown here how, extrapolating 
from today's knowledge of the frequency and architectures of multiple-planet 
systems, Gaia could detect and measure several hundred such systems, and 
perform a significant coplanarity analysis in a few hundred cases (see Table~\ref{nmult}).
These data, combined with those
available from Doppler measurements and transit photometry and transit timing
(e.g., Miralda Escud\'e 2002; Holman \& Murray 2005; Agol et al. 2005),
would then allow to put studies of the dynamical evolution of
planetary systems on firmer grounds.

\subsubsection{What are the phase functions and light curves of gas giant planets?}

The combination of high-cadence, milli-mag photometric 
and 1-5 m s$^{-1}$ precision radial-velocity measurements of transiting planet systems 
provides the fundamental observational data (planetary mass, radius, density, and 
gravity) needed for a meaningful comparison with structural models of hot Jupiters 
(e.g., Burrows 2005, and references therein). The special geometry of a transiting planet 
also permits a number of follow-up studies, which in particular have enabled direct observation of their
transmission spectra and emitted radiation (Charbonneau et al. 2007, and references therein). 
These data provide the first observational constraints on atmospheric models of these 
extrasolar gas giants (Burrows 2005, and references therein). 

The next logical step, the direct detection of extrasolar giant planets using high-contrast imaging 
instruments, requires that their dim light be separated from under the glare of their bright
parent stars. Several theoretical studies (Hubbard et al. 2002; Baraffe et al. 2003; 
Sudarsky et al. 2005; Dyudina et al. 2005; Burrows et al. 2004, 2007) 
have discussed exoplanet apparent brightness in reflected host
star light (expressed in units of the planet/host star flux ratio 
$\log(F_\mathrm{pl}/F_\mathrm{star}$) as functions of orbit geometry, orbital phase, cloud cover, 
cloud composition, mass and age. In particular, orbit and orientation of 
an extrasolar planet play a crucially important role in its flux at 
the Earth and in its interpretation, with strong dependence on eccentricity 
and inclination (Burrows et al. 2004). Depending upon $e$ and $i$, $\log(F_\mathrm{pl}/F_\mathrm{star})$ 
can be essentially constant (in case of $e\simeq 0.0$, $i\simeq 0^{\degr}$, for example), or vary by over an 
order of magnitude (in case of $e\simeq 0.6$, $i\simeq 90^{\degr}$ for example) along the orbit of an exoplanet, 
and this can induce significant changes in the chemical composition of its atmosphere 
(e.g., from cloudy to cloud-free). As for the knowledge of the actual mass of the planet, 
particularly at young ages theory predicts changes in intrinsic luminosity by a factor of nearly 100 can 
occur between objects in the mass range $1\,M_J\lesssim M_p\lesssim 5\,M_J$. The few wide-separation 
sub-stellar companions detected to-date by means of direct imaging techniques (Chauvin et al. 2005a, 2005b; 
Neuh\"auser et al. 2005; Biller et al. 2006), have planetary-mass solutions within their error bars, 
but these mass estimates rely upon so far poorly calibrated theoretical mass-luminosity relationships, 
thus their actual nature (planets or brown dwarfs) remains highly uncertain. 
It is then clear how accurate knowledge of all orbital parameters and actual mass 
are essential for understanding the thermophysical conditions on a planet and determining its visibility. 
Recently, the first prediction of epoch and location of maximum brightness 
was derived for the giant planet orbiting $\epsilon$ Eridani using HST/FGS astrometry 
in combination with high-precision radial-velocities (Benedict et al. 2006). As of today, 
there are some 20 RV-detected exoplanets with $M_p\sin i > 1$ $M_J$, $P > 1$ yr and projected 
separations $> 0.1$ arcsec (the typical size of the Inner Working Angle of coronagraphic instruments 
presently under study) for which Gaia could provide information on where and when to observe, and 
presumably several tens more will be discovered in the next several years by Doppler surveys and by 
Gaia itself. Gaia's ability to accurately measure orbital parameters (including inclination) and actual mass 
of a planet through high-precision astrometric measurements would then provide 
important supplementary data to aid in the interpretation of direct detections of exoplanets.
 
\subsubsection{How common are the terrestrial planets?}

With the advent of the new generation of ultra-high precision spectrographs 
such as HARPS (e.g., Pepe et al. 2004), radial-velocity programs 
achieving $\lesssim 1$ m s$^{-1}$ measurement precision have begun 
detecting around nearby M dwarfs close-in planets with $M_p\sin i\simeq 5-10$ $M_\oplus$ 
(Rivera et al. 2005; Lovis et al. 2006; Udry et al. 2007), 
so-called `super-Earths', likely to be mostly `rocky' in composition. One of them, GJ 581d (Udry et al. 2007), 
may orbit within the Habitable Zone of the parent star, depending on the assumed exoplanet atmosphere 
(Selsis et al. 2007; von Bloh et al. 2007). 
The announcement of the discovery of a short-period habitable terrestrial planet around 
a low-mass star might well be just around the corner. However, the strongest statistical 
constraints (including bona-fide detections) on the frequency of Earth-sized habitable 
planets orbiting Sun-like stars will likely come from currently operating and upcoming space-borne 
observatories devoted to ultra-high precision transit photometry, such as 
CoRot (Baglin et al. 2002) and Kepler (Borucki et al. 2003), and very high-precision 
narrow-angle astrometry, such as SIM (Beichman et al., 2007, and references therein).

The next challenging step will be to directly detect and characterize terrestrial, 
habitable planets orbiting stars very close ($d\lesssim 25$ pc) to our Sun, searching 
for elements in their atmospheres that can be interpreted as `bio-markers' 
(Hitchcok \& Lovelock 1967; Des Marais et al. 2002; Seager et al. 2005; Tinetti et al. 2007; 
Kaltenegger et al. 2007), implying the likely existence of a complex biology on the surface. 
Imaging terrestrial planets is presently the primary science goal of the 
coronagraphic and interferometric configurations of the Terrestrial Planet Finder 
(TPF-C \& TPF-I) and Darwin missions (Beichman et al. 2007, and references therein). 
Ultimately, the final list of targets will be formulated taking into account constraints coming 
from the knowledge of $f_p$ in the terrestrial mass regime, potential stellar 
host characteristics (spectral type, binarity, surface activity), and environment. 
In this respect, Gaia astrometry of all nearby stars, including the large 
numbers of M dwarfs, within 25 pc from the Sun 
will be an essential ingredient in order to provide Darwin/TPF with a comprehensive database of 
F-G-K-M stars with and without detected giant planets orbiting out to several AUs from
which to choose additional targets based on the presence or absence of Jupiter
signposts (Sozzetti et al. 2003b). Such measurements
will uniquely complement ongoing and planned radial-velocity programs 
and exo-zodiacal dust emission observations from the ground with Keck-I, LBTI, and VLTI.

\section{Summary and conclusions}

We have presented results from an extensive program of 
double-blind tests for planet detection and measurement with Gaia. 
The main findings obtained in this study include: 
a) an improved, more realistic assessment of the detectability
and measurability of single and multiple planets under a variety of
conditions, parametrized by the sensitivity of Gaia, and 
b) an assessment of the impact of Gaia in critical areas of
planet research, in dependence on its expected capabilities. 

Overall, the results of our earlier works (Lattanzi et al. 2000a; 
Sozzetti et al. 2001, 2003a) are essentially confirmed, with a 
fundamental improvement due to the successful development of 
independent orbital fitting algorithms applicable to real-life data 
that do not utilize any a priori knowledge of the orbital parameters 
of the planets. In particular, the results of the T1 test (planet detection) indicate that planets down to 
astrometric signatures $\alpha\simeq 25$ $\mu$as, corresponding to 
$\sim 3$ times the assumed single-measurement error, can be detected 
reliably and consistently, with a very small number of false positives 
(depending on the specific choice of the threshold for detection). 

The results of the T2 test (single-planet orbital solutions) indicate that: 
1) orbital periods can be retrieved with very good accuracy (better than 
10\%) and small bias in the range $0.3\lesssim P\lesssim 6$ yrs, and 
in this period range the other orbital parameters and the planet 
mass are similarly well estimated. The quality of the solutions degrades 
quickly for periods longer than the mission duration, and in particularly 
the fitted value of $P$ is systematically underestimated; 
2) uncertainties in orbit parameters are well understood; 
3) nominal uncertainties obtained from the fitting procedure are a good
estimate of the actual errors in the orbit reconstruction.  Modest
discrepancies between estimated and actual errors arise only for planets
with extremely good signal (errors are overestimated) and for planets
with very long period (errors are underestimated); such discrepancies
are of interest mainly for a detailed numerical analysis, but they do
not touch significantly the assessment of Gaia's ability to find planets
and our preparedness for the analysis of perturbation data. 

The results of the T3 test (multiple-planet orbital solutions) 
indicate that 1) over 70\% of the simulated orbits under the conditions of the T3 test 
(for every two-planet system, periods shorter than 9 years and 
differing by at least a factor of two, $2\leq\alpha/\sigma_\psi\leq 50$, 
$e\leq 0.6$) are correctly identified; 
2) favorable orbital configurations (both planet with 
periods $\leq 4$ yr and astrometric signal-to-noise ratio $\alpha/\sigma_\psi\geq 10$, 
redundancy of over a factor of 2 in the number of observations) 
have periods measured to better than 10\% accuracy $> 90\%$ 
of the time, and comparable results hold for other orbital 
elements; 3) for these favorable cases, only a modest 
degradation of up to $10\%$ in the fraction of well-measured orbits 
is observed with respect to single-planet solutions with comparable 
properties; 4) the overall results are mostly insensitive to the mutual 
inclination of pairs of planetary orbits; 5) over 80\% of the favorable 
configurations have $i_\mathrm{rel}$ measured to better 
than $10^{\degr}$ accuracy, with only mild dependencies on its 
actual value, or on the inclination angle with respect to 
the line of sight of the planets; 
6) error estimates are generally accurate, particularly for 
fitted parameters, while modest discrepancies (errors are 
systematically underestimated) arise between 
formal and actual errors on $i_\mathrm{rel}$. 

Then, we attempted to put Gaia's potential for planet detection 
and measurement in context, by identifying several areas of 
planetary science  in which Gaia can be expected, on the basis of 
our results, to have a dominant impact, and by delineating a 
number of recommended research programs that can be conducted successfully by the mission as planned. 
In conclusion, Gaia's main strength continues to be the unbiased 
and complete magnitude limited census of stars of all ages, spectral types, 
and metallicity in the solar neighborhood that will be screened for new planets, 
which translates into the ability to measure actual masses and orbital parameters for possibly 
thousands of planetary systems. 

The Gaia data have the potential to a) significantly refine our understanding of the statistical 
properties of extrasolar planets: the predicted database of several 
thousand extrasolar planets with well-measured properties will allow 
for example to test the fine structure of giant planet parameters 
distributions and frequencies, and to investigate their possible changes 
as a function of stellar mass with unprecedented resolution; 
b) help crucially test theoretical models of 
gas giant planet formation and migration: for example, specific predictions 
on formation time-scales and the role of varying metal content in the 
protoplanetary disk will be probed with unprecedented statistics thanks 
to the thousands of metal-poor stars and hundreds of young stars 
screened for giant planets out to a few AUs ; c) improve our comprehension of the 
role of dynamical interactions in the early as well as long-term 
evolution of planetary systems: for example, the measurement of orbital parameters 
for hundreds of multiple-planet systems, including meaningful coplanarity 
tests will allow to discriminate between various proposed mechanisms 
for eccentricity excitation; d) aid in the understanding of 
direct detections of giant extrasolar planets: for example, actual mass estimates and full orbital 
geometry determination (including inclination angles) for suitable systems will 
inform direct imaging surveys about where and when to point, in order to 
estimate optimal visibility, and will help in the modeling and interpretation 
of giant giant planets' phase functions and light curves;  
e) provide important supplementary data for the optimization of the selection 
of targets for Darwin/TPF: for example, all F-G-K-M stars within the useful volume ($\sim 25$ pc) 
will be screened for Jupiter- and Saturn-sized planets out to several AUs, and 
these data will help to probe the long-term dynamical stability of their 
Habitable Zones, where terrestrial planets may have formed, and maybe found. 

\begin{table}[tbh]
\begin{minipage}[t]{\columnwidth}
%\begin{center}
\caption{Number of single- and multiple-planet systems detected and measured by Gaia as a 
   function of $\sigma_\psi$.}
\label{sigdegrad}
\centering
   \renewcommand{\footnoterule}{}  % to avoid a line before footnotes
   \setlength\tabcolsep{5pt}
   \begin{tabular}{rrrrrrr}
       \hline\hline\noalign{\smallskip}
       $\sigma_\psi$\footnote{Single-measurement precision} ($\mu$as)& 
       $N_\star$\footnote{Number of stars within the 
   useful distance, assumed to scale with the cube of the radius (in pc) of a 
   sphere centered around the Sun} & 
       $N_{\rm d}$\footnote{Number of single-planet systems detected} &  
       $N_{\rm m}$\footnote{Number of single-planet 
       systems whose astrometric orbits are measured to better than 15\% accuracy} & 
       $N_{\rm d,\,mult}$\footnote{Number of multiple-planet systems detected} &  
       $N_{\rm m,\,mult}$\footnote{Number of multiple-planet systems with orbits 
       measured to better than 15\%-20\% accuracy} & 
       $N_{\rm copl}$\footnote{number of multiple-planet systems for which successful 
   coplanarity tests (with $i_\mathrm{rel}$ known to better than $10^{\degr}$ accuracy) 
   can be carried out.}\\
              \noalign{\smallskip}
       \hline
       \noalign{\smallskip}

       8    &   500\,000 &   8\,000  &  4\,000  &  1\,000  &   500   &  159   \\
      12    &   148\,148 &   2\,370  &  1\,185  &   296  &   148   &   47   \\
      16    &   62\,500  &  1\,000   &  500   &  125   &   62    &  19    \\
      24    &   18\,519  &   296   &  148   &   37   &   18    &   5    \\
      40    &   4\,000   &   64    &  32    &   8    &   4     &  1     \\
      80    &   500    &   8     &  4     &  1     &  0      & 0      \\
     \hline\noalign{\smallskip}
   \end{tabular}
%\end{center}
\end{minipage}
\end{table}

We conclude by providing a word of caution, in light of the possible degradations in the 
expected Gaia astrometric precision on bright stars ($V < 13$). Indeed, refinements in the 
overall Gaia error model (which includes centroiding as well as systematic 
uncertainties due to a variety of calibration errors) are still possible, and a better understanding 
of some of the many effects that need to be taken into account may help reduce the present-day 
end-of-mission scientific contingency margin of $\sim 20\%$ which is included to account for 
discrepancies that may occur between the simplified error-budget assessment performed now 
and the true performances on real data. However, if ultimately a degradation of $35\%-40\%$ in 
the single-measurement precision on bright stars were to be confirmed, the Gaia science case 
for exoplanets would be affected to some degree of relevance. For example, by simply scaling 
with the value of the astrometric signal needed for detection and measurement of the orbital 
parameters to 15\%-20\% ($\alpha/\sigma_\psi\sim 3-5$, see Figure~\ref{sign_per}), as $\sigma_\psi$ 
increases the same type of system (same stellar mass, same planet mass, same orbital period) 
would be characterized at increasingly shorter distances. A comparison between numbers of 
detectable and measurable single- and multiple-planet systems as a function of increasing 
Gaia single-measurement error is presented in Table~\ref{sigdegrad}. 
Assuming that the number of objects scales with the cube of the radius (in pc) of a 
sphere centered around the Sun (with no distinction of spectral types), 
if $\sigma_\psi$ degrades from 8 $\mu$as to $12$ $\mu$as (closer to the present-day estimate)
then this would correspond to a reduction of a factor $\sim 2$ in the distance limit and in 
a corresponding decrease in the number of stars available for investigation from $\sim 5\times 10^5$ 
to $\sim 1.5\times 10^5$. If $\sigma_\psi$ were to worsen by a factor 2, the number of 
stars available for planet detection and measurement ($\sim 6\times 10^4$) would be reduced by 
about an order of magnitude. Accordingly, the expected numbers of giant planets detected and 
measured would decrease from $\sim 4000$ to $\sim 1200$ and $\sim 500$, respectively, and 
the number of multiple systems for which coplanarity could be established would diminish from 
$\sim 160$ to $\sim 50$ and $\sim 20$, respectively. We conclude that a factor 2 degradation 
in astrometric precision would severely impact most of Gaia exoplanet science case. 
We are aware that, instead of using simple scaling laws, one should provide 
more quantitative statements based on new simulations. 
However, this activity will necessarily be tied to further developments of 
the understanding of the technical specifications of Gaia and its instruments, and of its
observation and data analysis process; 
therefore, we plan to revisit these issues as needed in the future, depending 
on the actual evolution of the knowledge of the Gaia measurement process.

\begin{acknowledgements}

We are indebted to an anonymous referee for a very careful, critical reading 
of the manuscript, and many useful comments and suggestions which helped to greatly 
improve it. This research has made use of NASA's Astrophysics
Data System Abstract Service and of the SIMBAD database, operated at
CDS, Strasbourg, France. M.G.L. acknowledges support from STScI through the 
Institute's Visitor Program for 2007. We gratefully acknowledge INAF 
(P. Vettolani, Projects Department) for its continued support of the Italian 
participation to the Gaia mission. 

\end{acknowledgements}

\onecolumn
\appendix

\section{The Simulated Model}

The code for the generation of synthetic Gaia observations of 
planetary systems is run by the Simulators group.

We start by generating spheres of $N$ targets. Each target's
two-dimensional position is described in the ecliptic reference
frame via a set of two coordinates $\lambda_b$ and $\beta_b$,
called here barycentric coordinates. We linearly update the
barycentric position as a function of time, accounting for the
(secular) effects of proper motion (two components, $\mu_\lambda$
and $\mu_\beta$), the (periodic) effect of the parallax $\pi$, and
the (Keplerian) gravitational perturbations induced on the parent
star by one or more orbiting planets (mutual interactions between
planets are presently not taken into account). The model of motion
can thus be expressed as follows:

\begin{equation}\label{uno}
\mathbf{x}_\mathrm{ecl} = \mathbf{x}_\mathrm{ecl}^0+
\mathbf{x}_\mathrm{ecl}^{\pi,\mu} +
 \sum_{j=1}^{n_p}\mathbf{x}_\mathrm{ecl}^\mathrm{K,j}
\end{equation}
Where:
\begin{displaymath}
{\rm x}_\mathrm{ecl}^0 = \pmatrix{\cos\beta_b\cos\lambda_b
\cr
         \cos\beta_b\sin\lambda_b \cr
     \sin\beta_b              \cr}
\end{displaymath}
is the initial position vector of the system barycenter. The various
perturbative effects are initially defined in the tangent plane.
The parallax and proper motion terms are contribute as:
\begin{displaymath}
{\rm{\bf x}}_{\pi,\mu} = \pmatrix{\mu_\lambda t + \pi F_\lambda
\cr
         \mu_\beta t +  \pi F_\beta \cr
             0             \cr}
\end{displaymath}
Where the parallax factors are defined utilizing the classic
formulation by Green (1985):
\begin{eqnarray}
F_\lambda & = & -\sin(\lambda_b-\lambda_\odot)\nonumber \\ F_\beta
& = & -\sin\beta_b\sin(\lambda_b-\lambda_\odot)\nonumber
\end{eqnarray}
and $\lambda_\odot$ is the sun's longitude at the given time $t$.
The term describing the Keplerian motion of the j-th planet in the
tangent plane is:
\begin{displaymath} {\rm{\bf x}}_\mathrm{K,j} =
\pmatrix{\mathrm{x}_\mathrm{K,j} \cr
         \mathrm{y}_\mathrm{K,j} \cr
            0             \cr} =
\pmatrix{\varrho_\mathrm{j}\cos\vartheta_\mathrm{j} \cr
        \varrho_\mathrm{j}\sin\vartheta_\mathrm{j} \cr
            0             \cr},
\end{displaymath}
where $\varrho_\mathrm{j}$ is the separation and
$\vartheta_\mathrm{j}$ the position angle. The two coordinates
$\mathrm{x}_\mathrm{K,j}$ and $\mathrm{y}_\mathrm{K,j}$ are
functions of the 7 orbital elements:
\begin{eqnarray}
\mathrm{x}_\mathrm{K,j} & = & a_\mathrm{j}(1-e_\mathrm{j}\cos
E_\mathrm{j})(\cos(\nu_\mathrm{j}+\omega_\mathrm{j})\cos\Omega_\mathrm{j}
- \sin(\nu_\mathrm{j}+\omega_\mathrm{j})\sin\Omega_\mathrm{j}\cos
i_\mathrm{j}) \\ \mathrm{y}_\mathrm{K,j} & = &
a_\mathrm{j}(1-e_\mathrm{j}\cos E_\mathrm{j})
(\cos(\nu_\mathrm{j}+\omega_\mathrm{j})\sin\Omega_\mathrm{j} +
\sin(\nu_\mathrm{j}+\omega_\mathrm{j})\cos\Omega_\mathrm{j}\cos
i_\mathrm{j}),
\end{eqnarray}
where $i_\mathrm{j}$ is the inclination of the orbital plane,
$\omega_\mathrm{j}$ is the longitude of the pericenter,
$\Omega_\mathrm{j}$ is the position angle of the line of nodes,
$e_\mathrm{j}$ is the eccentricity, $a_\mathrm{j}$ is the apparent
semi-major axis of the star's orbit around the system barycenter,
i.e. the {\it astrometric signature}. For what concerns
E$_\mathrm{j}$, the eccentric anomaly, is the solution to Kepler's
Equation:
\begin{equation}
E_\mathrm{j} - e_\mathrm{j}\sin E_\mathrm{j} = M_\mathrm{j},
\end{equation}
with the mean anomaly M$_\mathrm{j}$, expressed in terms of the
orbital period $P_\mathrm{j}$ and the epoch of the pericenter
passage $\tau_\mathrm{j}$:
\begin{equation}
M_\mathrm{j} = \frac{2\pi}{P_\mathrm{j}}(t - \tau_\mathrm{j})
\end{equation}
Finally, the true anomaly $\nu_\mathrm{j}$ is a function of the
eccentricity and the eccentric anomaly:
\begin{equation}
\nu_\mathrm{j} =
2\arctan\left\{\left(\frac{1+e_\mathrm{j}}{1-e_\mathrm{j}}\right)^{1/2}\tan
E_\mathrm{j}/2\right\}
\end{equation}
We then rotate on the ecliptic reference frame by means of the
transformation matrix:
\begin{displaymath}
{\rm R}(\lambda_b,\beta_b) = \pmatrix{-\sin\lambda_b &
-\sin\beta_b\cos\lambda_b &
          \cos\beta_b\cos\lambda_b \cr
          \cos\lambda_b & -\sin\beta_b\sin\lambda_b &
          \cos\beta_b\sin\lambda_b \cr
          0 & \cos\beta_b & \sin\beta_b \cr}
\end{displaymath}
The other two vectors in Eq.~\ref{uno} are thus defined as:
\begin{eqnarray}
{\rm{\bf x}}_\mathrm{ecl}^\mathrm{K,j} & = & {\rm
R}(\lambda_b,\beta_b)\cdot{\rm{\bf x}}_\mathrm{K,j} \nonumber \\ &
= &
\pmatrix{-\sin\lambda_b\varrho_\mathrm{j}\cos\vartheta_\mathrm{j}
         -\sin\beta_b\cos\lambda_b\varrho_\mathrm{j}\sin\vartheta_\mathrm{j} \cr
          \cos\lambda_b\varrho_\mathrm{j}\cos\vartheta_\mathrm{j}
         -\sin\beta_b\sin\lambda_b\varrho_\mathrm{j}\sin\vartheta_\mathrm{j} \cr
          \cos\beta_b\varrho_\mathrm{j}\sin\vartheta_\mathrm{j} \cr}\nonumber
\end{eqnarray}
\begin{eqnarray}
{\rm{\bf x}}_\mathrm{ecl}^{\pi,\mu} & = & {\rm
R}(\lambda_b,\beta_b)\cdot{\rm{\bf x}}_{\pi,\mu} \nonumber \\ & =
& \pmatrix{-\sin\lambda_b\{\mu_\lambda t + \pi F_\lambda\}
                -\sin\beta_b\cos\lambda_b\{\mu_\beta t + \pi F_\beta\}  \cr
                 \cos\lambda_b\{\mu_\lambda t + \pi F_\lambda\}
                -\sin\beta_b\sin\lambda_b\{\mu_\beta t + \pi F_\beta\}  \cr
                 \cos\beta_b\{\mu_\beta t + \pi F_\beta\}               \cr}\nonumber
\end{eqnarray}
This allows us to write Eq.~\ref{uno} in the form:
\begin{eqnarray}
\pmatrix{x_\mathrm{ecl} \cr
                        \cr
         y_\mathrm{ecl} \cr
            \cr
         z_\mathrm{ecl} \cr} & = &
 \pmatrix{ \cos\beta_b\cos\lambda_b
      -\sin\lambda_b\varrho_\mathrm{j}\cos\vartheta_\mathrm{j}
      -\sin\beta_b\cos\lambda_b\sum_{j=1}^{n_p}\varrho_\mathrm{j}
      \sin\vartheta_\mathrm{j} \cr
      -\sin\lambda_b\{\mu_\lambda t + \pi F_\lambda\}
      -\sin\beta_b\cos\lambda_b\{\mu_\beta t + \pi F_\beta\} \cr
           \cos\beta_b\sin\lambda_b
      +\cos\lambda_b\sum_{j=1}^{n_p}\varrho_\mathrm{j}\cos\vartheta_\mathrm{j}
      -\sin\beta_b\sin\lambda_b\sum_{j=1}^{n_p}\varrho_\mathrm{j}
      \sin\vartheta_\mathrm{j} \cr
      +\cos\lambda_b\{\mu_\lambda t + \pi F_\lambda\}
      -\sin\beta_b\sin\lambda_b\{\mu_\beta t + \pi F_\beta\} \cr
      sin\beta_b +\cos\beta_b\sum_{j=1}^{n_p}\varrho_\mathrm{j}
      \sin\vartheta_\mathrm{j}
      +\cos\beta_b\{\mu_\beta t + \pi F_\beta\}\cr}\nonumber
\end{eqnarray}
Finally, a rotation to the local reference frame defined by 
the Instantaneous Great Circles is made
by means of the transformation matrix (e.g., ESA 1997):
\begin{equation}
{\rm{\bf x}}_\mathrm{IGC} = {\rm
R}(\lambda_p,\beta_p)\cdot{\rm{\bf x}}_\mathrm{ecl},
\end{equation}
where:
\begin{displaymath}
{\rm R}(\lambda_p,\beta_p) = \pmatrix{-\sin\lambda_p &
\cos\lambda_p & 0 \cr
         -\sin\beta_p\cos\lambda_p & -\sin\beta_p\sin\lambda_p & \cos\beta_p \cr
         \cos\beta_p\cos\lambda_p & \cos\beta_p\sin\lambda_p & \sin\beta_p \cr}
\end{displaymath}
and $\lambda_p$, $\beta_p$ are the coordinates of the pole of the
IGC at any given time. The resulting vector can be expressed in
terms of the two angular coordinates $\psi$ and $\eta$:
\begin{displaymath}
{\rm {\bf x}}_\mathrm{IGC} = \pmatrix{x_\mathrm{IGC} \cr
         y_\mathrm{IGC} \cr
     z_\mathrm{IGC} \cr} =
\pmatrix{\cos\psi\cos\eta \cr
         \cos\eta\sin\psi \cr
     \sin\eta         \cr}
 \end{displaymath}
By now expanding in Taylor Series to first order the IGC cartesian
position vector of each target, it is possible to derive a set of
linearized equations of condition expressing only the observed
abscissa $\psi$ as a function of all astrometric parameters and
orbital elements. We formally have:
\begin{equation}
\delta{\rm {\bf x}}_\mathrm{IGC} =
\sum_{m=1}^{n}\frac{\partial{\rm {\bf x}}_\mathrm{IGC}} {\partial
a_m}{\rm d}a_m
\end{equation}
The $n$ unknowns $a_m$ represent positions, proper motions,
parallax, and the $7\star n_p$ orbital elements (if the star is
not single). Now consider that:
\begin{eqnarray}
\delta{\rm {\bf x}}_\mathrm{IGC} & = & \delta(x_\mathrm{IGC},
y_\mathrm{IGC}, z_\mathrm{IGC}) = (\delta(\cos\psi\cos\eta),
\delta(\sin\psi\cos\eta), \delta\sin\eta)\nonumber \\ & = &
(-\sin\psi\cos\eta {\rm d}\psi - \sin\eta\cos\psi {\rm d}\eta,
\cos\psi\cos\eta {\rm d}\psi - \sin\eta\sin\psi {\rm d}\eta,
\cos\eta {\rm d}\eta)\nonumber \\ & = & (-\sin\psi\cos\eta {\rm
d}\psi, \cos\psi\cos\eta {\rm d}\psi, 0)\nonumber \\ && + (-
\sin\eta\cos\psi {\rm d}\eta, - \sin\eta\sin\psi {\rm d}\eta,
\cos\eta {\rm d}\eta)\nonumber \\ & = & \cos\eta(-\sin\psi,
\cos\psi,0){\rm d}\psi + (-\sin\eta\cos\psi, -\sin\eta\sin\psi,
\cos\eta){\rm d}\eta\nonumber \\ & = & \cos\eta {\rm d}\psi{\rm
{\bf e}}_\psi + {\rm d}\eta{\rm {\bf e}}_\eta\nonumber
\end{eqnarray}
where $\mathbf{e}_\eta$ and $\mathbf{e}_\psi$ constitute the pair
of orthogonal unit vectors in the directions parallel to $\psi$
and $\eta$, as defined in the tangent plane. We then have:

\begin{equation}
\cos\eta {\rm d}\psi{\rm {\bf e}}_\psi + {\rm d}\eta{\rm {\bf
e}}_\eta = \sum_{m=1}^{n}\frac{\partial{\rm {\bf
x}}_\mathrm{IGC}}{\partial a_m}{\rm d}a_m
\end{equation}
By taking the scalar product with $\mathbf{e}_\psi$, we obtain the
following scalar expression:

\begin{equation}
\cos\eta {\rm d}\psi =
(-\sin\psi)\sum_{m=1}^{n}\frac{\partial{x_\mathrm{IGC}}}{\partial
a_m}{\rm d}a_m +
(\cos\psi)\sum_{m=1}^{n}\frac{\partial{y_\mathrm{IGC}}}{\partial
a_m}{\rm d}a_m
\end{equation}
If we now define:
\begin{equation}
c_{a_m} = (-\sin\psi)\frac{\partial{x_\mathrm{IGC}}}{\partial a_m}
+ (\cos\psi)\frac{\partial{y_\mathrm{IGC}}}{\partial a_m},
\end{equation}
then the linearized condition equation takes the form:
\begin{equation}
\cos\eta {\rm d}\psi = \sum_{m=1}^{n} c_{a_m}{\rm d}a_m =
F(\lambda,\beta,\mu_\lambda,\mu_\beta,\pi,a_\mathrm{j},P_\mathrm{j},
\tau_\mathrm{j},\omega_\mathrm{j},\Omega_\mathrm{j},e_\mathrm{j},i_\mathrm{j}),
\hskip 0.2cm j=1,\dots,n_P \label{cond}
\end{equation}
For each given target, there will be as many equations of this form as the number of observation
epochs. The quantity ${\rm d}\psi = \psi_\mathrm{obs} - \psi_\mathrm{cat}$ is defined as the
difference between the observed and catalog abscissa.


\begin{thebibliography}{}

\bibitem[Agol et al. 2005]{agol05}
Agol, E., Steffen, J., Sari, R., \& Clarkson, W. 2005, \mnras, 359, 567
\bibitem[Alibert et al., 2005]{alibert05}
Alibert, Y., Mordasini, C., Benz, W., \& Winisdoerffer, C. 2005,
\aap, 434, 343
\bibitem[Armitage et al., 2002]{armitage02}
Armitage, P. J., Livio, M., Lubow, S. H., \& Pringle, J. E. 2002, \mnras, 334, 248
\bibitem[Bakos et al., 2007]{bakos07}
Bakos, G. \'A., et al. 2007, \apj, 671, L173
\bibitem[Baglin et al., 2002]{baglin02}
Baglin, A., et al. 2002, in Radial and Nonradial Pulsations as Probes
of Stellar Physics, eds. C. Aerts, T. R. Bedding, \& J. Christensen-Dalsgaard,
ASP Conf. Ser., 259, 626
\bibitem[Baraffe et al., 2003]{baraffe03}
Baraffe, I., Chabrier, G., Allard, F., Hauschildt, P.H. 2003, \aap, 402, 701
\bibitem[Barnes \& Quinn, 2004]{barnes04}
Barnes, R., \& Quinn, T. 2004, \apj, 611, 494
\bibitem[Bean et al., 2007]{bean07}
Bean, J. L., et al. 2007, \aj, 134, 749 
\bibitem[Beaulieu et al., 2006]{beaulieu06}
Beaulieu, J.-Ph., et al. 2006, \nat, 439, 437
\bibitem[Beichman et al., 2007]{beichman07}
Beichman, C. A., Fridlund, M., Traub, W. A., Stapelfeldt, K. R., Quirrenbach, A., 
\& Seager, S. 2007, in Protostars and Planets V, B. Reipurth, D. Jewitt, and K. Keil (eds.), 
University of Arizona Press, Tucson, 915
\bibitem[Benedict et al., 2002]{benedict02}
Benedict, G. F., et al. 2002, \apjl, 581, L115
\bibitem[Benedict et al., 2006]{benedict06}
Benedict, G. F., et al., 2006, \aj, 132, 2206
\bibitem[Bienaym\'e et al., 1987]{bienayme87}
Bienaym\'e, O., Robin, A. C., \& Cr\'ez\'e, M. 1987, \aap, 180, 94
\bibitem[Biller et al., 2006]{biller06}
Biller, B. A., Kasper, M., Close, L. M., Brandner, W., \& Kellner, S. 2006, \apj, 641, L141
\bibitem[Bond et al., 2004]{bond04}
Bond, I., et al., 2004, \apj, 606, L155
\bibitem[Bonfils et al., 2005]{bonfils05}
Bonfils, X., et al. 2005, \aap, 443, L15
\bibitem[Borucki et al., 2003]{borucki03}
Borucki, W. J., et al. 2003, in Future EUV/UV and Visible Space
Astrophysics Missions and Instrumentation, eds. J. C. Blades \& O. H. W.
Siegmund, Proc. SPIE, 4854, 129
\bibitem[Boss 2000]{boss00}
Boss, A. P. 2000, \apjl, 536, L101
\bibitem[Boss, 2001]{boss01}
Boss, A. P. 2001, \apj, 563, 367
\bibitem[Boss 2002]{boss02}
Boss, A. P. 2002, \apjl, 567, L149
\bibitem[Boss, 2005]{boss05}
Boss, A. P. 2005, \apj, 629, 535
\bibitem[Boss, 2006]{boss06}
Boss, A. P. 2006, \apj, 643, 501
\bibitem[Bouchy et al., 2004]{bouchy04}
Bouchy, F., Pont, F., Santos, N. C., Melo, C., Mayor, M., Queloz,
D., \& Udry, S. 2004, \aap, 421, L13
\bibitem[Bouchy et al., 2005]{bouchy05}
Bouchy, F., et al. 2005, \aap, 444, L15
\bibitem[Burkert \& Ida, 2007]{burkert07}
Burkert, A., \& Ida, S. 2007,\apj, 660, 845
\bibitem[Burrows et al., 2004]{burrows04}
Burrows, A., Sudarsky, D., \& Hubeny, I. 2004, \apj, 609, 407
\bibitem[Burrows, 2005]{burrows05}
Burrows, A. 2005, \nat, 433, 261
\bibitem[Burrows et al., 2007]{burrows07}
Burrows, A., Budaj, J., \& Hubeny, I. 2007, \apj, accepted (arXiv:0709.4080)
\bibitem[Butler et al., 2004]{butler04}
Butler, R. P., Vogt, S. S., Marcy, G. W., Fischer, D. A.,
Wright, J. T., Henry, G. W., Laughlin, G., \& Lissauer, J. J. 2004,
\apj, 617, 580
\bibitem[Butler et al., 2006]{butler06}
Butler, R. P., et al. 2006, \apj, 646, 505
\bibitem[Casertano et al., 1996]{caser96}
Casertano, S., Lattanzi, M. G., Perryman, M. A. C., \& Spagna, A. 1996,
\apss, 241, 89
\bibitem[Casertano \& Sozzetti, 1999]{caser99}
Casertano, S., \& Sozzetti, A. 1999, in in Working on the Fringe: Optical and IR
Interferometry from Ground and Space, ed. S. Unwin \& R. Stachnik,
ASP Conf. Ser. 194, 171
\bibitem[Catanzarite et al., 2006]{catanzarite06}
Catanzarite, J., Shao, M., Tanner, A., Unwin, S., \& Yu, J. 2006, \pasp, 118, 1319
\bibitem[Charbonneau et al., 2007]{charbonneau07}
Charbonneau, D., Brown, T. M., Burrows, A., \& Laughlin, G. 2007, 
in Protostars and Planets V, B. Reipurth, D. Jewitt, and K. Keil (eds.), 
University of Arizona Press, Tucson, 701
\bibitem[Chauvin et al., 2005a]{chauvin05a}
Chauvin, G., et al. 2005a, \aap, 438, L25
\bibitem[Chauvin et al., 2005b]{chauvin05b}
Chauvin, G., et al. 2005b, \aap, 438, L29
\bibitem[Collier Cameron et al., 2006]{collier06}
Collier Cameron, A., et al. 2006, \mnras, 375, 95
\bibitem[Correia et al. 2005]{correia04}
Correia, A. C. M., Udry, S., Mayor, M., Laskar, J., Naef, D.,
Pepe, F., Queloz, D., \& Santos, N. C. 2005, \aap, 440, 751
\bibitem[de Felice et al., 1998]{defelice98}
de Felice, F., Lattanzi, M. G., Vecchiato, A., \& Bernacca, P. L.
1998, \aap, 332, 1133
\bibitem[de Felice et al., 2001]{defelice01}
de Felice, F., Bucciarelli, B., Lattanzi, M. G., \& Vecchiato, A.
2001, \aap, 373, 336
\bibitem[de Felice et al., 2004]{defelice04}
de Felice, F., Crosta, M. T., Vecchiato, A.,
Lattanzi, M. G., \& Bucciarelli, B. 2004, \apj, 607, 580
\bibitem[Delplancke et al., 2006]{delplancke06}
Delplancke, F., et al. 2006, in Advances in Stellar Interferometry, 
eds. J.D. Monnier, M. Sch\"oller, \& W.C. Danchi, \procspie, 6268, 27
\bibitem[Des Marais et al. 2002]{desmarais02}
Des Marais, D. J., et al. 2002, Astrobiology, 2, 153
\bibitem[Durisen et al., 2007]{durisen07}
Durisen, R. H., Boss, A. P., Mayer, L., Nelson, A. F., Quinn, T., \& 
Rice, W. K. M. 2007, in Protostars and Planets V, B. Reipurth, D. Jewitt, and K. Keil (eds.), 
University of Arizona Press, Tucson, 607
\bibitem[Dyudina et al., 2005]{dyudina05}
Dyudina, U. A., Sackett, P. D., Bayliss, D. D. R., Seager, S., Porco, C. C.,
Throop, H. B., \& Dones, L. 2005, \apj, 618, 973
\bibitem[Endl et al., 2006]{endl06}
Endl, M., Cochran, W. D., K\"urster, M., Paulson, D. B., Wittenmyer, R. A., 
MacQueen, P. J., \& Tull, R. G. 2006, \apj, 649, 436
\bibitem[Eriksson \& Lindegren, 2007]{eriksson07}
Eriksson, U., \& Lindegren, L. 2007, \aap, 476, 1389
\bibitem[ESA, 1997]{esa97}
ESA 1997, The Hipparcos and Tycho Catalogues, ESA SP-1200
\bibitem[Fischer \& Valenti 2005]{valenti05}
Fischer, D. A., \& Valenti, J. 2005, \apj, 622, 1102
\bibitem[Fischer et al., 2005]{fischer05}
Fischer, D. A., et al. 2005, \apj, 620, 481
\bibitem[Ford \& Tremaine, 2003]{ford03}
Ford, E. B., \& Tremaine, S. 2003, \pasp, 115, 1171
\bibitem[Ford, 2004]{ford04}
Ford, E. B. 2004, \pasp, 116, 1083
\bibitem[Ford, 2006]{ford06}
Ford, E. B. 2006, \pasp, 118, 364
\bibitem[Ford \& Gregory, 2007]{fordgreg07}
Ford, E. B., \& Gregory, P. C. 2007, in Statistical Challenges in Modern Astronomy IV, 
G.J. Babu and E.D. Feigelson (eds.), ASP Conf. Ser., 371, 189 
\bibitem[Ford \& Rasio, 2007]{ford07}
Ford, E. B., \& Rasio, F. A. 2007, \apj, submitted (astro-ph/0703163)
\bibitem[Frink et al., 2002]{frink02}
Frink, S., Mitchell, D. S., Quirrenbach, A., Fischer, D. A., Marcy, G. W., 
\& Butler, R. P. 2002, \apj, 576, 478
\bibitem[Galland et al., 2005]{galland05}
Galland, F., Lagrange, A.-M., Udry, S., Chelli, A., Pepe, F., Beuzit, J.-L., \& 
Mayor, M. 2005, \aap, 444, L21
\bibitem[Gaudi et al., 2005]{gaudi05}
Gaudi, B. S., Seager, S., \& Mall{\' e}n-Ornelas, G. 2005, \apj, 623, 472
\bibitem[Gonzalez 1997]{gonzalez97}
Gonzalez G., 1997, \mnras, 285, 403
\bibitem[Gould et al., 2006]{gould06}
Gould, A., et al. 2006, \apj, 644, L37
\bibitem[Go\'zdziewski \& Konacki, 2004]{godz04}
Go\'zdziewski, K., \& Konacki, M. 2004, \apj, 610, 1093
\bibitem[Green, 1985]{green85}
Green, R. M. 1985, Spherical Astronomy, Cambridge University Press
\bibitem[Hatzes et al., 2005]{hatzes05}
Hatzes, A. P., Guenther, E. W., Endl, M., Cochran, W. D., 
Döllinger, M. P., \& Bedalov, A. 2005, \aap, 437, 743
\bibitem[Hitchcock \& Lovelock, 1967]{hitch67}
Hitchcock, D. R., \& Lovelock, J. E. 1967, Icarus, 7, 149
\bibitem[Holman \& Murray, 2005]{holman05}
Holman, M. J., \& Murray, N. W. 2005, Science, 307, 1288
\bibitem[Horne \& Baliunas, 1986]{horne86}
Horne, J. H., \& Baliunas, S. L. 1986, \apj, 302, 757
\bibitem[Hubbard et al., 2002]{hubbard02}
Hubbard, W. B., Burrows, A., \& Lunine, J. I. 2002, \araa, 40, 103
\bibitem[Ida \& Lin, 2004a]{ida04a}
Ida, S., \& Lin, D. N. C. 2004a, \apj, 604, 388
\bibitem[Ida \& Lin, 2004b]{ida04b}
Ida, S., \& Lin, D. N. C. 2004b, \apj, 616, 567
\bibitem[Ida \& Lin, 2005]{ida05}
Ida, S., \& Lin, D. N. C. 2005, \apj, 626, 1045
\bibitem[Ida \& Lin, 2008]{ida08}
Ida, S., \& Lin, D. N. C. 2008, \apj, 673, 487
\bibitem[Johnson et al., 2006]{johnson06}
Johnson, J. A., Marcy, G. W., Fischer, D. A., Henry, G. W., 
Wright, J. T., Isaacson, H., \& McCarthy, C. 2006, \apj, 652, 1724
\bibitem[Johnson et al., 2007a]{johnson07a}
Johnson, J. A., et al. 2007a, \apj, 655, 785
\bibitem[Johnson et al., 2007b]{johnson07b}
Johnson, J. A., Butler, R. P., Marcy, G. W., Fischer, D. A., 
Vogt, S. S., Wright, J. T., \& Peek, K. M. G. 2007b, \apj, 670, 833
\bibitem[Ji et al. 2003]{ji03}
Ji, J., Liu, L., Kinoshita, H., Zhou, J., Nakai, H., \& 
Li, G. 2003, \apj, 591, L57
\bibitem[Jones et al. 2005]{jones05}
Jones, B. W., Underwood, D. R., \& Sleep, P. N. 2005, \apj, 622, 1091
\bibitem[Kaltenegger et al., 2007]{kaltenegger07}
Kaltenegger, L., Traub, W. A., \& Jucks, K. W. 2007, \apj, 658, 598
\bibitem[Kasting et al., 1993]{kasting93}
Kasting, J. F., Whitmire, D. P., \& Reynolds, R. T. 1993,
Icarus, 101, 108
\bibitem[Kiseleva-Eggleton et al. 2002]{kiseleva02}
Kiseleva-Eggleton, L., Bois, E., Rambaux, N., \& Dvorak, R. 2002,
\apj, 578, L145
\bibitem[Klioner \& Kopeikin, 1992]{klioner92}
Klioner, S. A., \& Kopeikin, S. M. 1992, \aj, 104, 897
\bibitem[Klioner, 2003]{klioner03}
Klioner, S. A. 2003, \aj, 125, 1580
\bibitem[Klioner, 2004]{klioner04}
Klioner, S. A. 2004, Phys. Rev. D, 69, No. 124001
\bibitem[Konacki et al., 2002]{konacki02}
Konacki, M., Maciejewski, A. J., \& Wolszczan, A. 2002, \apj, 567, 566
\bibitem[Konacki et al. 2003]{konacki03}
Konacki, M., Torres, G., Jha, S., \& Sasselov, D. D. 2003, \nat, 421, 507
\bibitem[Konacki et al., 2005]{konacki05}
Konacki, M., Torres, G., Sasselov, D. D., \& Jha, S. 2005, \apj, 624, 372
\bibitem[Kornet \& Wolf, 2005]{kornwolf05}
Kornet, K., \& Wolf, S. 2005, \aap, 454, 989
\bibitem[Kornet et al., 2005]{kornet05}
Kornet, K., Bodenheimer, P., R\'ozyczka, M., \& Stepinski, T. F. 2005, \aap, 430, 1133
\bibitem[Kornet et al., 2006]{kornet06}
Kornet, K., Bodenheimer, P., \& R\'ozyczka, M. 2006, \aap, 458, 661
\bibitem[Kov\'acs et al., 2007]{kovacs07}
Kov\'acs, G., et al. 2007, \apj, 670, L41
\bibitem[Lattanzi et al., 1997]{latt97}
Lattanzi, M. G., Spagna, A., Sozzetti, A., \& Casertano, S. 1997,
in Proc. of the ESA Symposium `Hipparcos - Venice '97', ESA SP-402, 755
\bibitem[Lattanzi et al., 2000a]{latt00a}
Lattanzi, M. G., Spagna, A., Sozzetti, A., \& Casertano, S. 2000a,
\mnras, 317, 211
\bibitem[Lattanzi et al., 2000b]{latt00b}
Lattanzi, M. G., Sozzetti, A., \& Spagna, A. 2000b,
in From Extra-solar Planets to Cosmology: The VLT
Opening Symposium, ed. J. Bergeron \& A. Renzini (Berlin:
Springer-Verlag), 479
\bibitem[Lattanzi et al., 2002]{latt02}
Lattanzi, M. G., Casertano, S., Sozzetti, A., \& Spagna, A. 2002,
in GAIA: A European Space Project, eds. O. Bienaym\'e \& C. Turon,
EAS Publications Series (EDP Sciences), 2, 207
\bibitem[Lattanzi et al., 2005]{latt05}
Lattanzi, M. G., Casertano, S., Jancart, S., Morbidelli, R.,
Pannunzio, R., Pourbaix, D., Sozzetti, A., \& Spagna, A. 2005,
in The Three-Dimensional Universe with Gaia,  
eds. C. Turon, K.S. O'Flaherty, \& M.A.C. Perryman, ESA SP-576, 251
\bibitem[Laughlin et al., 2004]{laughlin04}
Laughlin, G., Bodenheimer, P., \& Adams, F. C. 2004, \apjl, 612, L73
\bibitem[Laughlin et al., 2005]{laughlin05}
Laughlin, G., Butler, R. P., Fischer, D. A., Marcy, G. W., 
Vogt, S. S., \& Wolf, A. S. 2005, \apj, 622, 1182
\bibitem[Lindegren \& de Bruijne, 2005]{linde05}
Lindegren, L., \& de Bruijne, J. H. J. 2005, in Astrometry in the Age of the 
Next Generation of Large Telescopes, ASP Conf. Ser. 338, 25
\bibitem[Lissauer \& Stevenson, 2007]{lissa07}
Lissauer, J. J., \& Stevenson, D. J. 2007, 
in Protostars and Planets V, B. Reipurth, D. Jewitt, and K. Keil (eds.), 
University of Arizona Press, Tucson, 591
\bibitem[Livio \& Pringle, 2003]{livio03}
Livio, M., \& Pringle, J. E. 2003, \mnras, 346, L42
\bibitem[Lovis et al., 2006]{lovis06}
Lovis, C., et al. 2006, \nat, 441, 305
\bibitem[Lovis \& Mayor, 2007]{lovis07}
Lovis, C., \& Mayor, M. 2007, \aap, 472, 657
\bibitem[L\'opez-Santiago et al., 2006]{lopez06}
L\'opez-Santiago, J., Montes, D., Crespo-Chac\'on, I., \& 
Fern\'andez-Figueroa, M. J. 2006, \apj, 643, 1160
\bibitem[Mandushev et al., 2007]{mandushev07}
Mandushev, G., et al. 2007, \apj, 667, L195
\bibitem[Marcy et al., 2005]{marcy05}
Marcy, G. W., Butler, R. P., Fischer, D. A., Vogt, S. S., Wright, J. T.,
Tinney, C. G., \& Jones, H. R. A 2005, Progress of Theoretical Physics Supplement, 
158, 24
\bibitem[Mayer et al. 2002]{mayer02}
Mayer L., Quinn T., Wadsley J., \& Stadel J., 2002, Science, 298, 1756
\bibitem[Mayor \& Queloz 1995]{mayor95}
Mayor, M., \& Queloz, D. 1995, \nat, 378, 355
\bibitem[McArthur et al., 2004]{mcarthur04}
McArthur, B. E., et al. 2004, \apjl, 614, L81
\bibitem[Menou \& Tabachnik 2003]{menou03}
Menou, C., \& Tabachnik, S. 2003, \apj, 583, 473
\bibitem[Miralda-Escud\'e 2002]{miralda02}
Miralda-Escud\'e, J. 2002, \apj, 564, 1019
\bibitem[Neuh\"auser et al., 2005]{neuhauser05}
Neuh\"auser, R., Guenther, E. W., Wuchterl, G., Mugrauer, M., 
Bedalov, A., \& Hauschildt, P. H. 2005, \aap, 435, L13
\bibitem[Niedzielski et al., 2007]{nied07}
Niedzielski, A., et al. 2007, \apj, 669, 1354
\bibitem[Pepe et al., 2004]{pepe04}
Pepe, F., et al. 2004, \aap, 423, 385
\bibitem[Perryman et al., 1997]{perryman97}
Perryman, M. A. C., et al. 1997, \aap, 323, L49
\bibitem[Perryman et al., 2001]{perryman01}
Perryman, M. A. C., et al. 2001, \aap, 369, 339
\bibitem[Pollack et al. 1996]{pollack96}
Pollack, J. B., Hubickyj, O., Bodenheimer, P., Lissauer, J. J.,
Podolack, M., \& Greenzweig, Y. 1996, Icarus, 124, 62
\bibitem[Pont et al., 2004]{pont04}
Pont, F., Bouchy, F., Queloz, D., Santos, N. C, Melo, C., Mayor,
M., \& Udry, S. 2004, \aap, 426, L15
\bibitem[Pont et al., 2007]{pont07}
 Pont, F., et al. 2007, \aap, submitted (arXiv:0710.5278)
\bibitem[Pourbaix, 2002]{pourbaix02}
Pourbaix, D. 2002, \aap, 385, 686
\bibitem[Rice et al., 2003a]{rice03a}
Rice, W. K. M., Armitage, P. J., Bate, M. R., \& Bonnell, I. A. 2003a, \mnras, 338, 227
\bibitem[Rice et al., 2003b]{rice03b}
Rice, W. K. M., Armitage, P. J., Bonnell, I. A., Bate, M. R., Jeffers, S. V.,
\& Vine, S. G. 2003b, \mnras, 346, L36
\bibitem[Rivera et al. 2005]{rivera05}
Rivera, E., Lissauer, J. J., Butler, R. P., Marcy, G. W.,
Vogt, S. S., Fischer, D. A., Brown, T., Laughlin, G., \&
Henry, G. W. 2005, \apj, 634, 625
\bibitem[Robin \& Cr\'ez\'e, 1986]{robin}
Robin, A. C., \& Cr\'ez\'e, M. 1986, \aap, 157, 71
\bibitem[Santos et al., 2004]{santos04}
Santos, N. C., Israelian, G., \& Mayor, M. 2004, \aap, 415, 1153
\bibitem[Sato et al., 2003]{sato03}
Sato, B., et al. 2003, \apj, 597, L157
\bibitem[Seager et al., 2005]{seager05}
Seager, S., Turner, E. L., Schafer, J., \& Ford, E. B. 2005, Astrobiology, 5, 372
\bibitem[Selsis et al., 2007]{selsis07}
Selsis, F., Kasting, J. F., Levrard, B., Paillet, J., Ribas, I., \& 
Delfosse, X. 2007, \aap, 476, 1373
\bibitem[Setiawan et al., 2005]{setiawan05}
Setiawan, J., Rodmann, J., da Silva, L., Hatzes, A. P., 
Pasquini, L., von der L\"uhe, O., de Medeiros, J. R., D\"ollinger, M. P., 
\& Girardi, L. 2005, \aap, 437, L31
\bibitem[Setiawan et al., 2007a]{setiawan07a}
Setiawan, J., Weise, P., Henning, Th., Launhardt, R., Müller, A., \& Rodmann, J. 2007a, 
\apj, 660, L145
\bibitem[Setiawan et al., 2008]{setiawan08}
Setiawan, J., et al. 2008, in Proceedings of the ESO Workshop 
``Precision Spectroscopy in Astrophysics'', eds. L. Pasquini, M. Romaniello, N.C. Santos, \& A. Correia, 201
\bibitem[Sozzetti et al., 2001]{sozzetti01}
Sozzetti, A., Casertano, S., Lattanzi, M. G., \& Spagna A. 2001, \aap, 373, L21
\bibitem[Sozzetti et al., 2002]{sozzetti02}
Sozzetti, A., Casertano, S., Brown, R. A., \& Lattanzi, M. G. 2002, \pasp, 114, 1173
\bibitem[Sozzetti et al., 2003a]{sozzetti03a}
Sozzetti, A., Casertano, S., Lattanzi, M. G., \& Spagna A. 2003a,
in Towards Other Earths: DARWIN/TPF and the Search for Extrasolar Terrestrial Planets,
eds. M. Fridlund \& T. Henning, ESA SP-539, 605
\bibitem[Sozzetti et al., 2003b]{sozzetti03b}
Sozzetti, A., Casertano, S., Brown, R. A., \& Lattanzi, M. G. 2003b, \pasp, 115, 1072
\bibitem[Sozzetti, 2005]{sozz05}
Sozzetti, A. 2005, \pasp, 117, 1021
\bibitem[Sozzetti et al., 2006]{sozz06}
Sozzetti, A., Torres, G., Latham, D. W., Carney, B. W., Stefanik, R. P., 
Boss, A. P., Laird, J. B., \& Korzennik, S. G. 2006, \apj, 649, 428
\bibitem[Sudarsky et al., 2005]{sudarsky05}
Sudarsky, D., Burrows, A., Hubeny, I., \& Li, A. 2005, \apj, 627, 520
\bibitem[Tabachnik \& Tremaine, 2002]{tabachnik02}
Tabachnik, S., \& Tremaine, S. 2002, \mnras, 335, 151
\bibitem[Takeuchi et al., 2005]{takeuchi05}
Takeuchi, T., Velusamy, T., \& Lin, D. N. C. 2005, \apj, 618, 987
\bibitem[Tinetti et al., 2007]{tinetti07}
Tinetti, G., Rashby, S., \& Yung, Y. L. 2007, \apj, 644, L129
\bibitem[Thommes \& Lissauer, 2003]{thommes03}
Thommes, E. W., \& Lissauer, J. J. 2003, \apj, 597,566
\bibitem[Udalski et al., 2005]{udalski05}
Udalski, A., et al. 2005, \apj, 628, L109
\bibitem[Udalski et al., 2007]{udalski07}
Udalski, A., et al. 2007, \aap, submitted (arXiv:0711.3978)
\bibitem[Udry et al., 2007]{udry07}
Udry, S., et al., 2007, \aap, 469, L43
\bibitem[Unwin et al., 2007]{unwin07}
Unwin, S. C., et al. 2007, \pasp, submitted (arXiv:0708.3953)
\bibitem[Vecchiato et al., 2003]{vecchiato03}
Vecchiato, A., Lattanzi, M. G., Bucciarelli, B.,
Crosta, M., de Felice, F., \& Gai, M. 2003, \aap, 399, 337
\bibitem[von Bloh et al., 2007]{vonbloh07}
von Bloh, W., Bounama, C., Cuntz, M., \& Franck, S. 2007, \aap, 476, 1365
\bibitem[Wright et al., 2007]{wright07}
Wright, J. T., et al. 2007, \apj, 657, 533
\bibitem[Zuckermann \& Song, 2004]{zuckermann04}
Zuckerman, B., \& Song, I. 2004, \araa, 42, 685
\end{thebibliography}
\end{document}